\documentclass[bibtotoc,12pt,BCOR1cm]{scrbook}
\usepackage{graphicx}
\usepackage{amssymb}
\usepackage{natbib}
\bibpunct{(}{)}{;}{a}{}{,}
\usepackage[onefloatperpage]{plates}  %NB che ho modificato
\usepackage{hyperref}
\usepackage{rotating}

\usepackage{afterpage}
\newcommand{\OddPageProcessPlates}{%
  \afterpage{\clearpage%
           \ifthenelse{\isodd{\value{page}}}%
               {\afterpage{\ProcessPlates}}%
               {\ProcessPlates}}%
}  % pare funzionare cmq

\setcapindent{0pt}

\newcounter{stampafigureacolori}
\newcommand{\MyProcessPlates}{\ifthenelse{\value{stampafigureacolori} = 1}{%
   \OddPageProcessPlates}{}}

\newcounter{altaris}
\def\degr{\hbox{$^\circ$}}
\def\arcmin{\hbox{$^\prime$}}
\def\arcsec{\hbox{$^{\prime\prime}$}}
\DeclareRobustCommand{\ion}[2]{%
\relax\ifmmode
\ifx\testbx\f@series
{\mathbf{#1\,\mathsc{#2}}}\else
{\mathrm{#1\,\mathsc{#2}}}\fi
\else\textup{#1\,{\mdseries\textsc{#2}}}%
\fi}
\newcommand{\object}[1]{#1}
\def\S{Sect.}
\def\aj{AJ}%
\def\araa{ARA\&A}%
\def\apj{ApJ}%
\def\apjl{ApJ}%
\def\apjs{ApJS}%
\def\apss{Ap\&SS}%
\def\aap{A\&A}%
\def\aaps{A\&AS}%
\def\mnras{MNRAS}%
\newcommand{\Hii}{\ion{H}{ii} }
\newcommand{\Log}{\mbox{Log}}
\newcommand{\XMM}{XMM-{\em Newton}}
\newcommand{\xmm}{XMM-{\em Newton}}
\newcommand{\chandra}{{\em Chandra}}
\newcommand{\spitzer}{{\em Spitzer}}

\newcommand{\lognlogs}{Log~$N$--Log~$S$ }
\newcommand{\lognlogsa}{Log~$N$--Log~$S$}
\newcommand{\sun}{\odot}

\newcommand{\FIR}{{\rm FIR} }
\newcommand{\SFR}{{\rm SFR} }
\newcommand{\LX}{L_{\rm X} }
\newcommand{\Lsx}{L_{0.5-2} }
\newcommand{\Lhx}{L_{2-10} }
\newcommand{\de}{{\rm d}}
\newcommand{\ergscmq}{erg s$^{-1}$ cm$^{-2}$}
\newcommand{\e}[1]{\cdot 10^{#1}}

\begin{document}

\title{ Emissione di alta energia da galassie starburst \\
\em (High energy emission from starburst galaxies) }

\author{Ph.D. thesis by Piero Ranalli\\
\\
\\
 Universit\`a di Bologna,\\
  Dipartimento di Astronomia,\\
  via Ranzani 1, I--40127 Bologna, Italy\\
 piero.ranalli@bo.astro.it\\
\\
\\
Supervisors:\\
Prof. Giancarlo Setti\\
Dr. Andrea Comastri
}
%\and 
%  INAF -- Osservatorio Astronomico di Bologna,
%  via Ranzani 1, I--40127 Bologna, Italy
%}

\date{Discussed on 1/4/2004}

%% {\bf ABSTRACT}\\
%%  Radio and far infrared luminosities of star-forming
%% galaxies follow a tight linear relation. Making use of ASCA and
%% BeppoSAX observations of a well-defined sample of nearby star-forming
%% galaxies, we argue that tight linear relations hold between the X-ray,
%% radio and far infrared luminosities. The effect of intrinsic
%% absorption is investigated taking NGC\,3256 as a test case.  It is
%% suggested that the hard X-ray emission is directly related to the Star
%% Formation Rate.       Star formation processes may also account for most of
%% the 2-10 keV emission from LLAGNs of lower X-ray luminosities (for the
%% same FIR and radio luminosity).   Deep {\em Chandra} observations of a
%% sample of radio-selected star-forming galaxies in the Hubble Deep
%% Field North show that the same relation holds also at high
%% ($0.2\lesssim z\lesssim 1.3$) redshift. The X-ray/radio relations also
%% allow a derivation of X-ray number counts up to very faint fluxes from
%% the radio \lognlogsa, which is consistent with current limits and
%% models. Thus the contribution of star-forming galaxies to the X-ray
%% background can be estimated.

%\keywords{X-rays: galaxies -- radio continuum: galaxies --
%galaxies: high-redshift -- infrared: galaxies --
%%missions: ASCA, {\em Chandra} \ }
%galaxies: fundamental parameters -- galaxies: starburst }

%}

\maketitle 

\tableofcontents
\newpage

\addchap{Preface}

The star formation (SF) can be regarded as an engine powering the light
emission of spiral galaxies at several wavelengths.  Massive, young
stars emit the bulk of their light in the UV band; a large fraction of the
UV radiation is absorbed by cold ($T\sim 40$K) dust and reradiated in
the far infrared band. Massive stars explode as supernovae, and
supernova remnants accelerate the electrons which in turn power the
radio emission through the synchrotron mechanism. Binary systems
composed by a collapsed star with a massive companion, emit X-rays
through Comptonization of thermal photons coming from the accretion
disc. 

The star formation activity may go on for very long times at low
paces, or be significantly enhanced for short times as conquence
of perturbations in the gravity potential, which trigger the collapse
of gas clouds. Mergers, or cannibalization of small galaxies, are
often the reason for episodes of enhanced star formation.

The intensity of an episode of star formation can be classified on the
basis of its sustainability for long times. If we define the Star
Formation Rate (SFR, usually expressed in $M_\sun/$yr) as the mass of
gas converted into stars per unit time, we may also consider the
depletion time $\tau_\de = M_{\rm gas}/$SFR which measures the time
needed to convert into stars all the mass $M_{\rm gas}$ of gas present
in a galaxy if the star formation continues at a constant rate.  If the
depletion time is very short (much shorter than, say, a Hubble time)
we may call the present SF episode a {\em starburst}  episode.
The galaxies with an undergoing starburst episode can be called {\em
  starburst galaxies}. Note, however, that in the literature the
term ``starburst galaxy'' is also found to be loosely referred to
large galaxies with quiescent SF but large SFR values.

The star formation activity plays a significant role in the evolution
of galaxies. The production of heavy chemical elements results both in
the build up of different stellar populations and in the pollution of
the insterstellar medium (ISM). Intense starburst episodes also may form
bubbles of hot ionized gas, with a pressure larger than that of the
surrounding cold ISM. The bubbles may be able to expand in a direction
normal to the plane of the galaxy disc, eventually releasing their gas
content into the intergalactic medium (IGM). Depending on the strength
of gravitational potential of the galaxy, the gas may definitely
escape into the IGM or fall back (a {\em galactic fountain} is then
formed).

Star formation has also undergone a strong cosmological evolution,
since the average density of SFR was about ten times larger at
redshifts larger than 1, and has been declining since then
\citep{lilly96,madau96}. Thus the study of galaxies in the local
universe with a strong star formation activity may act as a guide to
understand the evolution of galaxies in the early universe.

In the following, we focus on the study of `normal', spiral, star
forming galaxies. The term `normal' will be referred to galaxies whose
energy output is dominated by all the processes related with star
formation and evolution, with no contribution from a possible active
galactic nucleus.

The first chapter is devoted to a review of the infrared, radio and
X-ray emission from star forming galaxies. In the second chapter we
explore the quantitative relationships between the total X-ray, radio
and infrared luminosities for a sample of 23 star forming galaxies. It
is found that linear correlations hold between the luminosities in the
above bands. In the third chapter we show, by analyzing a sample of 11
high redshift galaxies in the Hubble Deep Field, that the linear
correlations may be extended up to $z\sim 1$. In the fourth chapter we
turn to an analysis of the luminosity function and number counts of
star forming galaxies; we show that the number density of normal
galaxies at faint X-ray fluxes ($10^{-17}$--$10^{-15}$ \ergscmq) is very well
defined. % both by extrapolating the local ($z=0$) radio and far infrared
%luminosity functions an by considering 

The fifth chapter offers a first insight into a different aspect of
galaxy evolution, i.e.\ whether is it possible to determine the
metallicity enhancements in a starburst episode. An interesting, yet
problematic, picture of metal enrichment of the interstellar matter
arises from an analysis of the X-ray and near infrared spectra of M82.
Finally, a summary is offered of the work here described.

\chapter{Light emission from star forming galaxies}

%\dropping{3}{T}%
Two main stellar population are usually found in spiral galaxies: one
made up by old dwarf stars, and one made up by young, blue, massive
stars. While the former builds the bulk of the galaxy stellar mass and
of the energetic output in the optical band, the latter powers the
light emission in many different bands throughout the whole energy
spectrum. In the following, it will be shown how the processes of star
formation and death can be invoked to explain the light emission
properties of galaxies at radio, far infrared and X-ray wavelengths.

\section{Infrared emission from spiral galaxies}
\label{sec:introUV-FIR-Ha}

The far infrared luminosity ($\sim 3$--300$\mu$) of a star forming
galaxy is commonly interpreted as due to starlight reradiation by dust
grains.  The early evidence for a link between infrared emission and
dust was based on several arguments \citep{hp75,telesco80,devyoung}:
\begin{itemize}
\item  \Hii regions are associated with knots of infrared emission;
\item the far infrared luminosities may exceed the optical
  luminosities, consistent with the view that the FIR radiation is
  generated by heavily obscured star clusters;
\item the high luminosity-to-mass ratios derived from the comparison
  of FIR luminosities and dynamical masses, consistent with the view
  that the obscured star clusers are composed primarily of young
  massive stars;
\item the FIR luminosities are well correlated with masses of the
  molecular gas, which is the component of the interstellar medium
  from which stars form.
\end{itemize}
The observations made with the {\em Infrared Astronomical Satellite}
(IRAS) showed that infrared emission is ubiquitous among spiral
galaxies, thus reinforcing the view that the far infrared luminosity
of spiral galaxies is thermal dust emission. A comparison of the
H$\alpha$ and FIR luminosities in a sample of 124 galaxies
\citep{devyoung} showed that the mean ratio of extinction-corrected
H$\alpha$ to bolometric FIR luminosity in spiral galaxies is
comparable to that expected for \Hii regions powered by massive stars.
This led \citet{devyoung} to support the view that high mass stars
required to ionize the hydrogen gas can easily generate the far
infrared luminosities.

However, the presence of galaxies with a low level of infrared
emission that is energetically comparable to the optical luminosity
also led \citet{dejong84} to suggest later type stars as a significant
dust heating source, particularly in spirals with a cool
$S_{60\mu}/S_{100\mu}$ dust temperature. It was also suggested
\citep{lonsdale87,dejong87} that the emission at $60\mu$ and $100\mu$
measured by IRAS could be deconvolved in two components: a warm
component with temperatures of 50--60K, and a cool component (the {\em
  cirrus}) at 10--20K. An interpretation of the two-component model
which considered the warm component as radiation from dust heated by
young stars, while the cool component is heated by older, late type
stars, was considered as the possible explanation for a slight
non-linearity of the radio/FIR correlation (\citealt{fitt88};
however, the latest studies point toward almost exact linearity,
\citealt{yrc}; see \S~\ref{sec:introRadioFIR}).

\begin{figure*}[tp]    % NORMAN BAYESIAN - z DISTRIBUTION
  \begin{minipage}[t]{.49\textwidth}
    \centering
      \includegraphics[width=\textwidth]{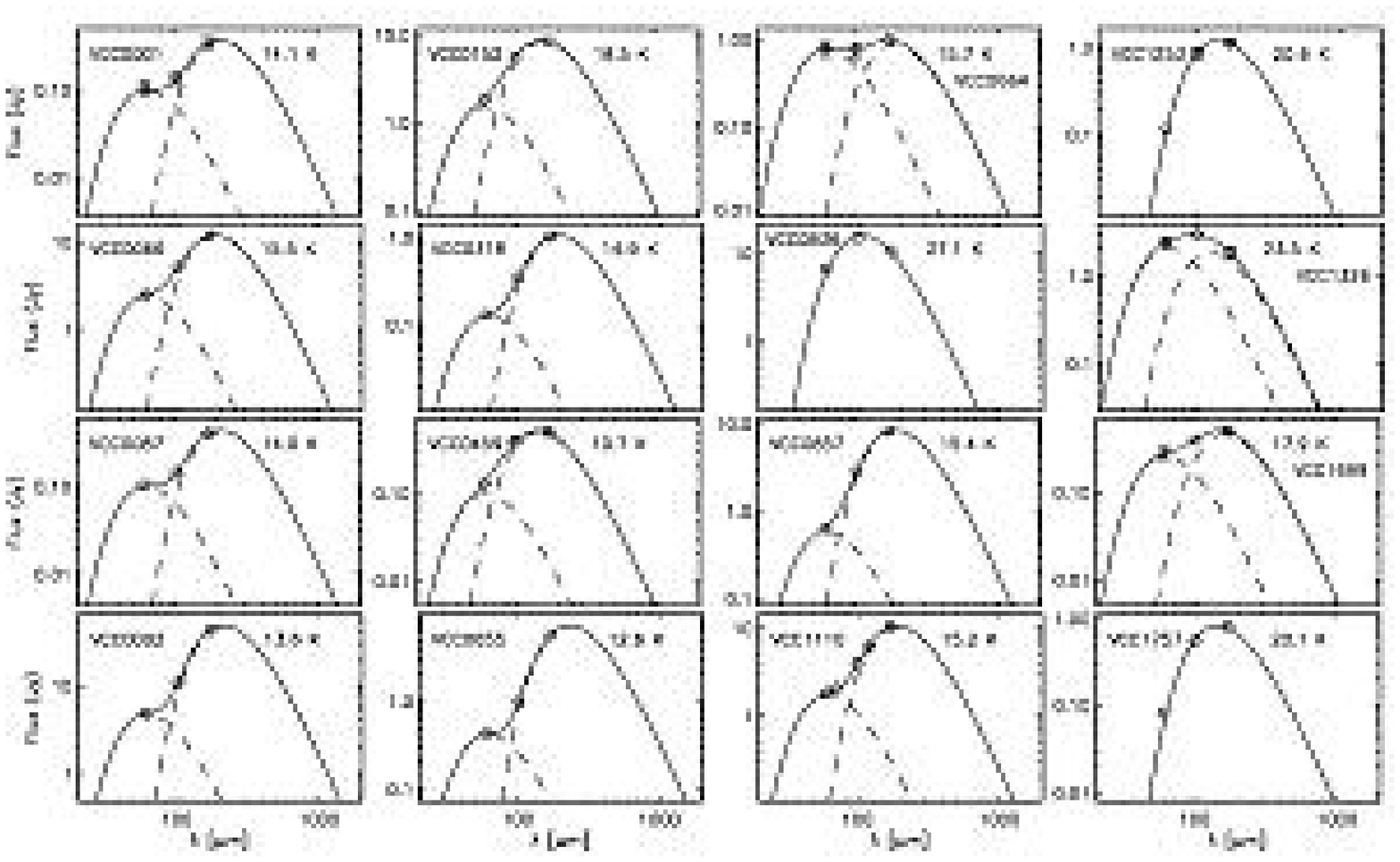}  
   \caption { Far infrared spectral energy distributions for a sample
      of spiral galaxies in Virgo, from \citet{popescu02}. The 
      points are ISOPHOT data, while the lines show the
      two-temperature model fit to the data (see text).
   \label{fi:twotempisophot} }
  \end{minipage}
  \hfill
  \begin{minipage}[t]{.49\textwidth}
    \centering
      \includegraphics[height=0.25\textheight]{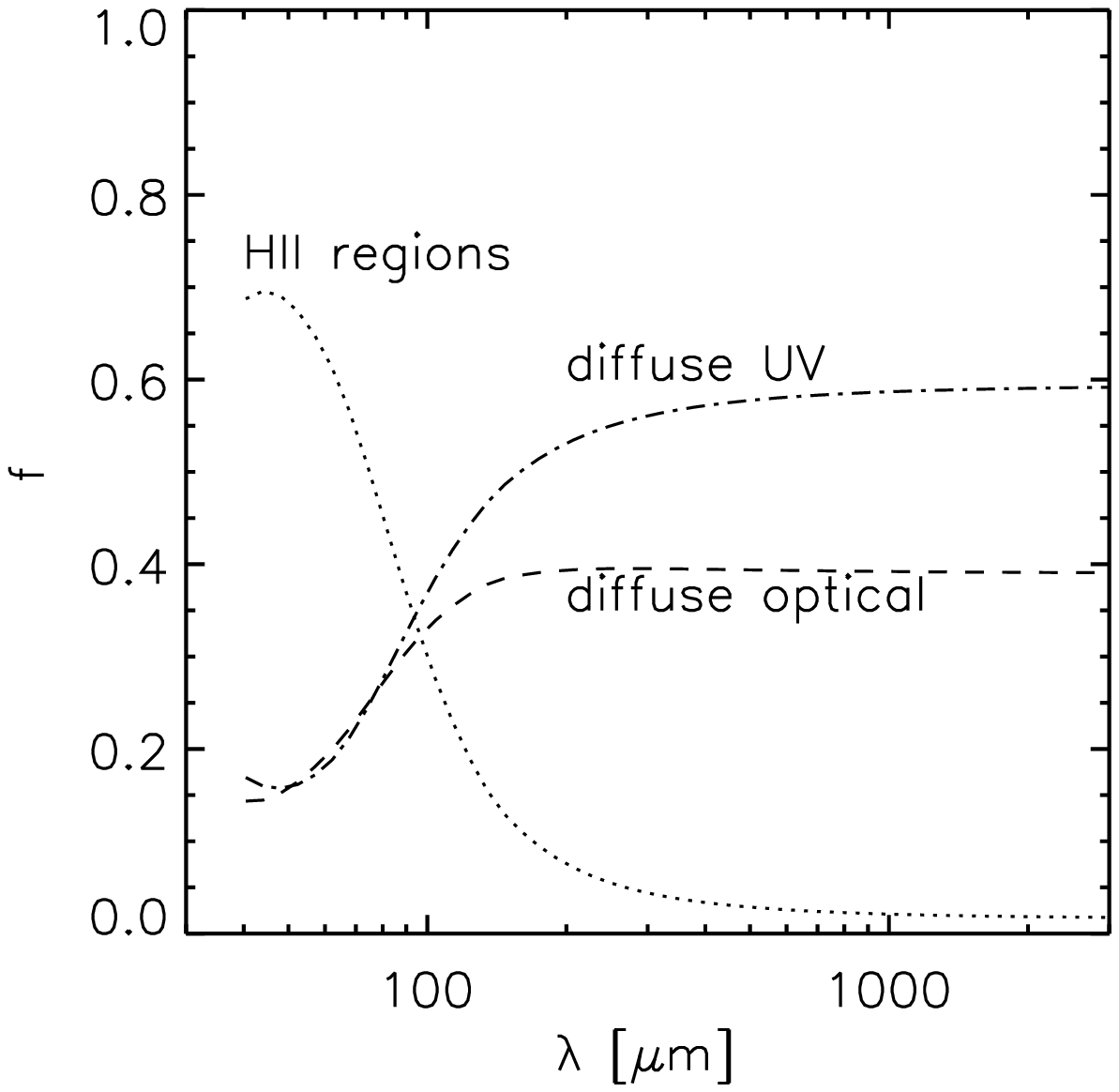}  
   \caption { Fractional contribution of ultraviolet light from
     different sources (\Hii regions and the diffuse old stellar
     component) to dust heating for the spiral galaxy NGC 891, from
     \citet{tuffspopescu02}.
   \label{fi:ngc891heating} }
  \end{minipage}
\end{figure*}

\begin{figure}[tp]
  \centering
  \ifthenelse{\value{altaris} = 1}{%
      \includegraphics[width=0.85\textwidth]{TuffsRJ1_2bw.eps}  
  }{  \includegraphics[width=0.85\textwidth]{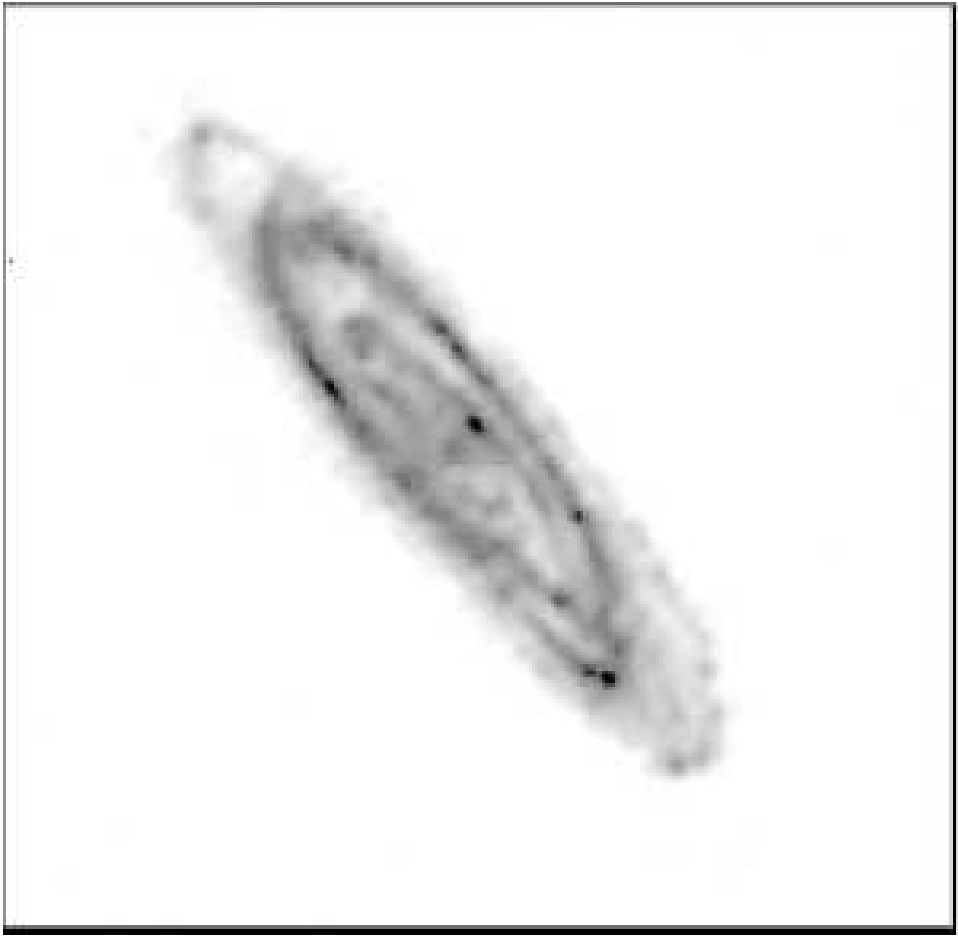}  
  }
  \caption{ISOPHOT mapping of the M31 galaxy at 170$\mu$, from
      \citet{haas98}. The darker areas show the emission due to the
      cold ($\sim 20$~K) dust. The angular resolution is $1.3\arcmin$, the
      field size $2.9\degr\times 2.9\degr$.}
  \label{fig:m31isophot}
\end{figure}

Since the peak in the thermal spectrum of dust with temperatures below
$\sim 30$K falls longwards of 100$\mu$ ---thus outside the IRAS
bands---, a clear separation of the warm and cool components was only
possible with the launch of the {\em Infrared Space Observatory}
(ISO), whose spectral coverage extended up to wavelengths of $\sim
180\mu$.  \citet{popescu02} studied a sample of 38 galaxies with the
ISOPHOT instrument at 60, 100 and 170$\mu$. Most of these galaxies
were discovered to contain a cold dust emission component which could
not have been detected by IRAS. The FIR Spectral Energy Distributions
(SEDs) of these objects were fitted with a combination of two modified
blackbody functions, one physically identified with the emissions from
a localised warm dust component, with a fixed temperature of 47K, and
one other with a diffuse cold dust component
(Figs.~\ref{fi:twotempisophot}, \ref{fi:ngc891heating}). The cold dust
temperatures were found to span a broad range, with a median of 18K.

Imaging of the cold component has been so far possible, with ISOPHOT,
only for very extended objects such as M31
(Fig.~\ref{fig:m31isophot}).  Detailed studies of the FIR emission
will be routinely possible in the immediate future with the \spitzer\ 
Space Telescope. Among the first images taken with this new
observatory, the pictures of M81 at different wavelengths
(Plates~\ref{plate:m81multiwave},\ref{plate:m81longw}) show the
details in the spiral structure of the cold dust at 170$\mu$.

\section{Radio emission}
\label{sec:introRadio}

\begin{figure}[t]
  \centering
  \ifthenelse{\value{altaris} = 1}{%
      \includegraphics[width=0.98\textwidth]{radiomorphs.eps}
  }{  \includegraphics[width=0.98\textwidth]{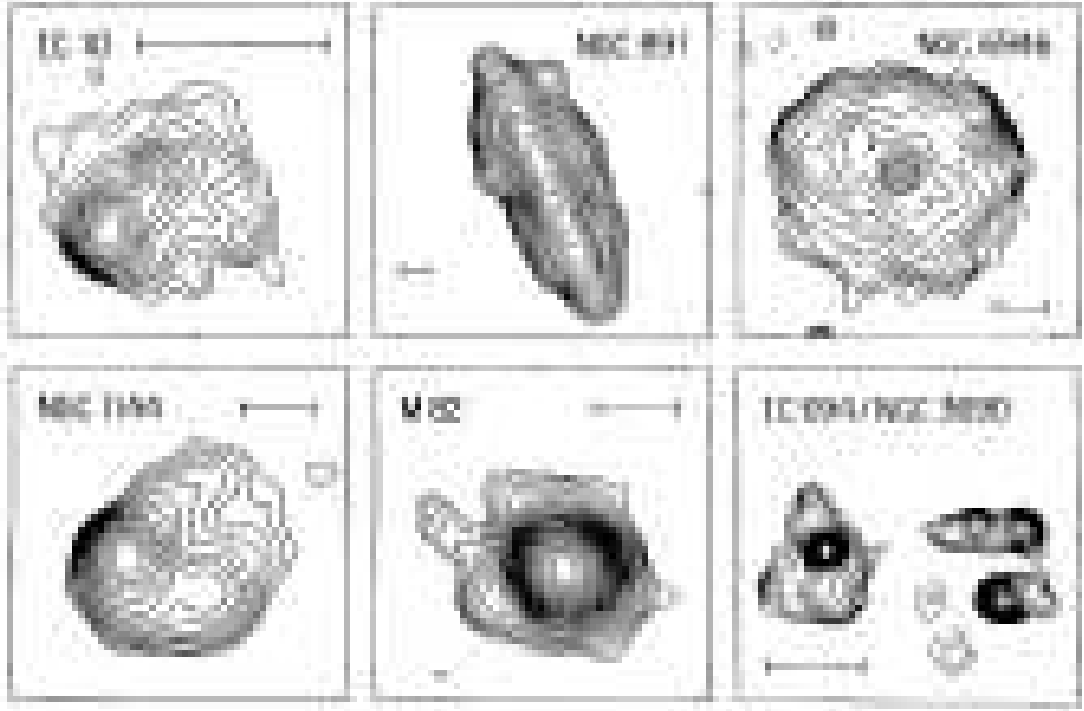}
  }
  \caption{Contour maps illustrating the range of radio source
    morphologies, sizes, and luminosities found in normal galaxies,
    from \citet{cond92}. The bars are 2$h^{-1}$ kpc long. The
    logarithmic contours are separated by $\sqrt{2}$ in brightness, and
    the 1.49 GHz brightness temperatures $T_{\rm b}$ of the lowest
    contours are 0.25K (IC10, NGC891, NGC6946), 0.5K (M82),
    8K(NGC1144), and 128K (IC694+NGC3690).}
  \label{fig:radiomorphs}
\end{figure}

Radio sources exist in nearly all normal spiral and dwarf irregular
galaxies, and in many peculiar or interacting systems. The morphology
of the radio emission found in normal galaxies is various \citep[][and
references therein]{cond92}. The faintest dwarf irregulars have radio
luminosities comparable with the Galactic Supernova Remnant (SNR)
Cassiopea A and little or no detectable emission extending beyond
known \Hii regions and SNRs. Their radio morphologies are lumpy and
irregular (see IC\,10 in Fig.~\ref{fig:radiomorphs}), and their radio
spectra are often relatively flat. The thick radio disk/halo of the
edge-on galaxy NGC891 and the fairly smooth radio disk with bright
spiral arms of the face-on galaxy NGC6946 are typical of the larger
spiral galaxies. The central radio sources in normal galaxies like
NGC6946 have a larger brightness but are usually much less luminous
than the disk-like sources. Luminous radio sources with complex
morphologies may be found in colliding galaxies (e.g.\ NGC1144). There
is a tendency for fairly compact (diameter $D\lesssim 1$ kpc) central
starbursts to dominate at higher radio luminosities, as in M82. The
most luminous radio sources in normal galaxies are frequently quite
compact ($D\sim 200$ pc), and confined to the nuclei of strongly
interacting systems (e.g.\ IC694+NGC3690). Numerical simulations show
that a collision involving a disk galaxy can drive about half of the
disk gas ($\lesssim 10^{10} M_\sun$) within 200 pc of the nucleus
\citep{hernquist89,barnes91} where the gas density becomes quite high
before a powerful starburst is triggered \citep{kennicutt89}. The
resulting massive stars and their SNRs then produce intense radio
emission.

\begin{figure}[p]
  \centering
  \ifthenelse{\value{altaris} = 1}{%
      \includegraphics[width=0.98\textwidth]{m82radio_hargrave.eps}\\
      \includegraphics[width=0.98\textwidth]{m82radio_muxlow.eps}
  }{  \includegraphics[width=0.98\textwidth]{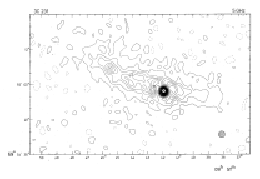}\\
      \includegraphics[width=0.98\textwidth]{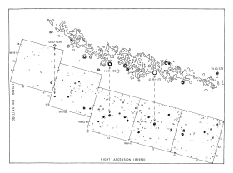}
  }
  \caption{Radio emission from M82. Upper panel: medium resolution image
      at 5GHz (Cambridge 5km radio telescope), from
      \citet{hargrave74}. Lower panel: high resolution \textit{Merlin}
      image at 5GHz, from \citet{muxlow94}. As normal with
      interferometer images, the lower the resolution, the higher the
      sensitivity to diffuse emission.}
  \label{fig:m82radio}
\end{figure}

\begin{figure}[t]
  \centering
      \includegraphics[width=.8\textwidth]{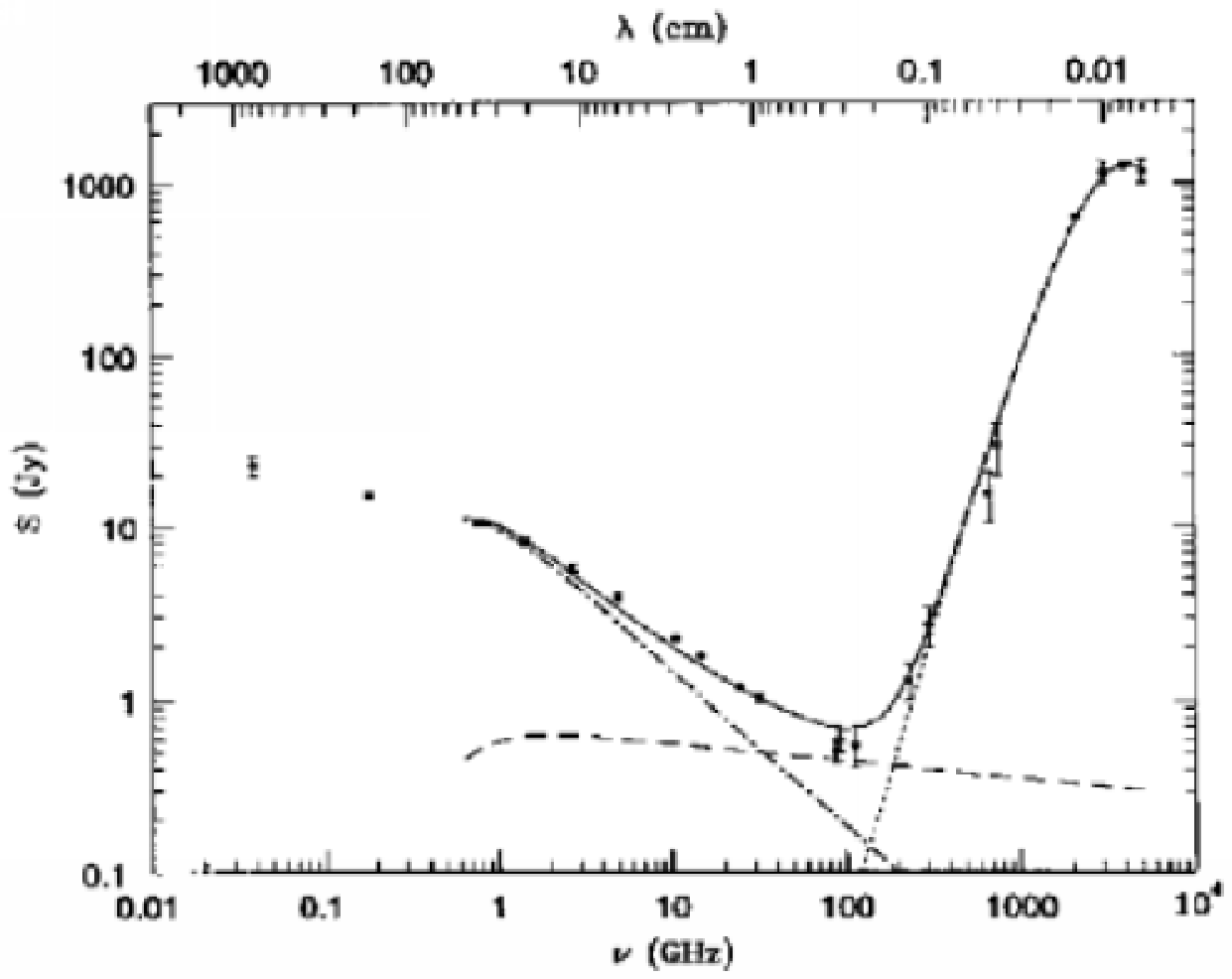}  
  \caption{Radio-FIR spectrum of M82, from \citet{cond92}. Dashed
      line: thermal contribution to the spectrum from
      the \Hii regions. Dot-dashed line: nonthermal
      emission. Dotted line: black body emission from the
      dust. The continous line shows the fitted spectrum resulting
      from the sum of the previous three components.}
  \label{fig:m82radiofir}
\end{figure}

Nearly all of the radio emission from normal galaxies is synchrotron
radiation from relativistic electrons and free-free emission from \Hii
regions \citep{cond92}. Thermal reradiation of starlight by dust
quickly overwhelms these components above $\nu\sim 200$GHz, defining a
practical separation between the radio and infrared bands. Typical
intensities of synchrotron radiation, free-free emission, and dust
reradiation are shown in the radio/FIR spectrum of M82
(Fig.~\ref{fig:m82radiofir}); the radio continuum only accounts for
$\lesssim 10^{-4}$ of its bolometric luminosity.

However, determining the amount of the thermal fraction is highly
controversial. In principle, the relatively flat spectrum
($S(\nu)\propto\nu^{-\alpha}$ with $\alpha\lesssim 0.1$) thermal
emission should be distinguishable from the steeper spectrum
($\alpha\sim 0.8$) nonthermal emission via total flux densities or
maps obtained at two or more frequencies. In practice, most normal
galaxies are not bright enough to be detected at frequencies much
higher than $\nu\sim 10$ GHz ($\lambda\sim 3$ cm).  Moreover, if
integrated flux densities at only three or four frequencies are used
to fit the (unknown) nonthermal spectral index and the thermal
emission simultaneously, the resulting nonthermal spectral index and
thermal fraction are strongly correlated. An upper limit $S_{\rm
  T}/S\lesssim 0.4$ to the thermal-to-total flux ratio was estimated
by \citet{klein81} and \citet{gioia82} from the lack of a flattening
of the radio spectrum in two samples of nearby spiral galaxies. The
thermal fraction is highly uncertain even in the brightest and best
observed galaxies such as M82 because the nonthermal component may not
have a straight spectrum. By assuming a power law spectrum for M82,
\citet{klein88} calculated a thermal flux density $S_{\rm T}=0.15$ Jy
at $\nu=32$~GHz. Using practically the same data but assuming that the
nonthermal spectrum steepens at high frequencies, \citet{carlstrom91}
obtained $S_{\rm T}=0.5$ Jy (and thus a thermal fraction of $\sim
70\%$, cfr.\ Fig.~\ref{fig:m82radiofir}) at $\nu=92$~GHz, where
$S_{\rm T}=0.17$~Jy (thermal fraction $\sim 25\%$) would be expected
by extrapolating Klein's estimate at 92GHz with a typical slope
$\alpha=0.1$. This simple consideration shows that the estimates of
the thermal fraction at high frequencies may have un uncertainty of
(at least, since M82 is among the best studied galaxies) a factor of
three.

Since the nonthermal spectrum has a steeper slope, the
uncertainty on the thermal fraction becomes a minor issue at lower
frequencies (e.g.\ at 1.4 GHz), where the majority of the radio
emission is of nonthermal origin (cfr.\ Fig.~\ref{fig:m82radiofir}).
The nonthermal radiation is usually explained as synchrotron radiation
from relativistic electrons.  Only stars more massive than $\sim
8M_\sun$ produce the core-collapse supernovae whose remnants (SNRs)
are thought to accelerate most of the relativistic electrons in normal
galaxies; these stars ionize the \Hii regions as well. Their lifetime
is $\lesssim 3\e7$ yr, while the relativistic electrons probably have
lifetimes $\lesssim 10^8$ yr.  Radio observations are therefore probes
of very recent star formation activity in normal galaxies.

Supernova remnants become radio sources about 50 years after the
explosion, as Rayleigh-Taylor instabilities develop in the boundary
between the shock and the ambient interstellar medium, and remain
visibile for hundreds or thousands of years. About 40 young SNRs are
conspicuos in high resolution maps of M82 at 5 GHz
\citep[][Fig.~\ref{fig:m82radio}]{kronberg85,muxlow94}.  Although SNRs
are probably responsible for cosmic ray acceleration, discrete
remnants themselves emit only a small fraction ($\lesssim 10\%$) of
the integrated flux \citep{pooley69,ilovaisky72,biermann76,helou}.
This also means that $\gtrsim 90\%$ of the nonthermal emission must be
produced long after the individual supernova remnants have faded out.
Most of the nonthermal emission is so smoothed by cosmic ray transport
that the spatial distribution of its sources cannot be deduced in
detail. Thus a various number of other possible sources for cosmic ray
acceleration has been proposed, but none of them was really convincing
\citep[a review is in][]{cond92}. A five-band (radio, FIR, near
infrared, blue, and X-ray) study of luminosity correlations in spirals
led \citet{fgt88} to conclude that the radio emission from starbursts
must originate in the young stellar population but no clear conclusion
can be drawn from the luminosity correlations in less active spiral
galaxies.

\section{The radio/FIR correlation}
\label{sec:introRadioFIR}

A correlation between the $\lambda=10\mu$ mid infrared and 1415 MHz
radio luminosities of Seyfert galaxies was discovered by \citet{vdK71}
and soon extended to normal spiral galaxies \citep{vdK73}. At
first both the infrared and radio emission were thought to be
synchrotron radiation from relativistic electrons accelerated by
nuclear monsters (e.g.\ massive black holes in Seyfert galaxies or
other AGN). Then \citet{hp75} proposed that the infrared is thermal
reradiation from dusty \Hii regions, while the 1415 MHz is dominated
by synchrotron radiation from relativistic electrons accelerated in
SNRs from the same population of massive stars that heat and ionize
the \Hii regions. This is still the current interpretation (cfr.\
\S~\ref{sec:introUV-FIR-Ha} and \ref{sec:introRadio}).

\begin{figure}[t]
  \centering
  \includegraphics[width=0.5\textwidth,height=0.4\textwidth,angle=-90]{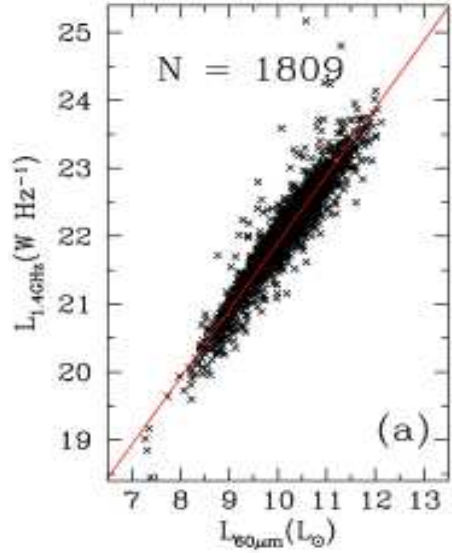}
  \caption{The radio/FIR correlation for a sample of 1809 galaxies
    from the IRAS 2~Jy catalogue, from \citet{yrc}. The correlation
    spans five orders of magnitude in luminosity with a best fit slope
    of $0.99\pm 0.01$.}
  \label{fig:radioFIR1809gal}
\end{figure}

The real significance of the FIR/radio correlation for normal galaxies
---its tightness and universality--- was not appreciated until the
large IRAS survey appeared \citep{dickey84,dejo85,helou}. The
correlation has a linear slope and spans $\sim 5$ orders of magnitude
in luminosity with less than 50\% dispersion \citep{yrc,wunder87},
making it one of the tightest correlations known in astronomy
(Fig.~\ref{fig:radioFIR1809gal}). It has been also tested for high
redshift objects, and holds up to $z\sim 1.3$ \citep{garrett02}.  The
correlation not only holds on global scales but is also found to hold
within the disks of galaxies, down to scales of the order of a few
hundreds of pc \citep{beck88}.

What is most extraordinary about this relationship is that its slope
is unity (the most up-to-date result gives $L_{1.4}\propto
L_{60\mu}^{0.99\pm 0.01}$, \citealt{yrc}), although it couples four
different radiative processes: i) thermal emission in the FIR from
warm dust around \Hii regions; ii) thermal emission in the FIR from
cool dust (the `cirrus') associated with dust heated by the old
stellar population; iii) synchrotron emission at radio wavelengths;
iv) thermal free-free emission at radio wavelengths.

Several studies have thus been devoted to investigate the correlation
by focusing on one or more of these processes. \citet{cox88}
correlated the luminosities at 60$\mu$ and 151 MHz (where the
contribution of the free-free emission to the radio luminosity should
be negligible) and found a non-linear slope of $L_\FIR\propto
L_{151}^{0.86\pm 0.03}$.  After deconvolving the FIR luminosity into
two temperature components \citep{fitt88}, the best fits yielded
$L_{\rm FIR, warm}\propto L_{151}^{0.97\pm0.06}$ for the warm
component, and $L_{\rm FIR, cool}\propto L_{151}^{\sim 0.6}$. Thus they
suggested that the warm dust component was correlated with the
nonthermal radio luminosity, while the link between the cool dust
and the synchrotron radiation remained unexplained.

However, decoupling the warm and cool components from IRAS data alone
should be taken with care, because the cool dust becomes visible only
at wavelengths longer than 100$\mu$. The PHOT instrument onboard the
ISO satellite was better suited for this task, because of the
availability of a $170\mu$ channel. \citet{pierini03} thus used PHOT
data taken at 60, 100 and $170\mu$ for a sample of 72 galaxies and
found (note that here `FIR' indicates the bolometric luminosity for
the dust)
\begin{equation}
  L_{1.4}\propto L_\FIR^{1.10\pm 0.03}\qquad
  L_{1.4}\propto L_{\rm FIR, warm}^{1.03\pm 0.03}\qquad
  L_{1.4}\propto L_{\rm FIR, cool}^{1.13\pm 0.04}
\end{equation}
thus confirming the view that the warm dust and the nonthermal radio
luminosity are linearly correlated. Also their best-fit slope for the
cool dust/radio correlation is consistent, within the errors, with
\citet{fitt88}.

%PER I MODELLI COPIARE DA GROVES ET AL.

\subsection*{Theoretical models for the radio/FIR correlation}

One of the earliest theories, the `optically-thick' or `calorimeter'
theory \citep{Volk89,VolkXu94,LisenVX96}, is based on three main
assumptions.  First, that all the far ultraviolet (i.e.\ $\sim
5$--13.6 eV) radiation from massive stars is absorbed by the dust
grains within a galaxy.  Second, that the energetic electrons
produced by the supernova explosions of these stars lose most of their
energy within the galaxy due to synchrotron and inverse Compton
process. As both these processes are proportional to the number of
massive stars, these calorimetric assumptions lead to the linear correlation.
Finally, the tightness of the correlation is provided by the third
assumption that the energy density of the interstellar radiation
field, $U_{\rm rad}$, is in a constant ratio with the magnetic field
energy density, $U_B$.

An alternative theory put forward by \citet{HelBic93} assumes the
opposite extreme, an `optically-thin' model, in which the cosmic rays
and UV photons both have high escape probabilities. To provide the
correlation in these optically- and cosmic ray thin galaxies they rely
on the assumptions that the UV or dust heating photons and
the radio emitting cosmic rays are created in constant proportion to
each other, which is again related to star formation.  Then, to
obtain a linear correlation, it is required by the theory that 
the magnetic field strength and the gas density are coupled.

A challenge to both these theories is the model by \citet{NikBec97}.
In this work they argue that observations indicate that within most
galaxies, the cosmic ray electrons lose very little energy before they
escape.  These same galaxies are optically thick to UV photons, thus
both the calorimetric and optically thin models are not supported by
the observations.  In the \citet{NikBec97} model, the controlling
factor they put forward for the correlation is the gas density: they
assume that both the star formation rate (and thus dust heating) and
the magnetic field strength (which determines the synchrotron
emission) are correlated with the gas volume density.

All these models require a sort of fine tuning between the magnetic
field strength and another parameter (the radiation field or the gas
density). As a way out from the fine tuning hypoteses, two mechanisms
have been proposed. \citet{bressan02} suggested that, in a `calorimeter'
theory, the radio/FIR correlation may arise if the SFR remains contant
for a time longer than the synchrotron-loss timescale. Under this
hypothesis, the nonthermal radio luminosity of a galaxy is
proportional to the integral of the synchrotron power over the
electron lifetime, and an increase of the former in a larger magnetic
field is compensated by a shortening of the latter.

Another mechanism was proposed by \citet{groves03}.  Following the
\citet{NikBec97} model, \citet{groves03} proposed magneto-hydrodynamic
turbulence as a possible mechanism to provide coupling between the gas
density and the magnetic field strength, under the requirement that
the radio and FIR emission is produced in the same volume element.

A `definitive' model has not yet emerged. Further work is needed, as
the current models do not explain the (non-linear) correlation between
the radio emission and the cool dust emission. Also, the possibility
that both the FIR and radio emission increase more than linearly with
increasing SFR has been invoked \citep{bell03}. Possible reasons might
be that low luminosity galaxies have significantly less dust
absorption (hence their infrared emission misses most of the star
formation), and that the thermal fraction of radio emission increases
with the SFR. Thus the radio/FIR correlation would arise as a sort of
conspiracy.

\begin{figure}[p]
  \centering
      \includegraphics[width=0.8\textwidth]{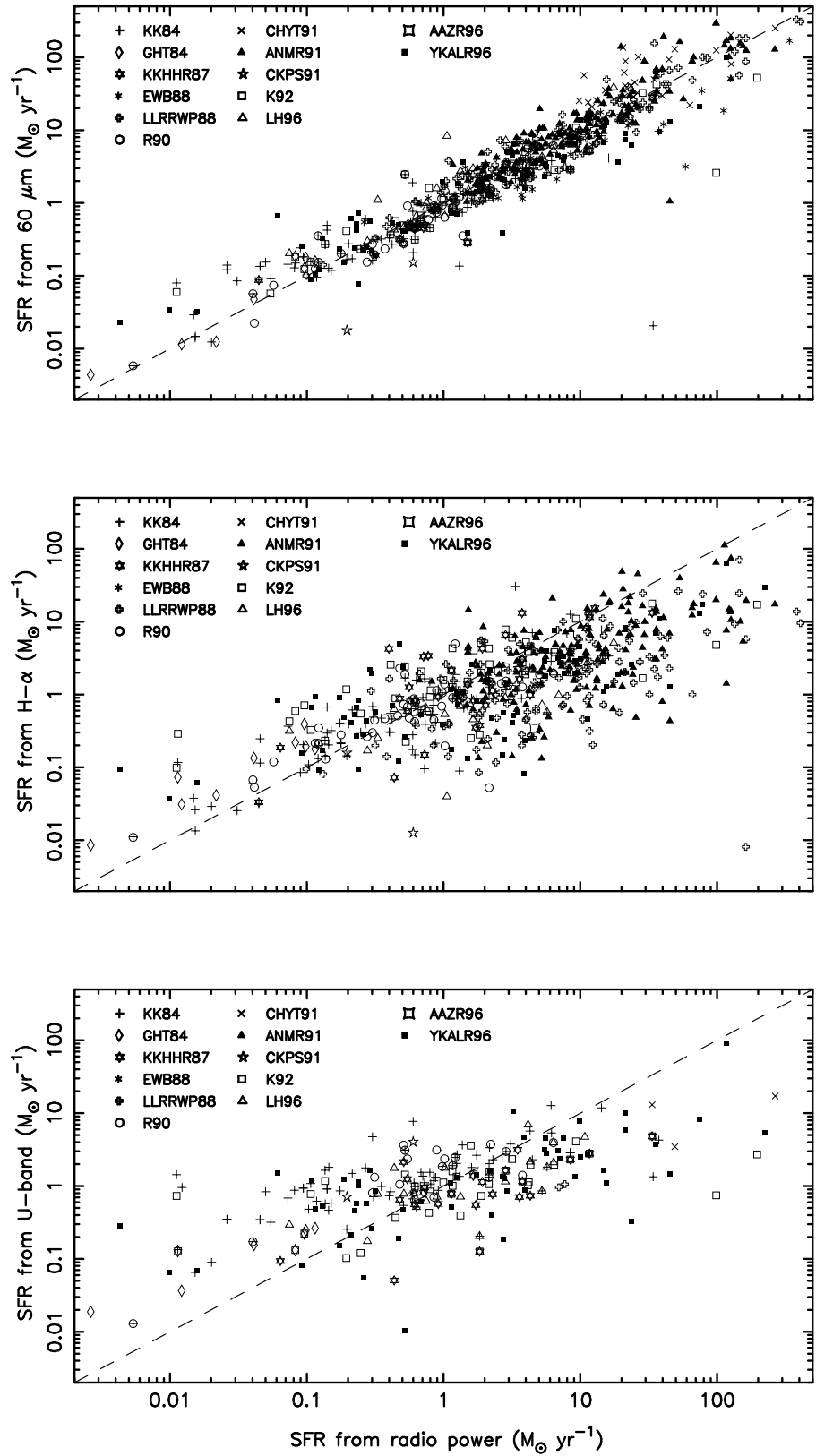}  
  \caption{ A comparison of SFR estimates from radio, infrared,
      H$\alpha$ and ultraviolet emission, from \citet{cram}. The
      radio/FIR correlation (upper panel) has the least scatter. }
  \label{fig:cram}
\end{figure}

\section{SFR indicators}
\label{sec:SFRindicators}

% FIXME - DISCORSO SULLO SCATTER CHE HANNO GLI INDICATORI DI SFR IN
% LETTERATURA? P.E. CRAM CHE È UN FATTORE DUE SOTTO, O BRESSAN CHE SI
% SCOSTA DI UN FATTORE TRE?
% BTW -- SCOPRO CHE C'È UNO SFR(FIR) ANCHE IN CONDON92

Here we will present some very simple considerations, on which the
models reviewed in \S~\ref{sec:introRadioFIR} are based, on how to
link the luminosity and the star formation rate.  If we assume that
the dust surrounding the SF region is optically thick to the UV
radiation emitted by young stars, and there is no other process
leading to far infrared emission, we may write
\begin{equation}
\label{eq:lumsfr}
L_\FIR \propto \int_{M_{\mathrm{l}}}^{M_{\mathrm{u}}}k\Psi (M)\,\tau (M)\,
L(M)\,  \de M
\end{equation}
where $\Psi(M)$ is the (time independent) Initial Mass Function (IMF),
i.e.\ the probability that a star of mass $M$ is formed in an episode
of SF, $k$ is a normalization for the IMF giving the actual number of
stars formed per unit time, and $\tau(M)$ and $L(M)$ are the lifetime
and UV luminosity of a star with mass $M$, respectively. Under the
hypothesis of continous star formation, or if a burst is long enough
to reach a steady state in the mass distribution of stars, then
$k\Psi(M)\tau(M)$ is the number of stars with mass $M$ actually
present. $M_{\rm l}$ may be taken as the lowest possible mass for a
star which emits most of its luminosity in the UV band ($6M_\sun$,
\citealt{devyoung}; however in the following the SFR calibrations will
be given for $M_{\rm l}=5 M_\sun$). $M_{\rm u}$ is upper bound of the
IMF; it is usually assumed $M_{\rm u}=100 M_\sun$ or $120 M_\sun$,
however for a Salpeter IMF ($\Psi(M)\propto M^{-2.35}$) the integral
is not really dependent on this parameter.
The life time for a main sequence star may be written as
\begin{equation}
\tau \propto \frac{q M X}{L}  
\end{equation}
where $q$ is the hydrogen fraction actually burned, and $X$ is the
hydrogen abundance. Thus $L(M)\tau(M)\propto M$. Since we define
\begin{equation}
\label{eq:sfrdef}
\mathrm{SFR}\, (M>M_{\mathrm{i}})=\int_{M_{\mathrm{i}}}^{M_{\mathrm{u}}}\, M\, k\Psi (M)\, dM
\end{equation}
then we get that $L_\FIR \propto \,$SFR.

Thus we may take \citep{cram}
\begin{equation}
  \label{eq:sfr60u}
  \SFR_{60\mu} (M>5M_\sun) = 2.0\e{-31} L_{60\mu}
\end{equation}
with $L_{60\mu}$ in erg s$^{-1}$ Hz$^{-1}$. Another calibration
  \citep{kenn98,cram}
\begin{equation}
\label{eq:SFRFIR}
\mathrm{SFR}_{FIR} (M>5M_\sun)=4.5\e{-44} L_{\mathrm{FIR}}\quad M_{\odot }\, \mathrm{yr}^{-1}
\end{equation}
 makes use of the $FIR$ parameter%\footnote{%
%In his paper, \citet{kenn98} uses a different definition of both
%the IMF and the $FIR$ parameter. Briefly, he takes $M_{\rm l}=0.1
%M_\sun$ and uses the ``extended $FIR$''
%\[ FIR=c(aS_{12.5}+bS_{25}+2.58S_{60\mu }+S_{100\mu })\quad \mathrm{erg}\, \mathrm{s}^{-1}\, \mathrm{cm}^{-2}.
%\] The calibrations shown in Eqs.~(\ref{eq:SFRFIR},\ref{eq:SFRHalfa})
% have been reported to
%our definition of $M_{\rm i}$ using a Salpeter IMF, and of $FIR$
%assuming a modified black body {\bf FIXME}.  }
 \citep{helou}
\begin{equation}
\label{eq:FIRdef}
FIR=1.26\cdot 10^{-11}(2.58S_{60\mu }+S_{100\mu })\quad \mathrm{erg}\, \mathrm{s}^{-1}\, \mathrm{cm}^{-2}
\end{equation}
with $S_{60\mu }$ and $S_{100\mu }$ in Jy. The $FIR$ parameter
estimates the flux which would be measured by a theoretical detector
with flat response between 42.5$\mu$ and 122.5$\mu$
if the underlying spectrum is a modified black body law
\begin{equation}
  \label{eq:greybody}
  B(\nu,T) = \frac{2h\nu^3}{c^2} \frac{1}{e^{\frac{h\nu}{kT}}-1}
  \left(1-e^{-\left(\frac{\nu}{\nu_0}\right)^\beta}\right)
\end{equation}
with $20{\rm K}\lesssim T\lesssim 80$K and $0\lesssim \beta\lesssim 2$.

If, on the other hand, we assumed that the dust and gas are optically
thin for the emission in the ultraviolet band and the H$\alpha$ line,
a similar reasoning might have been made by changing $L_\FIR$ with
$L_{\rm UV}$ or $L_{{\rm H}\alpha}$ in Eq.~(\ref{eq:lumsfr}). Balmer
line emission from star-forming galaxies is formed in \Hii regions
ionized by early-type stars. \citet{kennicutt83} has determined the
theoretical relationship between the H$\alpha$ luminosity and the current rate
of star formation in a galaxy in a form corresponding to 
\begin{equation}
\label{eq:SFRHalfa}
 \SFR_{{\rm H}\alpha} (M>5M_\sun)
=\frac{L_{\mathrm{H}\alpha }}{1.5\cdot 10^{41}\ \rm erg\ s^{-1}
}\quad M_{\odot }\, \mathrm{yr}^{-1}.
\end{equation}

The use of far-UV or U-band luminosities to infer the current SFR
rests on the idea that UV emission contains a substantial contribution
of light from the photospheres of young, massive stars
\citep{cowie97}.  However, the calibration of a UV indicator is less
straightforward than those listed above, because of the large numbers
of relatively old stars that also contribute to the actual U-band
luminosity of disk galaxies. To avoid this problem, \citet{cram} used
the relationship between SFR and far-UV luminosity given by
\citet{cowie97}, scaled from 250 nm to the U-band using relevant
synthesized spectra, to derive
\begin{equation}
\label{eq:SFRUV}
\SFR_{\rm UV} (M>5M_\sun)
=\frac{L_{\rm UV}}{1.5\e{29} \ \rm erg\ s^{-1}
}\quad M_{\odot }\, \mathrm{yr}^{-1}.
\end{equation}

The two hypotheses made above (the dust and gas being optically thick
or thin to UV radiation) are obviously at odds. Here we will not
venture into details, but rather remind that in a patchy distribution
of dust/gas clouds (a ``leaky cloud'' model,
\citealt{devyoung,boulanger88}), a fraction of the
ultraviolet/H$\alpha$ radiation produced in an \Hii region may still
be able to escape. Clearly, extinction corrections will be needed to
recover the SFR from UV/H$\alpha$ observations.  SFR estimates from UV
and H$\alpha$ vs.\ radio are shown in Fig.~(\ref{fig:cram}), middle
and lower panel respectively. If we assume that the radio luminosity
is a good SFR indicator (see next paragraph), then the high scatter
and the non-linearity of the (UV, H$\alpha$)-radio relation cleary
shows the effects of extinction. Note that Eq.~(\ref{eq:SFRHalfa})
incorporates a correction for the average extinction suffered in
spiral galaxies \citep{kennicutt83}, thus the scatter is entirely due
to the individual variations of the amount of absorption in individual
galaxies.

\smallskip 
The radio emission may be linked to the SFR with similar arguments
\citep{cond92}. If all stars with mass $M\geq M_{\rm sn}$ end their
life as core-collapse supernovae, then the supernova rate may be
defined as
\begin{equation}
  \nu_{\rm SN} = \int_{M_{\rm SN}}^{M_{\rm u}} k\Psi(M)\,\de M
\end{equation}
and, assuming that SNR are radio sources only during their adiabatic
lifetime, the nonthermal luminosity might be written as
\begin{equation}
  L_{\rm NT} \propto E^{-1/17} n^{-2/17} \nu_{\rm SN}.
  \label{eq:L_NTvssn_rate}
\end{equation}
where $E$ is the supernova explosion energy, and $n$ is the ambient
particle density.  However, it is impossible to reconcile the
supernova rates needed by Eq.~(\ref{eq:L_NTvssn_rate}) with those
observed in galaxies, since the former are about ten times larger.
Thus it was proposed \citep{pooley69,ilovaisky72} that SNRs may
accelerate the relativistic electrons producing the nonthermal radio
emission, but $\gtrsim 90\%$ of this emission must be produced long
after the individual supernova remnants have faded out and the
electrons have diffused throughout the galaxy. Thus it may be better
to use the empirical relation observed in the Milky Way
\begin{equation}
  \label{eq:radioNTinMW}
  L_{\rm NT} \sim 13 \, \nu^{-\alpha} \, \nu_{\rm SN}
\end{equation}
whit $L_{\rm NT}$ in units of $10^{29}$ erg/s, $\nu$ in GHz and
$\nu_{\rm SN}$ in yr$^{-1}$; the average $\alpha$ found in spiral
galaxies is $\sim 0.75$ \citep{cond92}. Eq.~(\ref{eq:radioNTinMW})
probably applies to most normal galaxies, since significant variations
in the ratio $L_{\rm NT}/\nu_{\rm SN}$ would violate the observed
radio/FIR correlation.  Also, \citet{volkkleinwiele89} argued that the
cosmic ray energy production per supernova is the same in M82 as in
the Galaxy.  Thus we have, in terms of the SFR:
\begin{equation}
  L_{\rm NT} \sim 5.3\e{28} \nu^{-\alpha} SFR.
\end{equation}

\section{X-ray emission: the \textit{Chandra} and XMM-\textit{Newton}
  revolution}
\label{sec:introX}
It has been known since the late eighties that star forming galaxies
are luminous sources of X-ray emission, due to a number of X-ray
binaries, young supernova remnants, and hot plasma associated to star
forming regions and galactic winds \citep{fabbiano89}.  The X-ray
emission is a key tool in understanding the star formation processes,
since it traces the most energetic phenomena (mass accretion,
supernova explosions, gas heating, cosmic rays acceleration), and it
is a probe into the dense and dusty star forming regions.

In the last few years ---i.e.\ after the launch of \chandra\ and
\xmm---, the `common picture' of normal galaxies underwent a dramatic
review, thanks to the quantum leap in sensitivity, angular and
spectral resolution achieved by the new observatories (cfr.\ 
Table~\ref{tab:telescopiX}).  As an example, in
\autoref{plate:antennaeX-ray}a the images of the Antennae galaxy
(NGC3038/9) taken with ROSAT and ASCA are shown beside an optical
image from the {\em Digitized Sky Survey}. The improvement provided by
\chandra\ (\autoref{plate:antennaeX-ray}b) is striking, since it
provides a ten-fold increase in angular resolution with respect to
ROSAT, while still offering a spectral resolution comparable to ASCA.

\begin{table}[t]
  \centering {\small
  \begin{tabular}{lrrrrr} 
    Telescope    &PSF  &Effective area  &Spectral resolution  &Bandpass\\
    (instrument) &(FWHM)  &(cm$^{-2}$ at 1 keV)  &($E/\Delta E$ at 1 keV) &(keV)\\
       \hline
    {\em Einstein} (IPC)  &$60\arcsec$   &100   &0.7   &0.4-3.5  \\ 
    ROSAT (PSPC-B)  &$25\arcsec$         &200   &2.5   &0.1--2.4   \\
    ROSAT (HRI)     &$3\arcsec$          &80    &---   &0.1--2.4    \\
    ASCA$^\star$    &$50\arcsec$         &420   &10    &0.5--10   \\
    BeppoSAX$^\dag$ &$90\arcsec$         &60    &5     &0.1--300    \smallskip \\
    \chandra\ (ACIS) &$0.5\arcsec$       &350   &10    &0.5--8      \\
    \xmm\ (EPIC)$^\ddag$  &$6\arcsec$    &1500  &13    &0.5--10  \\
    \xmm\ (RGS)     &($\sim 15\arcsec$)  &100   &300   &0.3--2.0        
  \end{tabular} }
  \caption{Characteristics of past and current X-ray telescopes.
      $^\star$effective area: GIS+SIS; energy resolution: SIS.
      $^\dag$effective area: LECS; PSF: MECS. 
      $^\ddag$effective area: MOS+PN;
      energy resolution: MOS.}
  \label{tab:telescopiX}
\end{table}

Point sources (X-ray binaries and young SNRs) have emerged as the
dominant contributors to the overall X-ray emission, especially in the
hard band (i.e.\ 2.0--10 keV); diffuse emission is
nonetheless significant in the soft band (0.5--2.0 keV) for the
galaxies with the largest SFRs, such as M82 \citep{grif00} where the
diffuse emission accounts for $\sim 30\%$ of the soft emission. In the
following, the main characteristics of the different kinds of sources
will be briefly discussed.

X-ray binaries are the brightest class of point sources in galaxies
\citep[reviews may be found in][]{white95,vanparadijs98,pr02}; they are
formed by a collapsed star (black hole or neutron star) with a normal
companion (main sequence or giant). Although the properties of
their X-ray emission (luminosity, variability, spectrum, whether the
emission is persistent or transient) depend on a number of parameters
(mainly the masses of the two stars and the accretion rate), an
essentially full characterization of X-ray binaries can be made on the
basis of the mass of the donor star. We have High Mass
X-ray Binaries (HMXB) when the donor is a main sequence star with
$M\gtrsim 8 M_\sun$, and Low Mass X-ray Binaries (LMXB) when the donor
is a post main sequence star with $M\lesssim 1 M_\sun$ (binary systems
with intermediate mass donor stars turn out not to be effective X-ray
emitters). This main distinction has to do with the nature of the mass
transer --- wind accretion in HMXB, transfer via Roche-lobe overflow in
LMXB. 

The spectral shape of X-ray binaries may be basically seen as the sum
of two components: a black body with $kT\gtrsim 1$ keV (`soft
component', associated with an accretion disc) and a power law
$S(E)\propto E^{1-\Gamma}$ with slope $\Gamma\sim 1.2$ for the HMXB and
$\sim 1.4$ for the LMXB (`hard component', arising from Comptonization of
the photons form the accretion disc); the relative weight of the two
components may vary with time, and one of the components may even
completely dominate the overall emission --- a transition occurs
between the so-called soft (or high) and hard (or low) states (cfr.\ 
\S~\ref{sect:gal-individuali}). The overall X-ray luminosity is
usually comprised in the $10^{37}$--$10^{39}$ erg/s interval, with the
HMXBs somewhat brighter on average, and depends for both types on how
close the normal star is to fill its Roche lobe: the closer, the
brighter.

Supernova remnants are moderately luminous sources ($\LX\sim 10^{37}$
erg/s) in the early part of their life, namely during the free
expansion phase and the beginning of the adiabatic expansion phase
\citep{woltjer72,chevalier77}. Their spectrum can be described as a
thermal plasma with decreasing temperature, from $kT\sim 40$ to $\sim
0.4$ keV in about $10^3$ yr, as the remnant expands (assuming a rather
dense ISM as should be the case in the most actively star forming
galaxies; namely $n\sim 10^3$ cm$^{-3}$, following
\citealt{chevalier01}; see also \citealt{drainewoods91}).  This rather
short lifetime makes SNRs a minor component among X-ray sources in
galaxies, accounting for $\lesssim 10\%$ of the total number of point
sources \citep{pence01,pr02}. It is nonetheless interesting to note
that since the radio-emitting phase only begins $10^4$ yr after the
supernova explosion, X-ray and radio SNRs represent two different
evolutive phases of the same population of objects; indeed the
comparison of \chandra\ and radio images has shown that sources
emitting in both bands are very rare \citep{chevalier01}. 
%FIXME-LA TROVO UNA REFERENZA SULLA RIGA DEL FERRO?

The interstellar medium (ISM) itself, when heated by supernova
explosions, turns into an X-ray source. Soft, diffuse emission in
galaxies was already observed with {\em Einstein}, and at least one
thermal component with a temperature around $\sim 1$ keV was required
to fit the ASCA and BeppoSAX data \citep{pta97,cappi99}.  Its spectrum
may be described as a Bremsstrahlung component plus a number of atomic
recombination lines.  Depending on the ISM density, the SFR, and the
gravity potential of the galaxy, the hot high-pressure plasma may
eventually find a way out of the galaxy disc and escape into the
intergalactic medium --- thus creating an outflow
(\autoref{plate:M82chandra}).  Observations conducted with the ROSAT
HRI instrument showed a spatial overlap of the X-ray emission with
H$\alpha$ emission from the outflow \citep{dellaceca96,dellaceca97}.
It was not clear, however, if the X-ray emission came from the
interface between the ambient material and the wind, or from the wind
itself. \chandra\ observations showed that the X-ray emission has a
conical, shell-like shape (\autoref{plate:NGC253xmmchandra}b;
\citealt{strickland00}). It can be explained as the hot rarefied and
expanding wind heats the surrounding material which in turn produces
the X-ray emission. \xmm\ observations
(\autoref{plate:NGC253xmmchandra}a) have shown that also the galactic
disc is an X-ray emitter. An interesting and yet unresolved question
is the fate of the metals produced in the supernova explosions:
whether they mix with the ISM or escape from the galaxy; cfr.\ 
Chap.~\ref{sec:m82intro}.

\begin{figure}[tp]
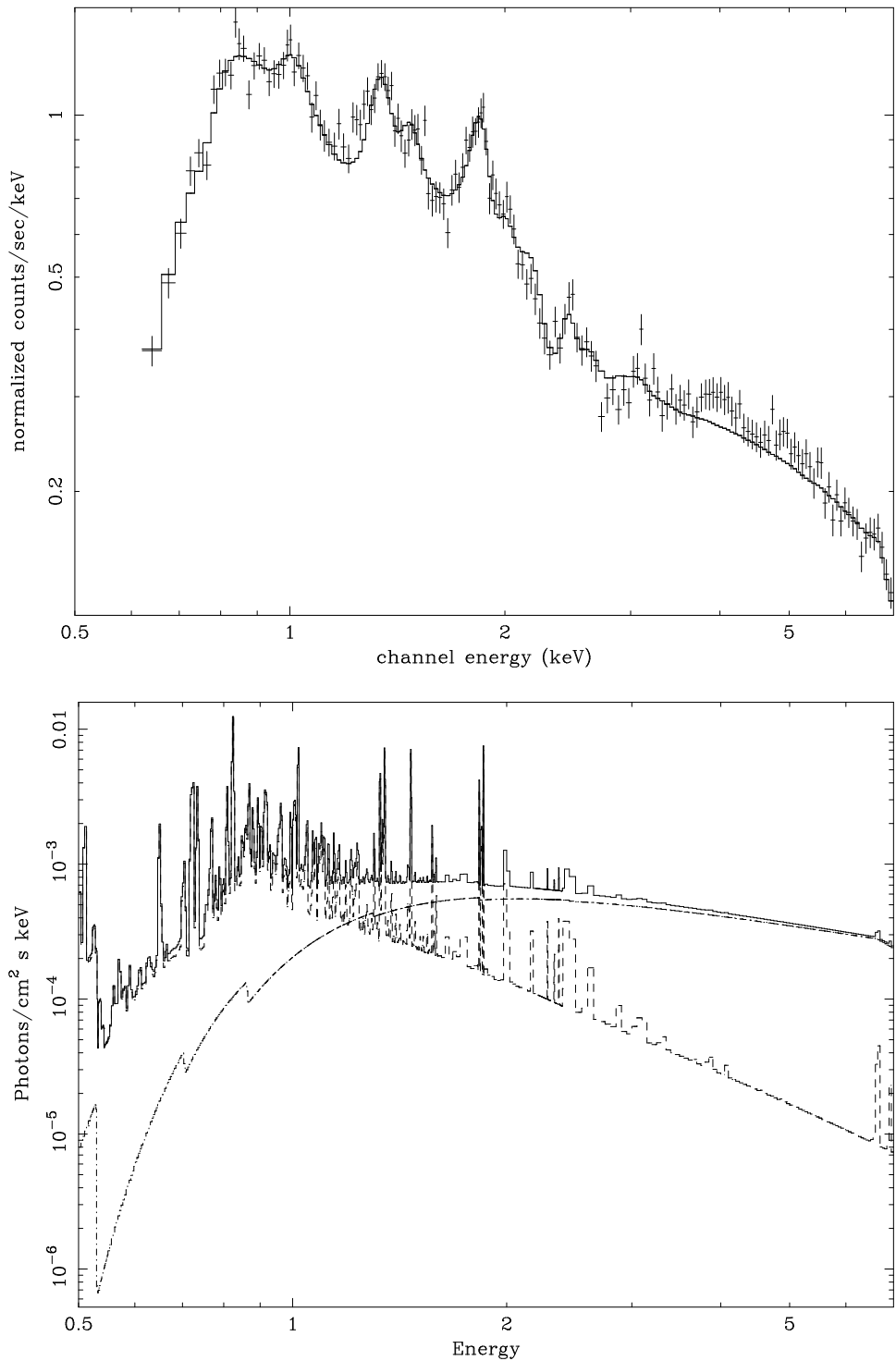

  \centering
      \includegraphics[height=0.85\textwidth,angle=-90]{m82spec_pn.ps}\bigskip\\
      \includegraphics[height=0.85\textwidth,angle=-90]{m82modelspec.ps}  
  \caption{X-ray spectrum of M82. Upper panel: observed data from the
    {\em pn} instrument onboard \xmm, rebinned to achieve a
    significance of at least $15\sigma$ per energy bin. The continous
    line shows the model, folded with the instrumental response. Lower
    panel: model spectrum (continous line) composed by a hot plasma
    (dashed line) plus an absorbed power law (dot-dashed line).}
  \label{fig:m82modelspec}
\end{figure}

The total spectrum of a star forming galaxy is thus made up by the sum
of a thermal spectrum from hot plasma and of a power law representing
the emission from the X-ray binaries. While a detailed discussion of
the relative contribution of the different classes of sources to the
total spectrum may be found in \citet{pr02}, we will focus here on 
two simple considerations: i) the power law component is often found
to be absorbed, with column densities aroud $10^{21-22}$ cm$^{-2}$
\citep{pta99,moran99}; ii) the thermal component has a steeper slope
than the power law. The joint effect of the steeper slope of the
thermal component, and of absorption on the power law, makes the two
components mainly emit in different energy bands: the plasma dominates
at softer energies ($\lesssim 1$--2 keV) and the binaries at harder
ones.  As an example, the spectrum of the galaxy M82 obtained with the
{\em pn} instrument onboard \xmm\ is shown in
Fig.~(\ref{fig:m82modelspec}).  It is interesting to notice that,
although the individual spectrum of X-ray binaries has a slope around
$\Gamma\sim 1.2$--1.4, the resulting total spectrum of a galaxy may
have a flatter slope if a number of binaries is present with different
absorbing column densities; cfr.\ the \chandra\ spectrum of the point
sources in M82, Fig.~\ref{fig:m82pointsrc}.  Conversely, if the
spectrum of the X-ray binaries is not heavily absorbed, the more
intense the plasma emission, the more steep the overall spectrum of a
galaxy. As a consequence, a relatively large spread may be found in
the average hard X-ray spectrum of starburst galaxies \citep{pta99}.

% \footnote{The effect at work here is, on a
%   smaller scale, the same that produces the flat spectrum of the
%   cosmic X-ray background \citep{sw89}.}

\smallskip An active galactic nucleus (AGN) may also be present in
star forming galaxies. While it has been proposed that nuclear
activity might be related to enhanced star formation (the
`starburst/AGN connection') as a consequence of the disposal of
large quantities of gas, nuclear activity is not truly a star
formation related process. If present, moreover, an AGN tends to
dominate over the X-ray emission from the rest of the galaxy. A
typical AGN luminosity ($\LX\gtrsim 10^{42}$ erg/s) exceeds by (at
least) a few orders of magnitude the emission from a normal galaxy
($\LX\sim 10^{40\pm 1}$ erg/s). Thus the main concern about AGN in
X-ray studies of starburst galaxies has been so far {\em to exclude}
them from the galaxy samples (cfr.\ \S~\ref{campione_section}).  The
high resolution of \chandra\ allows to resolve most nearby galaxies,
so that AGN may be identified and properly accounted for in data
analysis \citep{ho2001}. However, the problem still arises for more
distant, hence unresolved, galaxies.

Optical spectroscopy is the main tool used in the classification of
Seyfert vs.\ star forming galaxies. The equivalent widths of emission
lines from the centre of a galaxy are used in {\em diagnostic
  diagrams} to determine the excitation and ionization states of the
nebular gas, which in turn are determined by the nature of the energy
source \citep{vo87}.  Although a determination of the X-ray luminosity
and/or an optical spectrum usually suffice, this can prove to be a
difficult task, especially for luminous dusty objects (such as NGC
3256, NGC 3690, NGC 6240, Arp220) in which the overall luminosity is
still sustainable by star formation, and AGN signatures may be hidden
by absorption. NGC 3690 \citep{dellaceca02} and NGC 6240
\citep{vignati01} are emblematic cases: both of them are currently
undergoing a powerful starburst (100--200 $M_\sun$/yr), and the
optical spectrum does not show clear AGN signatures; however when
observed with the PDS instrument onboard BeppoSAX they have shown the
emergence at energies larger than 10 keV of a heavily absorbed
($N_{\rm H} \sim 10^{24}$ cm$^{-2}$) nuclear component.

At high redshifts ($z\sim 1$), the use the diagnostic diagrams in
separating Seyfert galaxies from star forming ones has to be treated
with caution because of two effects: a possibly lower signal/noise
ratio, and the shift of emission lines outside the observing spectral
band.  Moreover, because of the little angular size (a few arcsec) of
high redshift galaxies, it is not possible to obtain a spectrum of the
central regions only.  Thus the signatures of a low luminosity AGN may
be diluted in the overall radiation of the galaxy. With this caveat,
optical spectroscopy may still be regarded as an effective method in
selecting candidate star forming galaxies at high redshifts.

\chapter{The X-ray luminosity as a SFR indicator}

While radio continuum and far infrared (FIR) luminosities of
star-forming galaxies are known to show a {\em tight linear}
relationship spanning four orders of magnitude in luminosity and up to
a redshift $\sim 1.3$ (\S~\ref{sec:introRadioFIR}), a relation between
FIR and X-ray luminosities was found in the early nineties but its
details remained somewhat controversial. A non linear ($L_{\rm X}
\propto L_{\rm FIR}^{0.6}$) and highly scattered (dispersion of about 2
dex) relation was found between FIR and soft (0.5--3.0 keV) X-ray
luminosities of IRAS-bright and/or interacting/peculiar galaxies
measured by the {\em Einstein} satellite \citep{gp90}. A somewhat
different result was found by \citet{djf92}, i.e. a linear relation
between FIR and 0.5--4.5 keV luminosities for a sample of starburst
galaxies observed by {\em Einstein}. A large number of upper limits to
the X-ray flux (12 upper limits vs.~11 detections for \citealt{gp90})
along with high uncertainties in the X-ray and FIR fluxes may explain
this discrepancy. Moreover, these studies suffered by the lack of
knowledge about spectral shapes and internal absorption in star
forming galaxies caused by the limited sensitivity and spectral
capabilities of the IPC detector onboard {\em Einstein}.

%PAPER OUTLINE
In this chapter, with the high sensitivity and the broad-band spectral
capabilities of the ASCA and BeppoSAX satellites, we extend these
studies to the 2--10 keV band which is essentially free from
absorption. In the following paragraphs a sample of nearby star
forming galaxies is assembled (\S~\ref{campione_section}) and linear
relations among radio, FIR and both soft and hard X-ray luminosities
are found (\S~\ref{correlazioni_section}).  Possible biases are
discussed and the use of X-ray luminosities as a SFR indicator is
proposed (\S~\ref{biases_section}). In Chap.~\ref{hdf_chap} we present a
study of star-forming galaxies in the Hubble Deep Field North and test
the validity of the X-ray SFR law. Implications for the contribution
of star-forming galaxies to the X-ray counts and background are
discussed in Chap.~\ref{fondox_chap}.

% Throughout this thesis, we assume H$_0=50$ and q$_0=0.1$; we remind
% that an open universe with $q_0=0.1$ offers the best approximation to
% the current `concordance model' universe with $\Omega_{\rm M}=0.3$ and
% $\Omega_\Lambda=0.7$ while still retaining the `classical', simpler
% formulae.

Throughout this thesis we assume, unless otherwise stated, $H_0=70$, 
$\Omega_{\rm M}=0.3$ and $\Omega_\Lambda=0.7$.

\section{The local sample}       \label{campione_section}

The atlas of optical nuclear spectra by \citet{hfs97}
(hereafter HFS97) represents a complete spectroscopic survey of
galaxies in the Revised Shapley-Ames Catalog of Bright Galaxies
(RSA; \citealt{RSA}) and in the Second Reference Catalogue
of bright galaxies (RC2; \citealt{rc2}) with declination $\delta >
0\degr$ and magnitude $B_T<12.5$.  Optical spectra are classified in
HFS97 on the basis of line intensity ratios according to
\citet{vo87}; galaxies with nuclear line ratios typical of
star-forming systems are labeled as ``\Hii nuclei''.
This sample of \Hii galaxies contains only spirals and irregulars from Sa
to later types, except for a few S0 which were excluded from our
analysis since their properties resemble more those of elliptical
galaxies.

A cross-correlation of the HFS97 sample with the ASCA
archive gives 18 galaxies clearly detected in the 2--10
keV band with the GIS instruments. Since most of these
galaxies were observed in pointed observations (rather than in
flux-limited surveys) it is unlikely that the results hereafter shown
are affected by a Malmquist bias. Four additional objects in
the field of view of ASCA observations were not detected: the 2--10
keV flux upper limits are too loose to add any significant
information, and thus we did not include them in the sample.
The cross-correlation of the HFS97 sample with the BeppoSAX archive
does not increase the number of detections. When a galaxy was observed
by both satellites, the observation with better quality data was chosen.

Far infrared fluxes at 60$\mu$ and 100$\mu$ were taken from the IRAS
Revised Bright Galaxy Sample \citep[RBGS,][]{iras-rbgs} which is a
reprocessing of the final IRAS archive with the latest calibrations.
While the RBGS measurements should be more accurate, we checked that
the use of the older catalogue of IRAS observations of large optical
galaxies\footnote{Blue-light isophotal major diameter ($D_{25}$)
  greater than $8^\prime$.}  by \citet{rice88}, coupled with the Faint
Source Catalogue (FSC, \citealt{fsc}) for smaller galaxies, does not
significantly change our statistical analysis. FIR fluxes for
NGC\,4449 were taken from \citet{rush}.  Radio (1.4\,GHz) fluxes were
obtained from the \citet{cond90,cond96} catalogues (except for
NGC\,4449, taken from \citealt{haynes75}).  Distances were taken from
\citet{tully} and corrected for the adopted cosmology.

% flusso limite iras:
%  The FSC contains data for 173,044 point sources
%  in unconfused regions with flux densities typically above 0.2 Jy at
%  12, 25, and 60 microns, and above 1.0 Jy at 100 microns.

% ngc4449 non c'e' nel FSC (credo sia nel faint source reject ma e' un
% casino trovarlo) perche' per lo FSC servivano almeno 6 scan mentre lei
% ne ha solo 2
%
% cmq c'e' nel 12 micron sample di rush et al., ed il loro flusso infrarosso
% concorda con quello di mazzarella joe

Part of the X-ray data have already been published; in the cases where
published data were not available in a form suitable for analysis,
the original data were retrieved from the archive and reduced following
standard procedures and with the latest available calibrations.
Images and spectra were extracted from the pipeline-screened event
files.  The images were checked against optical (Digital Sky Survey)
and, where available, radio (1.4 GHz) images in order to look for
possible source confusions. Fluxes were calculated in the 0.5--2.0 and
2--10 keV bands from best-fit spectra for the GIS2 and GIS3
instruments and corrected for Galactic absorption only. The uncertainty
on the fluxes is of the order of 10$\%$. Depending on
the quality of data, the best-fit spectrum is usually represented by a
two-component model with a thermal plasma plus a power-law or just a
power-law.

The galaxies IC\,342 and M82 have shown some variability, mainly due
to ultraluminous X-ray binaries.  For the two of them we summarize in
\S~\ref{sect:gal-individuali} the results from several X-ray
observations and estimate time-averaged luminosities.

One object (\object{M33}) was not included in the sample since its
broad-band (0.5--10 keV) X-ray nuclear spectrum is dominated by a
strong variable source (\object{M33 X-8}) identified as a black hole
candidate
\citep{parmar01}. % Although M33 is identified by HFS97 as
%an \Hii nucleus, it has a very low SFR ($\sim 0.009$ M$\sun$/yr) so
%that the spectral signatures related to star formation can be easily
%hidden by a single powerful source.

% completezza
Therefore, the sample (hereafter local sample) consists of the 17
galaxies listed in Table \ref{campione}.  Since it is not complete in
a strict sense due to the X-ray selection, we have checked for its
representativeness with reference to the SFR. The median SFR value for
HFS97 was computed from the FIR luminosities and resulted SFR$_{\rm
  med} \sim$ 1.65 M$_\sun$/yr.  Considering objects with SFR $> {\rm
  SFR}_{\rm med}$ there are 14 galaxies in the local sample out of 98
in HFS97 (14$\%$), while there are 3 objects with SFR $<{\rm SFR}_{\rm
  med}$ (3$\%$). Thus the high luminosity tail is better sampled than
the low luminosity one.

%cross-correlated the HFS97 sample with the \citet{rice88} and the FSC
%catalogues, obtaining a complete homogeneous sample of 196 nearby
%($z<0.01$) star-forming galaxies with known SFR (notice that the FSC
%only covers the sky with galactic latitude $|b|>10\degr$ and is
%complete down to limiting fluxes of 0.2\,Jy at $60\mu$ and 1.0\,Jy at
%$100\mu$).
%
%In our X-ray-detected sample 14 galaxies out of 120 in the complete
%sample ($12\%$) have a ${\rm SFR}>1~ {\rm M}\sun/$yr, and 11 out of 72
%have a ${\rm SFR}>3~ {\rm M}\sun/$yr ($15\%$). 

%All
%statistical tests presented here are performed on this sample unless
%otherwise stated.

We also include data for 6 other well-known starburst
galaxies which were not in the HFS97 survey because they are in the
southern emisphere. On the basis of their line intensity ratios%
\footnote{References: NGC\,55 - \citet{webster83}; NGC\,253, 1672 \&
1808 - \citet{kewley}; NGC\,3256 - \citet{moran99}; Antennae -
\citet{rubin70}, \citet{crimea}.  } they should be classified as \Hii
nuclei.  In Table \ref{campione} we label them as supplementary
sample.

% ~~~~~~~~~~~~~~~~~~~~~~~~~~~~~
%  TABELLA: DATI LOCAL SAMPLE
% ~~~~~~~~~~~~~~~~~~~~~~~~~~~~~
\begin{table*}[p]       %e' un po' piu' larga e va centrata
\hskip-.8em\vbox{\small \centering\begin{tabular}{lcccccccccc}
\multicolumn{11}{c}{\sc Fluxes and Luminosities: Main Sample} \\
\\
%  & &\multicolumn{2}{c}{\sc 0.5--2.0 keV} &\multicolumn{2}{c}{\sc 2.0--10 keV} 
%&\multicolumn{2}{c}{\sc FIR} &\multicolumn{2}{c}{\sc 1.4 GHz} \\
{\sc Galaxy} &$D$ &$F_{0.5-2}$ &$L_{0.5-2}$ &$F_{2-10}$ &$L_{2-10}$ &$F_{\rm FIR}$ &$L_{\rm FIR}$ &$F_{1.4}$ &$L_{1.4}$  &{\sc Refs.} \\
\hline
%
%  H_0 = 70
M82*     &5.6   &97    &3.6    &290   &11      &67    &25     &7.7    &2.9      &1      \\                
M101     &5.7   &5.4   &0.22   &6.8   &0.27    &6.0   &2.4    &0.75   &0.30     &this work \\             
M108     &15    &4.4   &1.2    &6.0   &1.6     &2.0   &5.4    &0.31   &0.83     &this work       \\       
NGC891   &10    &8.3   &0.99   &19    &2.3     &4.5   &5.4    &0.70   &0.84     &this work\smallskip\\    
NGC1569  &1.7   &5.4   &0.019  &2.2   &0.0077  &2.5   &0.088  &0.41   &0.014    &2      \\                
NGC2146  &18    &8.2   &3.4    &11    &4.5     &7.3   &30     &1.1    &4.5      &3      \\                
NGC2276  &39    &2.1   &3.9    &4.4   &8.1     &0.85  &16     &0.28   &5.2      &this work       \\       
NGC2403  &4.5   &16    &0.39   &9.3   &0.23    &2.7   &0.65   &0.33   &0.080    &this work\smallskip\\    
NGC2903  &6.7   &7.9   &0.43   &7.0   &0.38    &3.7   &2.0    &0.41   &0.22     &this work       \\       
NGC3310  &20    &7.4   &3.5    &2.1   &1.0     &1.7   &8.1    &0.38   &1.8      &4      \\                
NGC3367  &46    &1.8   &4.5    &1.6   &4.0     &0.38  &9.5    &0.10   &2.5      &this work       \\       
NGC3690  &49    &5.7   &17     &11    &32      &5.3   &150    &0.66   &19       &4\smallskip \\           
NGC4449  &3.2   &8.3   &0.10   &4.8   &0.060   &1.9   &0.23   &0.6    &0.074    &5      \\                  
NGC4631  &7.1   &9.4   &0.57   &9.3   &0.57    &4.9   &3.09   &1.2    &0.73     &6      \\                  
NGC4654  &18    &0.6   &0.2    &0.9   &0.3     &0.93  &3.5    &0.12   &0.46     &this work       \\         
NGC6946  &5.9   &30    &1.2    &12    &0.49    &7.9   &3.2    &1.4    &0.57     &this work       \\        
IC342    &4.2   &18    &0.38   &110   &2.3     &11    &2.3    &2.3    &0.49     &this work   \\             
\hline
\\ \\
\multicolumn{11}{c}{\sc Supplementary Sample} \\
\\
%{\sc Galaxy} &{\sc Dist} &F052 &L052 &F210 &L210 &FIR &LFIR &S14 &L14 &{\sc Refs.} \\
\hline
NGC55    &1.4   &18    &0.040  &6.8   &0.015   &4.7   &0.10   &0.38   &0.0084 &6      \\       
NGC253*  &3.2   &25    &0.31   &50    &0.62    &49    &6.1    &5.6    &0.69   &1      \\     
NGC1672  &16    &5.8   &1.7    &6.1   &1.8    &2.3    &6.8    &0.45   &1.3    &this work\\  
NGC1808  &11    &6.5   &1.0    &7.6   &1.2    &5.3    &8.2    &0.52   &0.81   &this work\\  
NGC3256  &40    &9.0   &17     &6.2   &12     &4.8    &92     &0.66   &13     &7      \\  
Antennae &27    &7.2   &6.3    &5.3   &4.7    &2.6    &23     &0.57   &5.0    &8      \\  
\hline
\end{tabular}
\smallskip}
\caption{Data for galaxies in our local samples. All galaxies were observed with
ASCA, except those marked with * observed by BeppoSAX. Distances in Mpc;
X-ray fluxes in $10^{-13}$ erg s$^{-1}$ cm$^{-2}$, FIR fluxes in 
$10^{-9}$ erg s$^{-1}$ cm$^{-2}$ and radio fluxes in Jy; X-ray luminosities in
$10^{40}$ erg s$^{-1}$, FIR luminosities in $10^{43}$ erg s$^{-1}$
and radio luminosities in $10^{29}$ erg s$^{-1}$ Hz$^{-1}$. The uncertainty
on fluxes and luminosities is of the order of 10$\%$. \newline
References: 1 \citet{cappi99}; 2 \citet{dellaceca96};
3 \citet{dellaceca99}; 4 \citet{zezas98}; 5 \citet{dellaceca97};
6 \citet{dahlem98}; 7 \citet{moran99}; 8 \citet{sansom96}. \newline
\label{campione}
}
\end{table*}

\section{The radio/FIR/X-rays correlation}         \label{correlazioni_section}

\newcommand{\figurecorr}{%   in newcommand cosi' mi viene piu' facile spostarlo poi
 \begin{figure*}[p]    % firx e radiox
  \begin{center} 
      \includegraphics[height=0.45\textheight]{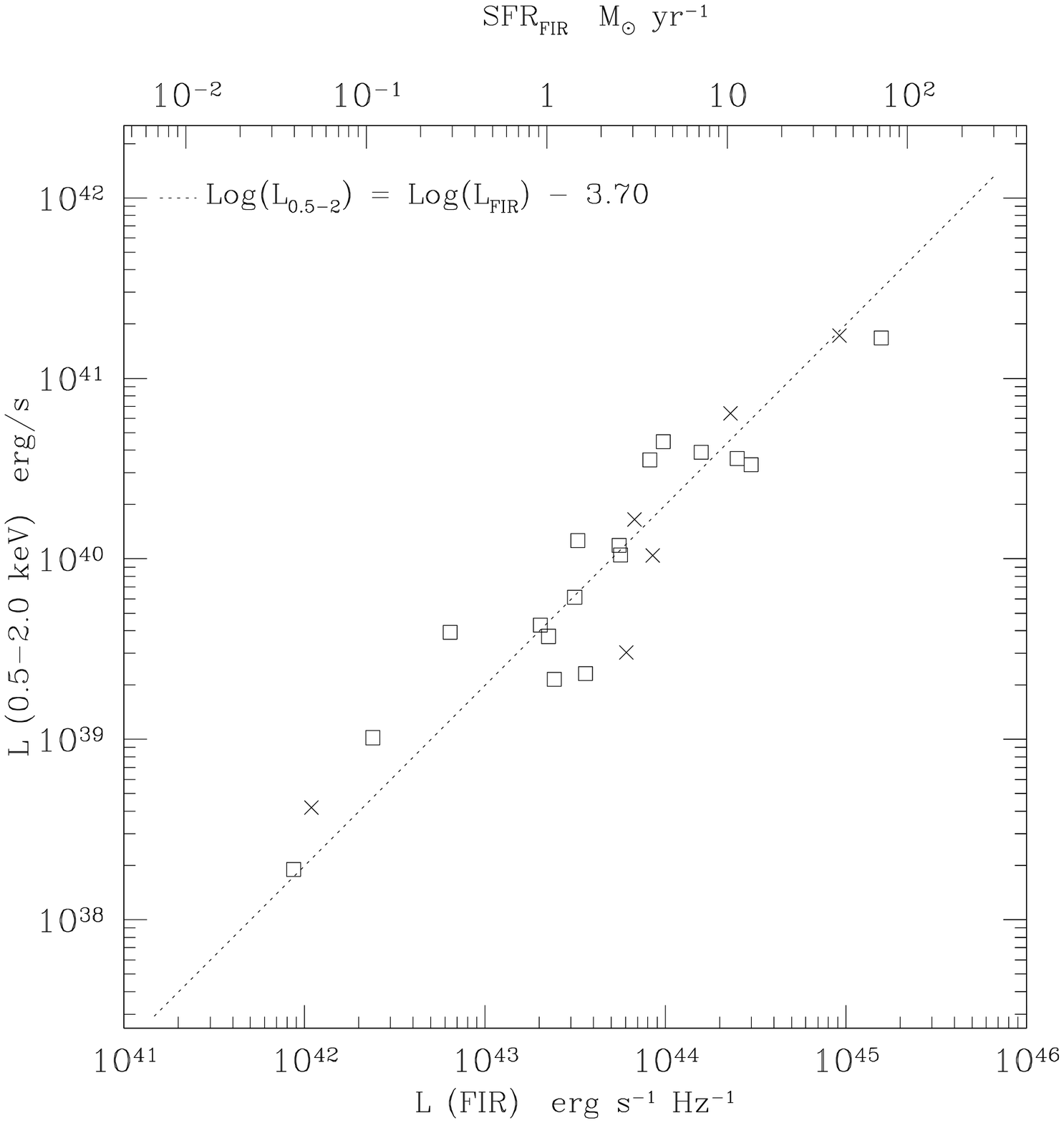}\bigskip\\
      \includegraphics[height=0.45\textheight]{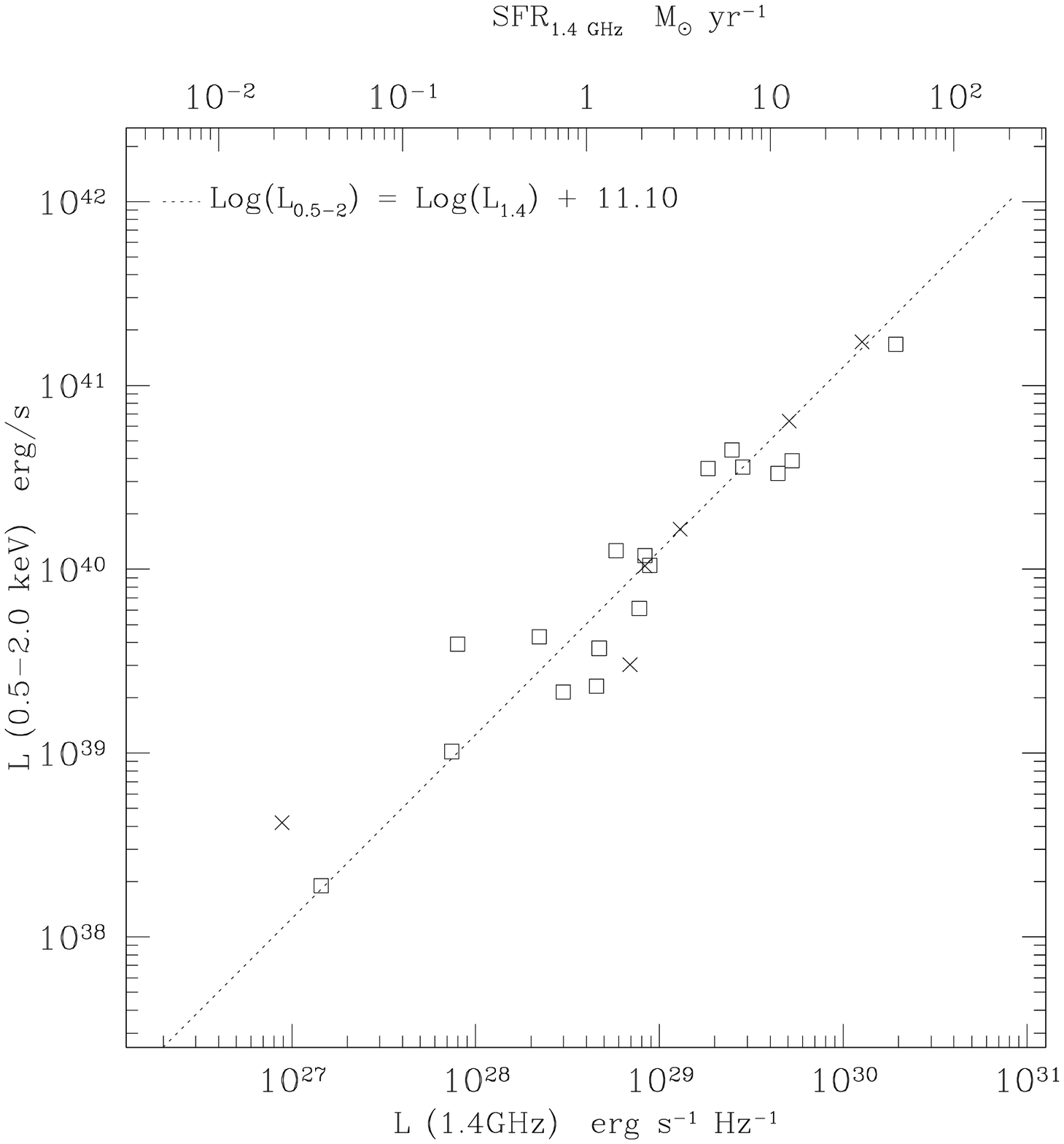}
  \end{center}
  \caption {The 0.5--2.0 keV luminosity of local star-forming galaxies
  vs.~radio and FIR ones. 
  Squares: local sample; crosses: supplementary sample;
    dotted lines: eqs.(\ref{eq:firx}, \ref{eq:radiox}).
  \label{pranalli-E3_fig:fig1}
  }
 \end{figure*}
 \begin{figure*}[p]    % firx e radiox
  \begin{center} 
      \includegraphics[height=0.45\textheight]{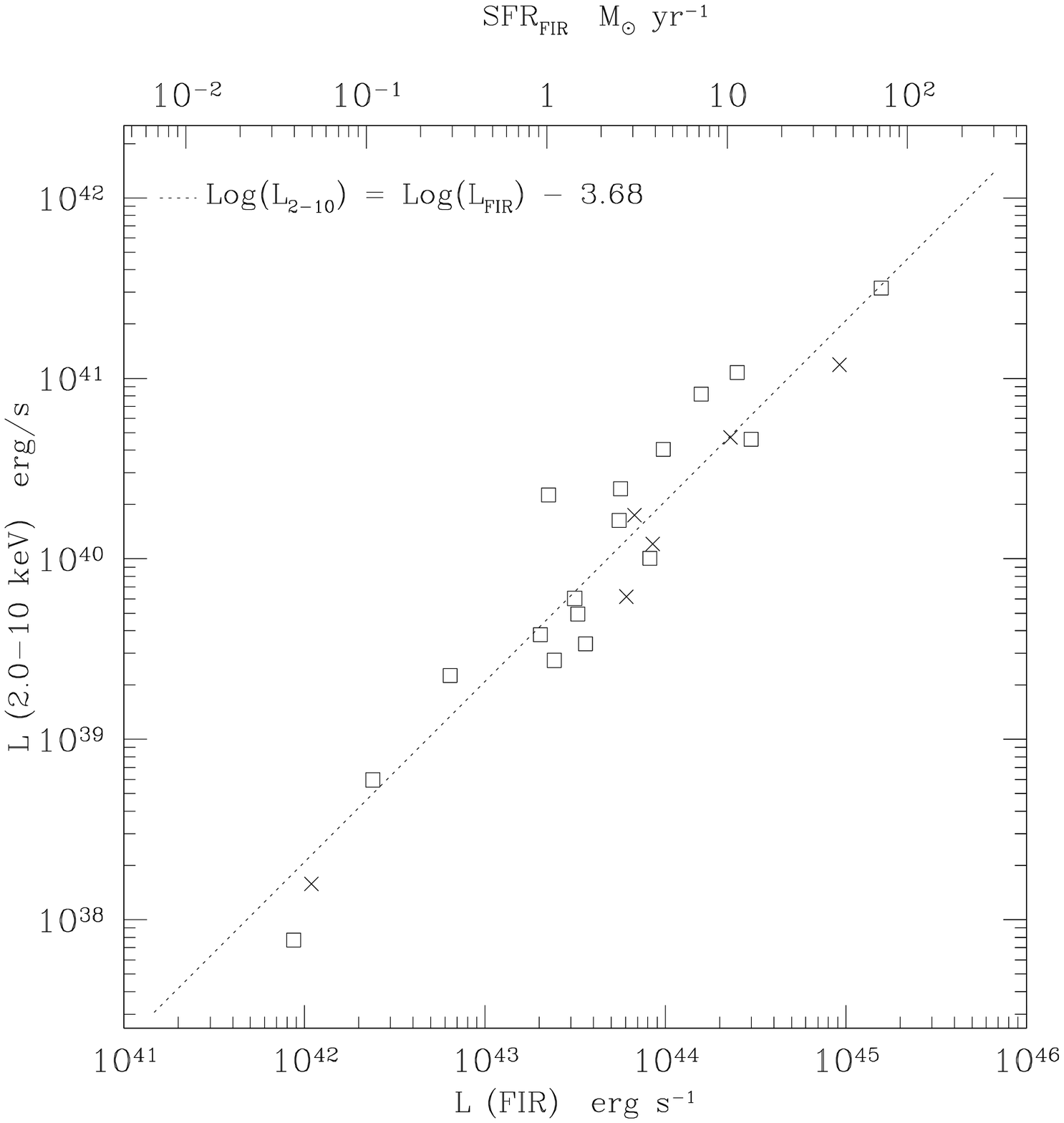}\bigskip\\
      \includegraphics[height=0.45\textheight]{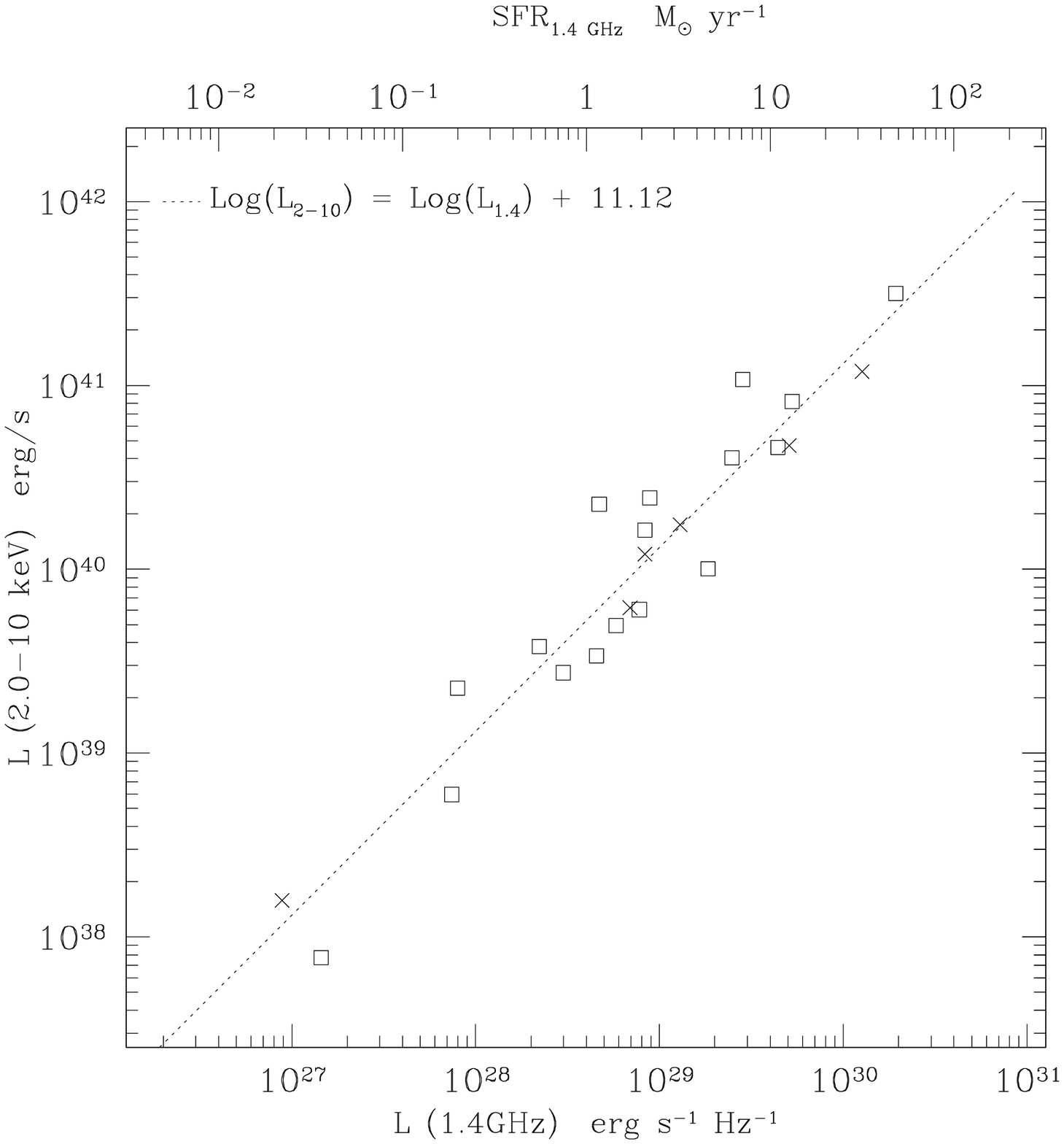}
  \end{center}
  \caption {The 2.0--10 keV luminosity of local star-forming galaxies
  vs.~radio and FIR ones. Symbols as in Fig.~(\ref{pranalli-E3_fig:fig1});
  dotted lines: Eqs.~(\ref{eq:firx2},\ref{eq:radiox2}).
  }
  \label{pranalli-E3_fig:fig2}
 \end{figure*}
}

As a preliminary test, we perform a least-squares analysis for the
well-known radio/FIR correlation, which yields
\begin{equation}\label{radiofir}
\Log (L_{\rm FIR}) = (0.98\pm 0.06) ~\Log (L_{\rm 1.4}) 
     +15.4\pm 1.6
\end{equation}

The dispersion around the best-fit relation is given as the
estimate $s$ of the standard deviation $\sigma$:
\begin{equation}
s = \frac{1}{N-\nu} \cdot \sqrt {\sum \left( \Log~ L_{\rm obs}-\Log~ L_{\rm pred}
\right)^2 }     \label{eqvarianza}
\end{equation}
where $\nu$ is the number of free parameters
and $N$ is the number of points in the fit),
$L_{\rm pred}$ is the luminosity expected from the best fit
relation and $L_{\rm obs}$ the observed one.
For the radio/FIR correlation
(Eq.~\ref{radiofir})  we find $s=0.18$.

Following \citet{helou} we also calculate the mean ratio
$q$ between the logarithms of FIR and radio fluxes, obtaining
$q\sim 2.2\pm 0.2$. This value is consistent
with the mean $q=2.34\pm 0.01$ for the 1809 galaxies in
the IRAS 2\,Jy sample by \citet{yrc}.

%\begin{equation}
%q=\Log ({\rm FIR}/3.75\cdot 10^{12}~{\rm Hz}/S_{1.4})
%\end{equation}
%where $3.75\cdot 10^{12}~{\rm Hz}$ is the frequency at $80\mu$, and
%$S_{1.4}$ is in erg/s/cm$^2$.

\subsection*{Soft X-rays}

A  test for the soft X-ray/FIR/radio relations
(Fig.~\ref{pranalli-E3_fig:fig1}) yields \\
\begin{eqnarray}
\Log (L_{0.5-2}) &= &(0.87\pm 0.08) ~\Log (L_{\rm FIR}) 
        +2.0\pm 3.7 \label{eq:firx1} \\
\Log (L_{0.5-2}) &= &(0.88\pm 0.08) ~\Log (L_{\rm 1.4})
         +14.6\pm 2.2 \label{eq:radiox1} \medskip
\end{eqnarray}\\
with $s\sim 0.26$ and 0.24 respectively.

Our result is consistent with the $L_{0.5-4.5}\propto L_{\rm
FIR}^{0.95\pm 0.06}$ relation found by \citet{djf92} for normal and
starburst galaxies from the IRAS Bright Galaxy Sample, but it is only
marginally consistent with the much flatter and more dispersed
relationship obtained by \citet{gp90} for a sample of IRAS selected
galaxies ($L_{0.5-3 \rm keV} \propto L_{60\mu}^{0.62\pm 0.14}$) and
for a sample of starburst/interacting galaxies ($L_{0.5-3 \rm keV}
\propto L_{60\mu}^{0.70\pm 0.12}$).

The inclusion of the objects of the supplementary sample (Table~%
\ref{campione}) does not significantly change
the slopes, i.e.  $L_{0.5-2.0 \rm keV}\propto L_{\rm
FIR}^{0.88\pm 0.07}$; likewise, if we use the $60\mu$ luminosity
instead of FIR, we obtain $L_{0.5-2.0 \rm keV}\propto
L_{60\mu}^{0.85\pm 0.07}$. 

By assuming an exactly linear slope, the best fit relations 
for the local (local+supplementary) sample become:\\
\begin{eqnarray}
\Log (L_{\rm 0.5-2}) &= &\Log (L_{\rm FIR}) - 3.68~~ (3.70) \label{eq:firx}\\
\Log (L_{\rm 0.5-2}) &= &\Log (L_{\rm 1.4}) + 11.08~~ (11.10) 
   \label{eq:radiox}
\end{eqnarray}
with $s \sim 0.27$ and 0.24 respectively.

%ora l'F-test
%in grassetto perche' e' post-referee
%{\bf  
%% nell'articolo non l'hanno voluto? e io lo rimetto per la tesi!!
% un attimo.. non mi torna più la formula..
% The F-test for an additional term in best-fits can be used to compare
% the significance of the free-slope and fixed-slope fits.  Under the
% assumption of equal uncertainties for each point in the fit, it can be
% shown that Fisher's $F$ parameter reduces to
% \begin{equation}
% F=\frac{s_1^2\frac{\nu+1}{\nu} - s_2^2}{s_2^2} \cdot \nu
% \end{equation}
% where $s_1^2$ and $s_2^2$ are the variance estimates for the
% free-slope and fixed-slope fits (eq.~\ref{eqvarianza}), and $\nu$ is
% the degrees of freedom for free-slope fit.
By applying the F-test we find that the free-slope fits are not
significantly better than those with the linear slope, the improvement
being significant only at the $1\sigma$ level.

%% \begin{figure*}[t]    % firsx e radiosx
%%   \begin{center} 
%%       \includegraphics[width=0.49\textwidth]{h3889f1a.eps}
%%       %\hskip.8cm
%%       \includegraphics[width=0.49\textwidth]{h3889f1b.eps}
%%   \end{center}
%%   \caption {The 0.5--2.0 keV luminosity of local star-forming galaxies
%%   vs.~radio and FIR ones. 
%%   Squares: local sample; crosses: supplementary sample;
%%     dotted lines: eqs.(\ref{eq:firx}, \ref{eq:radiox}).
%%   \label{pranalli-E3_fig:fig1}
%%   }
%% \end{figure*}
%% \begin{figure*}[t]    % firx e radiox
%%   \begin{center} 
%%       \includegraphics[width=0.49\textwidth]{h3889f2a.eps}
%%       %\hskip.8cm
%%       \includegraphics[width=0.49\textwidth]{h3889f2b.eps}
%%   \end{center}
%%   \caption {The 2.0--10 keV luminosity of local star-forming galaxies
%%   vs.~radio and FIR ones. Symbols as in Fig.~(\ref{pranalli-E3_fig:fig1});
%%   dotted lines: Eqs.~(\ref{eq:firx2},\ref{eq:radiox2}).
%%   }
%%   \label{pranalli-E3_fig:fig2}
%% \end{figure*}

\figurecorr

%% \afterpage{\clearpage%  non funziona granche' (se uso
%%                      %  begin{figure} fa a modo suo, se
%%                      %   non lo uso non i numera la figura)
%%            \ifthenelse{\isodd{\value{page}}}%
%%                {\afterpage{\figurecorr}}%
%%                {\figurecorr}}%
%% \clearpage

\subsection*{Hard X-rays}

In Fig.~(\ref{pranalli-E3_fig:fig2}) we plot 2--10 keV luminosities
versus FIR and radio ones.  Least-squares fits yield:\\
\begin{eqnarray}
\Log (L_{\rm 2-10}) &= &(1.08\pm 0.09) ~\Log (L_{\rm FIR}) 
        -7.1\pm 4.2 \label{bf210a} \\
\Log (L_{\rm 2-10}) &= &(1.08\pm 0.09) ~\Log (L_{\rm 1.4})
         +8.8 \pm 2.7 \label{bf210b}
\end{eqnarray}\\
with $s\sim 0.30$ and 0.29 respectively.  The linearity
and the dispersion are not significantly changed neither by the
inclusion of the supplementary sample ($L_{2-10}\propto L_{\rm
FIR}^{1.04\pm 0.07}$ and $L_{2-10}\propto L_{\rm 1.4}^{1.01\pm 0.07}$,
$s\sim 0.27$ and 0.26 respectively), nor by the use of the $60\mu$
luminosity ($L_{2-10}\propto L_{60\mu}^{1.01\pm 0.07}$).

By assuming an exactly linear slope, the best fit relations 
for the local (local+supplementary) sample become:\\
\begin{eqnarray}
\Log (L_{\rm 2-10}) &= &\Log (L_{\rm FIR}) - 3.62~~ (3.68) \label{eq:firx2}\\
\Log (L_{\rm 2-10}) &= &\Log (L_{\rm 1.4}) + 11.13~~ (11.12) 
   \label{eq:radiox2}
\end{eqnarray}
with $s \sim 0.29$ for both fits. There is no significant improvement
(less than $1\sigma$)
in the free-slope fits with respect to the linear slope ones.

\section{X-rays and the Star Formation Rate}      \label{biases_section}

The existence of a tight linear relation implies that the three
considered bands all carry the same information. Since the radio and
far infrared luminosities are indicators of the SFR, the 0.5--2 keV and
2--10 keV luminosities should also be SFR indicators.
However, before attempting to calibrate such relationships, we should    
consider the possible existence of selection effects.

\citet{hfs95} made a special effort in obtaining {\em nuclear} nebular
spectra, so that a reliable spectral classification of the central
engine could be derived.      The main concern is the possibility that
the \Hii galaxies in the HFS97 sample could host a Low Luminosity AGN
(LLAGN), which might significantly contribute to the overall energy
output. To check for this possibility, \citet{ulvestad} observed with
the VLA at 1.4 GHz a complete sample of 40 Sc galaxies in HFS97 with
\Hii spectra and did not find any compact luminous radio
core. Instead, they found that the radio powers and morphologies are
consistent with star formation processes rather than by
accretion onto massive black holes; thus they suggest that \Hii nuclei
intrinsically lack AGN. Therefore we believe that the
HFS97 classification is reliable and that our sample is not polluted
by AGN.

It is also worth noticing that the soft X-rays relationships may
involve some further uncertainties related to the possible presence of
intrinsic absorption (negligible in the 2--10 keV band for
column densities usually found in normal galaxies).  An example of
this effect is the southern nucleus of NGC\,3256 (\S~\ref{sec:n3256}), a
dusty luminous merger remnant with two bright radio-IR cores where
star formation is ongoing: while both of them fall on the
radio/hard X-ray relation, only the northern core is on the
radio/soft X-ray relation because the southern one lies behind a dust
lane which absorbes at all wavelengths from $\sim 1\mu$ to $\sim 2$
keV.  The quasi-linearity of the soft X-ray relations suggests that
absorption is unlikely to be relevant for the majority of the objects in
our sample; however this effect may become significant at
cosmological distances ($z\gtrsim 1-2$) where galaxies have more dust
and gas at their disposal to form stars.

Thus we feel confident to propose the use of X-ray luminosities
as SFR indicators. From eqs.(\ref{eq:firx},\ref{eq:radiox},\ref{eq:firx2},%
\ref{eq:radiox2}) and from the calibrations of SFR$_{FIR}$ and
SFR$_{\rm radio}$ (\S~\ref{sec:SFRindicators}) we derive:
\begin{eqnarray}
{\rm SFR}~=&2.2\cdot10^{-40} ~L_{\rm 0.5-2 keV} &M_\sun{\rm /yr}\label{eqsfrsoftX2}\\
{\rm SFR}~=&2.0\cdot10^{-40} ~L_{\rm 2-10 keV}  &M_\sun{\rm /yr}.\label{eqsfrhardX2}
\end{eqnarray}

     We also notice that there is growing evidence that star formation could play
a major role even among those objects classified as LLAGN. The preliminary results of the
\chandra\ LLAGN survey \citep{ho2001} show that only about one third
of LLAGN have a compact nucleus dominating the X-ray emission, while
in the remaining objects off-nuclear sources and diffuse emission 
significantly contribute to the overall emission.

\begin{figure}[t]    % firx e radiox
  \begin{center} 
      \includegraphics[width=0.8\textwidth]{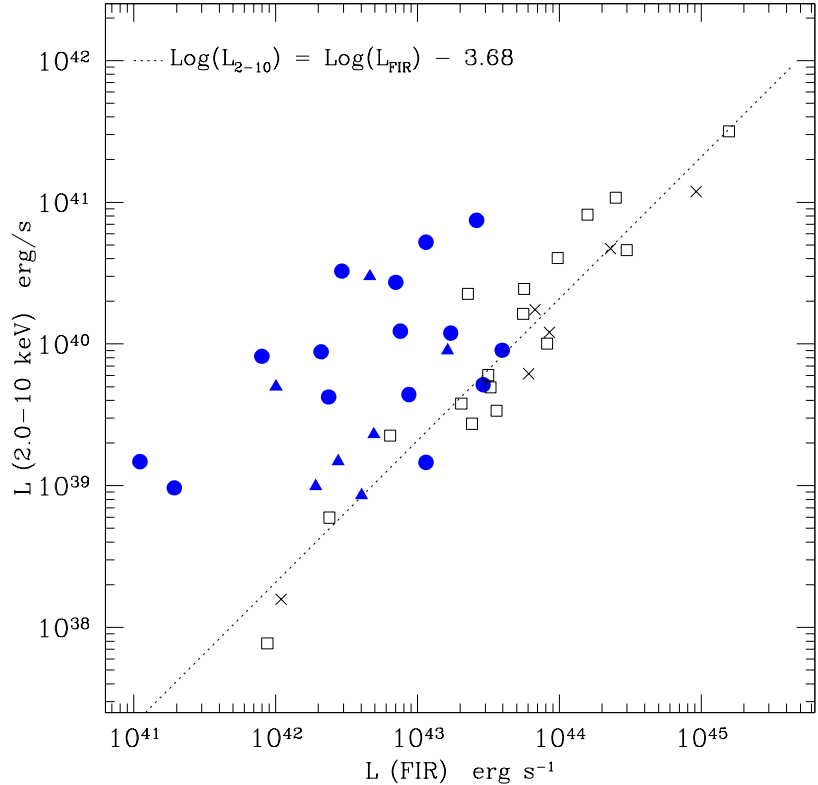}
  \end{center}
  \caption {2.0--10 keV vs.~FIR luminosities of star-forming galaxies
  and LLAGN. Filled triangles: LINERs, filled circles: Seyfert galaxies.
  Open symbols as in Fig.~(\ref{pranalli-E3_fig:fig1});
  line: Eq.~(\ref{eq:firx2}).
  }
  \label{figseyfirx}
\end{figure}

     Following this investigation, we have analyzed the relations
between radio/FIR/X-ray luminosities for the spiral galaxies in the
\citet{terashima} sample of LLAGN, drawn from HFS97 and observed with
ASCA, comprising 7 LINERs and 15 Seyfert's with $4\cdot 10^{39}\lesssim
L_{2-10}\lesssim 5\cdot 10^{41}$ erg s$^{-1}$.  We find that
the X-ray/FIR and X-ray/radio luminosity ratios generally exceed those
of star-forming galaxies, but about one third of the objects have ratios
falling on the same {\em locus} of the star-forming galaxies
(Fig.~\ref{figseyfirx}). Therefore, the nuclear X-ray emission of
these last LLAGN must be comparable to or weaker than the emission from
star formation related processes.  Moreover, the infrared (IRAS band)
colours of these objects are also similar to those of star-forming
galaxies, and completely different from those of QSOs,
thus suggesting that the FIR luminosities of LLAGN may be powered by
star formation.

\section{Comparison with other X-ray based SFR indicators}
\label{sec:XSFRcompare}

Our inference of using the 2--10 keV luminosity as a SFR indicator is
consistent with a study on Lyman-break galaxies by
\citet{nandra02} who have extrapolated the \citet{djf92} FIR/soft X-ray
relation to the hard X-ray band obtaining a SFR/2--10 keV luminosity
relation within $10\%$ of our Eq.~(\ref{eqsfrhardX2}). From
a stacking analysis of \chandra\ data for a sample of optically
selected Lyman-break and Balmer-break galaxies in the HDFN they find a
good agreement of the average SFR as estimated from X-ray and
extinction-corrected UV luminosities.

A similar derivation of an $\LX$-SFR relation has been performed by
\citet{bauer02}, using 0.5--8.0 keV and 1.4 GHz luminosities of a
joint sample of 102 nearby late-type galaxies observed with {\em
  Einstein} \citep{shapley01} and of 20 galaxies with emission line
spectra in the \chandra\ Deep Field North.

\citet{grimm02}; (see also \citealt{gilfanov04b}) have recently shown
that the luminosity function of High Mass X-ray Binaries (HMXB) can be
derived from a universal luminosity function whose normalization is
proportional to the SFR. They also show that, due to small numbers
statistics, a regime exists, in which the total X-ray luminosity grows
non-linearly with the number of point sources. This has the
consequence that also the total X-ray luminosity vs.\ SFR relationship
is non-linear in the same regime, namely for SFR below $4-5 M_\sun$/yr.
Although this might seem at odds with our findings of
\S~\ref{correlazioni_section}, it is not really so, since
\citet{grimm02} analysis only refers to the contribution from HMXB. It
has been shown \citep{gilfanov04a} that once the Low Mass X-ray
Binaries are taken into account, their result are in agreement with
the X-ray/radio/FIR correlations we found. However, the number of LMXB
should be related with the integrated star formation over a much
longer time than the HMXB (e.g.\ several Gyr) and thus not trace the
current SFR. Mantaining both the radio/FIR/X-ray correlations, and the
interpretation of the emission in these bands as powered by recent
star formation, should require that also the FIR and radio emission
are non-linearly related with the SFR; this has indeed been suggested
for the radio \citep{bell03}. As to the FIR emission, further
investigations will be required to clarify its relation with the warm
and dust component; thus at present the question stays open.

\begin{table} \label{tabsfrcomparison}
\centering\begin{tabular}{lll}
{\sc Author} &{\sc SFR estimate}  &{\sc Method}\\
\hline
\citet{rcs03}  &SFR$ = 2.0\e{-40} \Lhx$  &radio/FIR/X-ray correlations,\\
               &SFR$ = 2.2\e{-40} \Lsx$  &23 galaxies obs.\ w.\ ASCA\medskip\\
\citet{nandra02} &SFR$ = 1.8\e{-40} \Lhx$  &0.5--4.5 keV/FIR correlation\\
                 &                         &in \citet{djf92}\medskip\\
\citet{bauer02}  &SFR$ = (1.7\pm0.5)\e{-43}\cdot$ &102 galaxies with {\em Einstein}\\
                 &\hfill $\cdot L_{0.5-8}^{1.07\pm0.08}$ &+20 galaxies in the CDFN\medskip\\
\citet{grimm02}  &SFR$ = 1.5\e{-40} \Lhx$ &theory + 23 nearby \\
                 &\hfill ($M>4.5 M_\sun$/yr) &galaxies (mainly \chandra\\
                 &SFR$ = (3.8\e{-40} \Lhx)^{0.6}$ &data) \\
                 &\hfill ($M<4.5 M_\sun$/yr) &\\
\hline
\end{tabular}
\caption{A comparison of different SFR indicators based on the X-ray luminosity.}
\end{table}

\section{NGC\,3256: a case for intrinsic absorption}
\label{sec:n3256}

We present the test case of \object{NGC 3256}, a luminous dusty
merger remnant included in the supplementary sample.  Detailed studies
at several wavelengths (radio: \citealt{norrisforbes95}, IR:
\citealt{kotilainen96}, optical: \citealt{lipari}, X-ray:
\citealt{moran99}, \citealt{lira}) have shown that the energetic
output of this galaxy is powered by star formation occurring at
several locations, but mainly in the two radio cores discovered by
\citet{norrisforbes95} and also detected with {\em Chandra}
(\autoref{fig_3256}).

\begin{figure}[tp]  % kotilainen
  \begin{center} 
  \ifthenelse{\value{altaris} = 1}{%
      \includegraphics[width=\textwidth]{kotilainen.ps}  
  }{  \includegraphics[width=\textwidth]{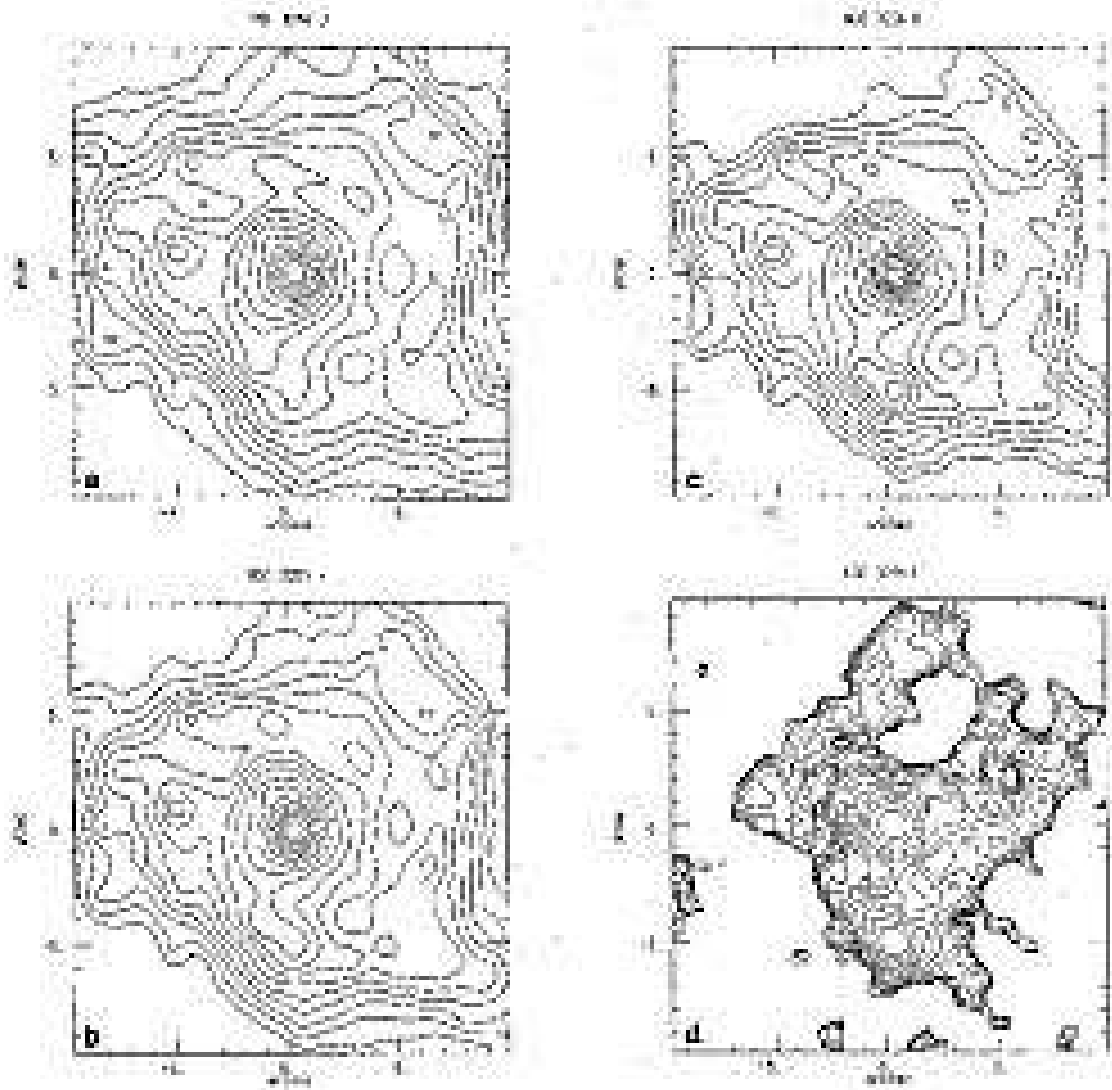}  
  }\\
  \end{center}
  \caption {A sequence of infrared images of the centre of NGC3256 at
    increasing wavelenghts, from \citet{kotilainen96}. The southern
    nucleus lies behind a dust lane, and becomes visible in the
    L$^\prime$ band. The two nuclei are the brightest sources in
    NGC3256 in the radio and 2.0--10 keV bands (cfr.\ \autoref{fig_3256}).}
  \label{fig:duenuclei}
\end{figure}

\begin{figure}[tp]    % firx e radiox per i due nuclei
  \begin{center} 
      \includegraphics[width=0.6\textwidth]{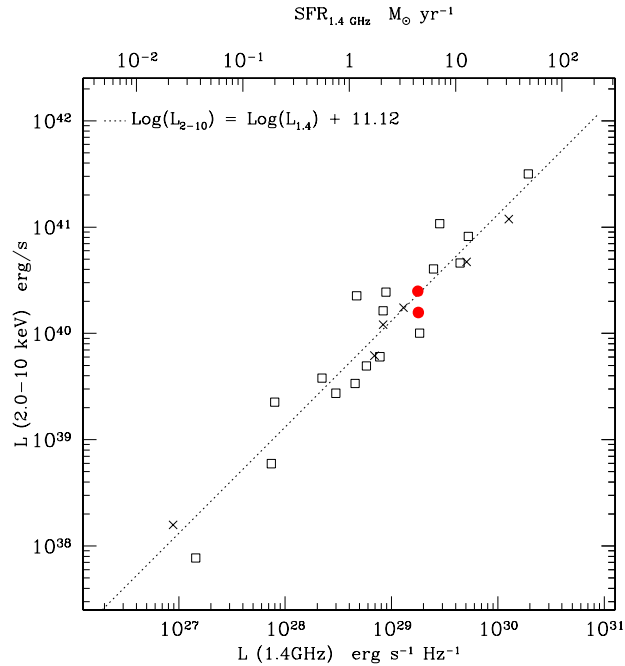}
  \end{center}
  \caption {2.0--10 keV vs.~radio luminosities of the two nuclei of
    NGC 3256 (filled circles). The upper and lower circles represent
    the north and south nucleus, respectively.  The other symbols
    (same as in Fig.~\ref{pranalli-E3_fig:fig1}) represent the other
    galaxies in the sample.  }
  \label{fig:duenuclei}
\end{figure}

The 3 and 6\,cm radio maps \citep{norrisforbes95} reveal two distinct,
resolved (FWHM $\sim1.2\arcsec$) nuclei and some fainter diffuse radio
emission. Separated by $5\arcsec$ in declination, the two cores
dominate the radio emission, the northern one being slightly (15\%)
brighter. They share the same spectral index ($\alpha\sim 0.8$).
\chandra\ observations (\autoref{fig_3256}) have shown that they also
have similar 2--10 keV fluxes.  Both of them follow the radio/hard
X-ray correlation (Fig.~\ref{fig:duenuclei}), while only the northern
one follows the radio/soft X-ray correlation.  At other wavelengths
the northern core is the brightest source in NGC 3256, while the
southern one lies behind a dust lane and is only detected in the far
infrared ($\lambda > 10\mu$m), as clearly shown in the sequence of
infrared images at increasing wavelengths in \citet{kotilainen96}.

Although the southern core appears as a bright source in the hard
X-rays ($E>2$ keV), there are not enough counts to allow an accurate
spectral fitting. However, it is still possible to constrain the
absorbing column density by assuming a template spectrum, such as a
simple power-law or the spectrum of the northern core, leading (after
standard processing of the {\em Chandra} archival observation of NGC
3256) to an intrinsic $N_{\rm H} = (2.2\pm 1.0) \cdot 10^{22}$
cm$^{-2}$ ($N_{\rm H, Gal}=9.5\cdot 10^{20}$ cm$^{-2}$), fully
consistent\footnote{Assuming $A_{\rm V}=4.5\cdot 10^{22} N_{\rm H}$
  (Galactic value).}  with the $A_{\rm V}=10.7$ estimated by
\citet{kotilainen96} from infrared observations. We note that 84\% of
the 2--10 keV flux and only 10\% of the 0.5--2 keV one are transmitted
through this column density. Thus, while the flux loss in the hard
band is still within the correlation scatter, the larger loss in the
soft band throws the southern nucleus off the correlation of
Eq.~(\ref{eq:radiox}).

\section{Effect of variability on the flux estimate}
\label{sect:gal-individuali}
\subsection*{IC\,342: variability of ULXs}
In the ASCA observations, the X-ray emission of \object{IC\,342} (a
face-on spiral galaxy at 3.9 Mpc) is powered by three main sources;
two of them (source 1 and 2 according to \citealt{fabtrinch87}) are
ultraluminous X-ray binaries (ULX) while source 3 is associated with
the galactic centre. Observations with higher angular resolution
(ROSAT HRI, \citealt{bregman93}) showed that sources 1 and 2 are
point-like while source 3 is resolved in at least three sources.  Our
main concern in determining the flux of this galaxy is the variability
of the two ULX, which was assessed by a series of observations
spanning several years: IC342 was first observed by {\em Einstein} in
1980 \citep{fabtrinch87}, then by ROSAT in 1991 \citep{bregman93},
by ASCA in 1993 \citep{okada98} and 2000 \citep{kubota01}, and by \XMM\ in
2001.

In Table \ref{ic342ulx} we report soft X-ray fluxes for sources 1 and
2. We have chosen the soft X-ray band due to the limited energy band
of both {\em Einstein} and ROSAT; the fluxes observed with these
satellites were obtained from the count rates reported in
\citet{fabtrinch87} and \citet{bregman93} assuming the powerlaw and
multicolor disk examined in \citet{kubota01} for source 1 and 2
respectively; we take ASCA 1993 and 2000 fluxes from \citet{kubota01}.
\XMM\ archival observations were reduced by us with SAS 5.3 and the
latest calibrations available.

Source 1 was in a low state ($F_{0.5-2} \sim 3-5\cdot 10^{-13}$ erg
s$^{-1}$) during the 1980, 1991, 2000 and 2001 observations, and in a
high state ($F_{0.5-2} \sim 16\cdot 10^{-13}$ erg s$^{-1}$) during the
1993 observation. The broad-band (0.5-10 keV) spectrum changed, its
best-fit model being a disk black-body in 1993 and a power-law in 2000
and 2001.  Source 2 has also shown variability, its 0.5-2.0 keV flux
oscillating between $0.52\cdot 10^{-13}$ (ASCA 2000) and $2.56\cdot
10^{-13}$ (ASCA 1993) erg s$^{-1}$; the main reason for this
variability being the variations in the strongly absorbing column
density, which was $9.9\cdot 10^{21}$ cm$^{-2}$ in 1993 and $18\cdot
10^{21}$ cm$^{-2}$ in 2000. The spectrum was always a power-law.

The high state for source 1 seems thus to be of short duration, and we
feel confident that its time-averaged flux may be approximated with
its low state flux. We thus choose to derive our flux estimate for
IC\,342 from the ASCA 2000 observation, estimating the variation for
the total flux of the galaxy caused by source 2 variability to be less
than 10\%.

% ~~~~~~~~~~~~~~~~~~~~~~~~~~~~
% TABELLA: SRC 1 & 2 IN IC 342
% ~~~~~~~~~~~~~~~~~~~~~~~~~~~~
\begin{table} \label{ic342ulx}
\centering\begin{tabular}{clccc}
& & \multicolumn{2}{c}{\sc Src 1 Flux} &{\sc Src 2 Flux} \\
{\sc Year} &{\sc Mission} &{\sc bb} &{\sc po} &{\sc po} \\
\hline
1980 &{\em Einstein}   &2.7 &4.1  &0.85   \\
1991 &ROSAT            &3.3 &2.7  &1.3    \\
1993 &ASCA             &16  &     &2.6    \\
2000 &ASCA             &    &5.0  &0.52   \\
2001 &XMM-{\em Newton} &    &4.1  &1.3    \\
\hline
\end{tabular}
\caption{0.5--2.0 keV fluxes (in $10^{-13}$ erg s$^{-1}$ cm$^{-2}$) for sources 1
and 2 in IC\,342 for a blackbody (bb) and a power-law (po) model.}
\end{table}

\subsection*{Variability in M82}
Hard (2--10 keV) X-ray variability in M82 was reported in two
monitoring campaigns with ASCA (in 1996, \citealt{ptak99}) and RXTE
(in 1997, \citealt{rephaeli02}). M82 was found in ``high state''
(i.e.~$4\cdot 10^{-11} \lesssim f_{2-10} \lesssim 7\cdot 10^{-11}$) in
three out of nine observations with ASCA and in 4 out of 31
observations with RXTE. In all the other observations it was in a ``low
state'' ($2\cdot 10^{-11}\lesssim f_{2-10} \lesssim 3.5\cdot
10^{-11}$). A low flux level was also measured during the observations with other
experiments: HEAO~1 in 1978 \citep{griffiths79};
{\em Einstein} MPC in 1979 \citep{watson84};
EXOSAT in 1983 and 1984; BBXRT
in 1990; ASCA in 1993 \citep{tsuru97}; BeppoSAX in 1997
\citep{cappi99}; \chandra\ in 1999 and 2000, \XMM\ in 2001.  No
variability was instead detected in the 0.5--2.0 keV band
\citep{ptak99}.

The high state has been
of short duration: less than 50 days in 1996, when it was observed by
ASCA, %(M82 was in low state on 23.3.1996, then in high state on 15.4,21.4,24.4,
%and in low state on 13.5)
 and less than four months in 1997.
%(RXTE detected M82 in low state on 21.7.1997, in high state on 13.8,24.9,2.11,25.11; BeppoSAX
%detected M82 again in low state on 6.12).
A monitoring campaign was also undertaken with {\em Chandra}, which
observed M82 four times between September 1999 and May 2000.
We reduced the archival data, and found that
the galaxy was always in a low state, with its flux slowly increasing
from $1.2\cdot 10^{-11}$ to $2.9\cdot 10^{-11}$.
%dire qualcosa sul fatto che comunque chandra non ha potuto determinare
%chi è a fare questa variabilità, anche se il candidato più probabile
%è comunque la sorgente più luminosa (quella di kaaret, e prima ancora
%di watson, e di collura)?   o va bene così?

We do not attempt a detailed analysis of the variability (see
\citealt{rephaeli02}); however, we feel confident that, given the
short duration of the high states and the fact that the difference
between high- and low state flux is about a factor 2, the
time-averaged flux of M82 can be approximated with its low state
flux. We thus choose to derive our flux estimate for M82 from the
BeppoSAX 1997 observation ($f_{2-10}=2.9\cdot 10^{-11}$ erg s$^{-1}$
cm$^{-2}$), estimating the uncertainty caused by
variability to be around $30\%$.

\chapter{Star-forming galaxies in the Hubble Deep Field}
\label{hdf_chap}

The X-ray and radio observations in deep fields reach limiting fluxes
deep enough to detect star-forming galaxies at large redshifts
$z\gtrsim1$. Thus they can be used to check whether the radio/X-ray
relation holds also for distant galaxies.

We consider the surveys performed in the {\em Hubble Deep Field North}
(HDFN), a region of sky at a high Galactic latitude which was selected
because of the very low extinction due to our Galaxy, centred at the
(J2000) coordinates $12^{\rm h}\ 36^{\rm m}\ 49.5^{\rm s}\ +62^\circ\ 
12^\prime\ 58\arcsec$. The \chandra\ survey centred on the HDFN region
(also called \chandra\ Deep Field North, CDFN) was initially performed
for an assigned time of 1 million seconds \citep{bran01} and then
extended with one other million secondes \citep{alexander03}. It is
customary to refer to the intermediate data release as the `1~Ms
survey' and to the final one as the `2~Ms survey'. The final survey
reaches a limiting flux of $2.5\e{-17}$ \ergscmq\ in the 0.5--2.0
keV band for sources in the centre of the field.  The HDFN was also
imaged at radio wavelengths, with the Very Large Array (VLA) at 8.4
GHz \citep{vla98} and at 1.4 GHz \citep{vla00}, and with the
Westerbork Synthesis Radio Telescope (WSRT) at 1.4 GHz \citep{wsrt}.
The limiting fluxes for these radio surveys are all around
0.05~mJy.

We searched for X-ray counterparts of radio sources in the
\citet{vla98} catalogue which contains optical and IR identifications
allowing the selection of candidate star-forming galaxies.  Our
selection criterium has been to include all galaxies with Spiral or
Irregular morphologies, known redshifts and no AGN signatures in their
spectra (from \citealt{vla98} or \citealt{cohen}).  The X-ray
counterparts were initially searched for in the 1 Ms catalogue
\citep{bran01}. When the final data release was made available, we
reduced the X-ray data, generated a catalogue, and we looked for
X-ray counterpars of the radio sources. Our data reduction method
followed rather closely the one described in \citet{bran01}. Unless
otherwise stated, all data reported here come from reduction of the
complete 2~Ms CDFN survey.

The mean positional uncertainties of both \chandra\ (for on-axis
sources) and the VLA are $\sim 0.3\arcsec$, which added in quadrature
give $\sim 0.5\arcsec$. Using this value as the encircling radius for
coordinate matching 5 galaxies were found in \citet{bran01}.  However,
there are two effects that may increase this value:
\begin{enumerate}
\item the shape and width of the \chandra\ PSF strongly depend on the
  off-axis and azimutal angles. Since the 2 Ms HDFN data consist of 20
  observations with different pointing directions and position angles,
  there is no unique PSF model even for near on-axis sources;% \citet{bran01}
%estimates a positional uncertainty of $\sim 0.6\arcsec$ for on-axies sources
%in the HDFN. Adding this value to the VLA one gives $0.7\arcsec$, within which
%6 galaxies are matched;
\item a displacement %of several kpc
  between the brightest radio and X-ray positions, induced e.g.~by an
  ultraluminous X-ray binary placed in a spiral arm and dominating the
  X-ray emission.
\end{enumerate}
Thus, by making cross-correlations with increasing encircling radii we found
that the number of coincidences increases up to a radius
of $1.0\arcsec$, yielding 7 matchings. There are no further coincidences
up to a radius of several arcsecs, indicating that the sample should not
be contamined by chance coincidences. 

By performing the same cross-correlation on the 2~Ms CDFN data, we
added 4 more galaxies with coincidences within $1.0\arcsec$, thus
totalling 11 sources.  One of these
the 11 galaxies was dropped since it is probably an AGN on the basis
of X-ray spectral and variability properties, as discussed below.  The
selected objects are listed in Table~\ref{deepnomi}.  Fluxes at
1.4\,GHz and spectral slopes were retrieved from \citet{vla00} in 9
cases, and from \citet{wsrt} in one case.

\begin{table}  
\centering\begin{tabular}{lc }
\hfil {\sc Chandra} &{\sc VLA} \\
\hline
\object{CXOHDFN J123634.4+621212}   (134) &3634+1212\\
\object{CXOHDFN J123634.5+621241}   (136) &3634+1240\\
\object{CXOHDFN J123637.0+621134}   (148) &3637+1135\\
\object{CXOHDFN J123651.1+621030}   (188) &3651+1030\\
%\object{CXOHDFN J123653.4+621139}   (194) &3653+1139\\
\object{CXOHDFN J123708.3+621055}   (246) &3708+1056\\
\object{CXOHDFN J123716.3+621512}   (278) &3716+1512\\
not present in \citet{bran01}             &3644+1249\\
{\em id.}                                 &3701+1146\\
{\em id.}                                 &3638+1116\\
{\em id.}                                 &3652+1354\\
%HDFN--Cardiel                             &???\\
\hline
\end{tabular}
\caption{Identification of candidate star-forming galaxies in the \citet{vla98}
catalogues with their X-ray counterpart \citep{bran01}. Where
applicable, the entry number in the \chandra\ catalogue is shown in
parenthesis.
\label{deepnomi}
}
\end{table}

\begin{figure}[t]    % deep field hardness ratios
  \begin{center} 
     \includegraphics[width=0.7\textwidth]{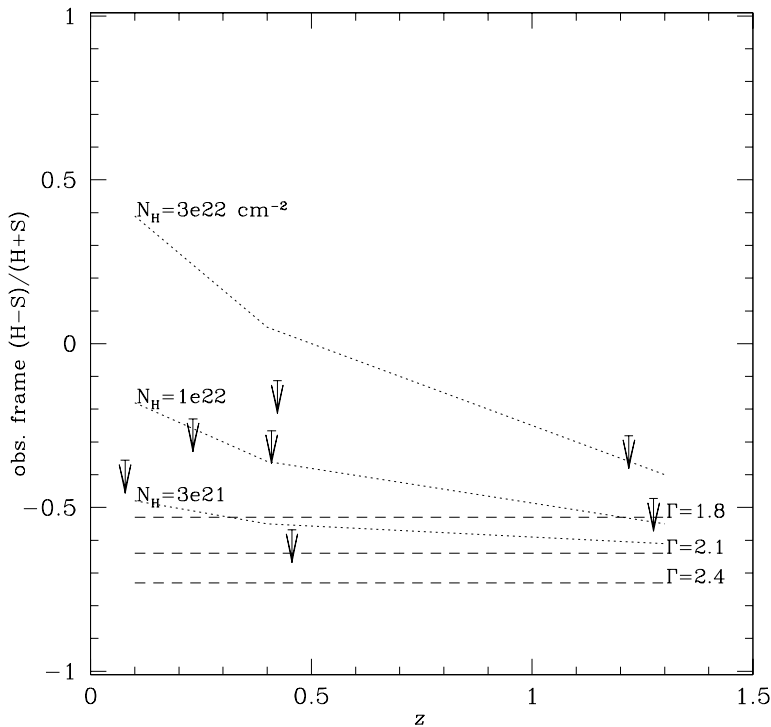}
  \end{center}
  \caption { Observer's-frame hardness ratios for the 7 galaxies in the CDFN
    catalogue by \citet{bran01}; there are only upper limits since
    none is detected in the 2--8 keV band.  The dotted lines show the
    {\em loci} for power-law spectra with slope $\Gamma=2.1$ and
    different intrinsic rest-frame absorption; the dashed lines are
    for spectra of different slopes and no intrinsic
    absorption. %Figure taken from \citet{rcs03}.
\label{fig_hr}
}
\end{figure}

While the rest-frame 0.5--2.0 and 2.0--10 keV luminosities could be
obtained by K-correcting the observer's frame counts, this would imply
the assumption of a spectral shape; but none of the %seven 
deep field galaxies is individually detected in the hard band, %FIXME-davvero?
so that any constraint on their spectra obtained with the use of a
hardness ratio diagram is too loose to be significant
(Fig.~\ref{fig_hr}).
%Thus, the derived luminosities would suffer of an additional uncertainty
%coming from a somewhat arbitrary assumption of the spectrum.
However this problem may be partially circumvented by resizing the
X-ray bands in the observer's frame according to the redshift of the
objects -- i.e.\ 
%However, it is possible to get rid of the K-correction if the redshift
%information is properly used: the bands from which counts are
%extracted can be resized according [...]    Thus
%   recupero un vecchio pezzo di testo, così mi spiego meglio
the counts extracted with this method correspond to the photons actually
emitted in the rest-frame 0.5--2.0 and 2--8 keV bands.
It is therefore possible to give better constraints to the
spectra of the deep field galaxies and derive better estimates of
their luminosities.

\begin{figure*}[tp]    % deep field lumin
  \begin{center} 
      \includegraphics[height=0.42\textheight]{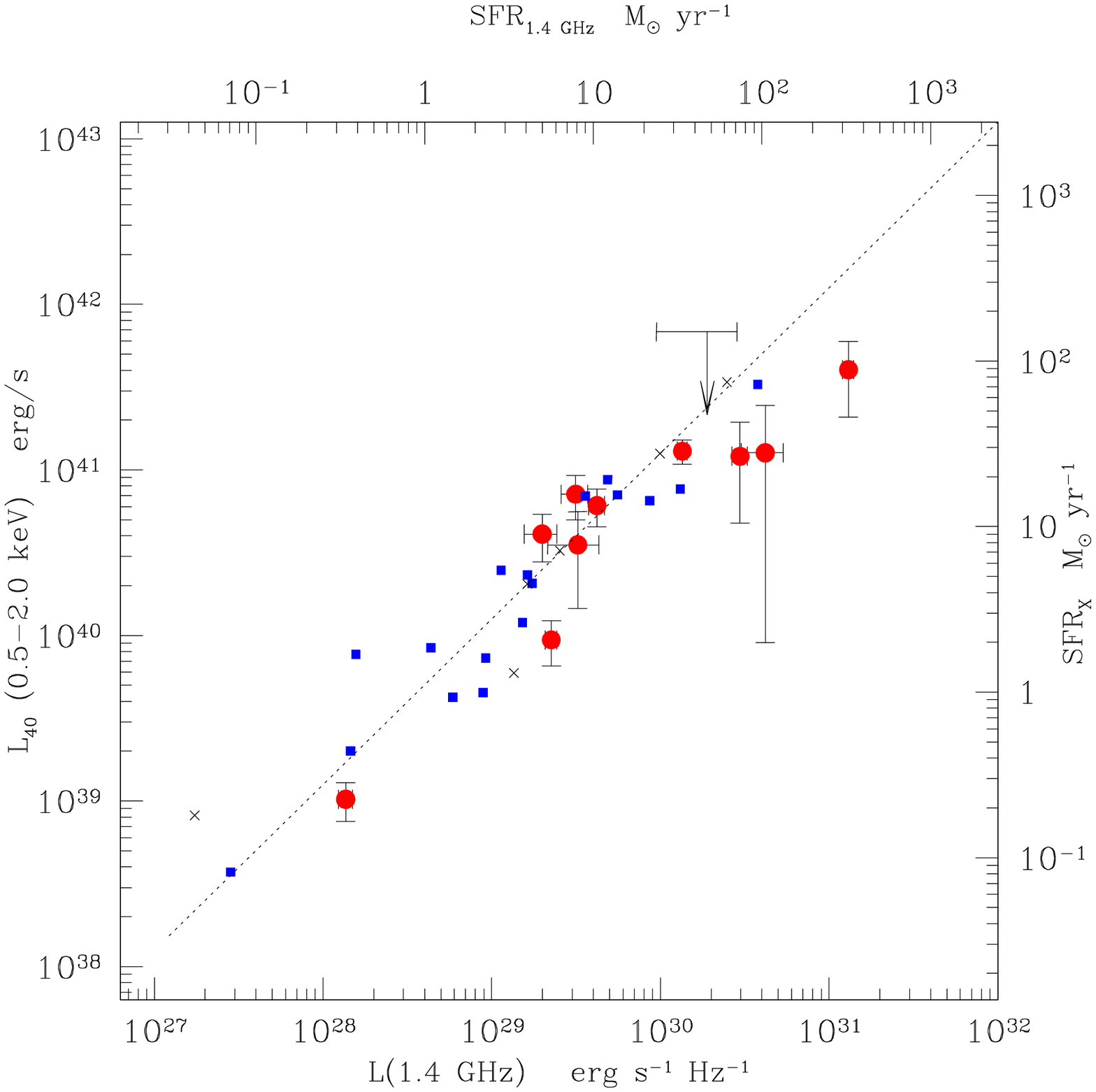}
      %\hskip.8cm
      \includegraphics[height=0.42\textheight]{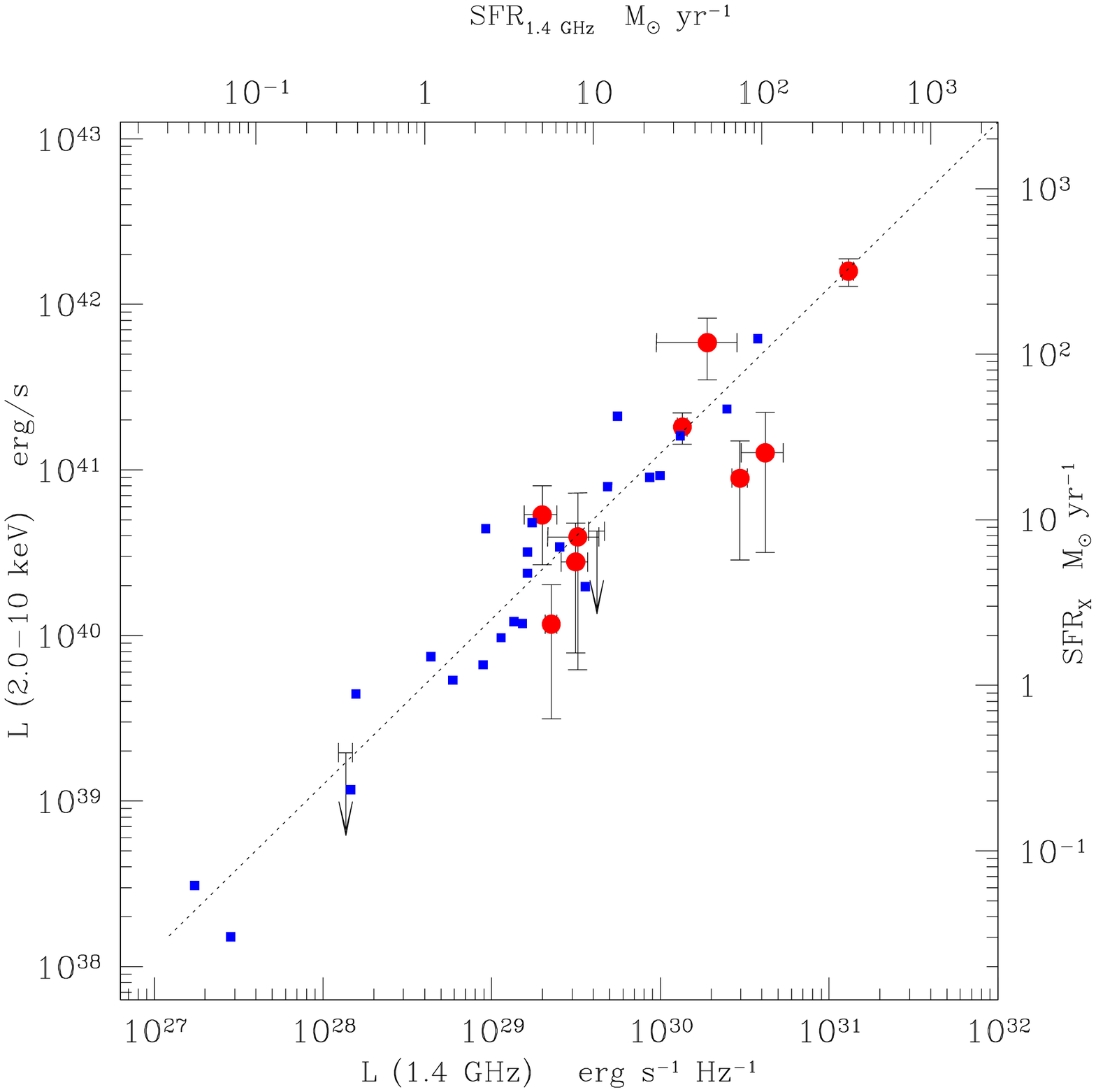}
  \end{center}
  \caption { The radio/X-ray luminosity relation for galaxies in the Hubble 
   Deep Field (filled circles). Squares: local galaxies
    as in Fig.~(\ref{pranalli-E3_fig:fig1}); dotted lines:
    linear fits for local galaxies (Eq.~% 
   \ref{eq:radiox}). Besides the 10 deep field galaxies discussed in
   the text, here is shown also the point relative to the galaxy
   hd2-264.1 (in the Hubble Deep Field, at $z=0.475$) which is the
   only galaxy from the sample of \citet{cardiel03} to have an X-ray
   detection ($\Lsx=1.5\e{41}$ erg/s, $\Lhx=5\e{40}$ erg/s,
   $L_{1.4}=6\e{29}$ erg/s/Hz). 
%% The two most luminous galaxies
%%    are at $z=1.275$ and 1.219 in order of increasing radio luminosity.
%%     The two upper limits (3$\sigma$ level) in the left panel refer to source \#194 (see text) 
%%   and show the unabsorbed luminosity as estimated with $\Gamma\sim 0.1$ and no intrinsic
%%    absorption (lower) and with $\Gamma\sim 2.1$ and $N_{\rm H}\sim 3.4
%%    \cdot 10^{22}$ cm$^{-1}$ (upper).
  \label{fig_deeplumin}
  }
\end{figure*}

%In order to obtain information about their spectra, for each source
%we extracted counts in {\em redshifted} bands corresponding to
%rest-frame 0.5--2.0 and 2.0--8.0 keV. The advantage is that for $z\sim
%1$ the hard band becomes $\sim$1-4 keV in observer's frame, which
%corresponds to the range of highest sensitivity of {\em Chandra}.
Thus, we have redefined the soft and hard%
\footnote{Since \chandra\ has very poor sensitivity between 8 and 10
  keV, the use of the reduced 2--8 keV band enhances the signal/noise
  ratio. Note that while counts are extracted in the 2--8 keV band,
  fluxes and luminosities are always extrapolated to the 2--10 keV
  band.} 
bands as the [0.5; $2.0/(1+z)$] and [$2.0/(1+z)$; $8/(1+z)$]
intervals, respectively.  Another advantage of this procedure is that
the higher the redshift, the more akin the new hard band is to the
zone of maximum sensitivity of \chandra\ ($\sim1-4$ keV). Note that
since the ACIS-I detector has almost no sensitivity below 0.5 keV, we
fixed this energy as the lower limit for count extraction. For the two
highest redshift galaxies, this reduces the soft band to 0.5--0.9 keV,
still significantly larger than the ACIS-I energy resolution
(FWHM~$\lesssim 100$ eV).

We extracted counts in circular regions (with radii of 5 pixel, i.e.\ 
$\sim 2.5\arcsec$) around our selected targets; the background counts
were taken in circular annuli surrounding the targets, with 
outer radii of 12 pixels ($\sim 6\arcsec$), respectively. % The net counts were
%converted in count rates with the exposure times listed in the
%\citet{bran01} catalogue. 
Counts and rates are reported in Table \ref{tab_deep}.  Best-fit
slopes reproducing the soft/hard count ratio were derived by assuming
a power-law spectrum with Galactic absorption. We find that all
objects have spectral slopes falling in the range $\Gamma\sim
1.5$--$2.7$ (Table \ref{tab_deep}).  To check whether these spectra
are consistent with those of the galaxies in the local sample we
calculated the observed soft/hard flux ratio for galaxies in the local
sample of \S~\ref{campione_section}: the median value for this flux
ratio is 0.95 leading to a slope $\Gamma\sim 2.1$.  The count ratios
for each of the six deep field galaxies are consistent within
1--2$\sigma$ with the $\Gamma\sim 2.1$ slope.

One of the sources (\#194 in \citealt{bran01}, not shown in Table~%
\ref{tab_deep}, with $z=1.275$) has an upper limit on the soft X-ray
counts and a count ratio not consistent at the $5\sigma$ level with
the unabsorbed $\Gamma\sim 2.1$ spectrum: it requires an inverted
spectrum ($\Gamma<0.1$) if no absorption is assumed, otherwise, if we
assume $\Gamma=2.1$, the intrinsic absorbing column has to be $N_{\rm
  H}\gtrsim 3.4\cdot 10^{22}$ cm$^{-2}$.  We also find that the flux
of this source is strongly variable, since all detected counts were
recorded in the first half of the survey, and no one in the second
half which was performed about one year later.  Thus, from its
spectral and variability properties, we classified this source as an
AGN and dropped it from the sample considered here.

With the best-fit slopes we derived soft and hard band fluxes and
luminosities (Fig.~\ref{fig_deeplumin}).
The linear radio/X-ray correlations hold also for the Deep Field galaxies;
the dispersion of the relations given in \S\ref{correlazioni_section}
is not changed by the inclusion of the deep field objects.

A similar result was also found by \citet{bauer02} by analyzing a
sample of 20 galaxies selected at 1.4 GHz in the \chandra\ Deep Field
North. Their sample is partly overlapping with the one here described
(not identical, since ours was selected at 8.4 GHz).

%  FIXME -- QUESTO PEZZO CHE ERA STATO CANCELLATO MAGARI VALE LA PENA
%  DI RECUPERARLO:

%: the best-fit
%relations for the deep field sample are:
%\begin{eqnarray}
%\Log (L_{\rm 0.5-2}) &= (0.9\pm 0.1) &\Log (L_{\rm 1.4}) + 14\pm 3 \\
%\Log (L_{\rm 2-10}) &= (1.0\pm 0.1) &\Log (L_{\rm 1.4}) + 11\pm 2.98 
%   \label{eq:radioxhdf}\qquad. 
%\end{eqnarray}
%Note that in these fits the two ($3\sigma$) upperlimits were treated
%as detections. 
%the slope of the soft X-ray/radio relation is still
%consistent with unity within $1\sigma$, despite 

% ~~~~~~~~~~~~~~~~~~~~~~~~~~~~~
%  TABELLA: CHANDRA DEEP FIELD
% ~~~~~~~~~~~~~~~~~~~~~~~~~~~~~
\begin{sidewaystable}%[p]
\hskip-2em\vbox{\centering\begin{tabular}{cccccccccccccc}
\multicolumn{12}{c}{\sc Fluxes and Luminosities: Deep Sample} \\
\\
& & &\multicolumn{3}{c}{\sc Soft X-rays}&\multicolumn{3}{c}{\sc Hard X-rays}&&\multicolumn{3}{c}{\sc Radio}\\
& & &\multicolumn{3}{c}{\downbracefill}&\multicolumn{3}{c}{\downbracefill}&&\multicolumn{3}{c}{\downbracefill}\\
\multicolumn{2}{c}{\sc Source} &$z$ &{\sc Net Cts.} &$F$ &$L$ &{\sc Net Cts.} &$F$ &$L$ &$\Gamma$ &$F$ &$L$ &$\alpha$  \\
\hline
%% 134 &0.456 &$31\pm  6$   &23   &29   &$22 \pm  6$  &28  &36   &$2.0_{-0.3}^{+0.4}$ & 210 &2.6  & 0.7\\%
%% 136 &1.219 &$5.9\pm 3$   &6.7  &95   &$22 \pm  5$  &19  &260  &$1.6_{-0.5}^{+0.9}$ & 180 &26   & 0.7\\%
%% 148 &0.078 &$14\pm  4$   &8.4  &0.23 &$< 2.1$     &$<4.1$ &$<0.11$ &$>2.6$ & 96  &0.027& 0.6 \\%
%% 188 &0.410 &$15\pm  4$   &13   &13   &$ 5.2 \pm  4$  &5.8 &5.8  &$2.7_{-0.6}^{+1.5}$ &  83 &0.82 & 0.6\\%
%% 194 &1.275 &$< 1.6$      &$<0.88$ &$<$14 &$23 \pm  5$  &20  &310  &$<0$&  60 &9.6  & 0.9\\%
%% 246 &0.423 &$ 6.2\pm 4$&3.7  &4.0  &$ 5.4 \pm  4$  &7.5 &8.1  &$1.7_{-0.6}^{+1.4}$ &  36 &0.39 & 0.4\\%
%% 278 &0.232 &$11\pm  4$ &5.8  &1.6  &$ 7.8 \pm  4$  &16  &4.4  &$1.5_{-0.4}^{+0.7}$ & 160 &0.44 & 0.2\\%
134 &3634+1212  &0.456 &55  &20   &  13  &39   &28     &   18  &1.9 &210 &1.4   &0.7\\
136 &3634+1240  &1.219 &11  &5.6  &  40  &44   &22     &   160 &1.4 &180 &13    &0.7\\
148 &3637+1135  &0.078 &27  &7.2  & 0.10 &---  &$<$4.6 &$<$0.065 &--- &96  &0.014 &0.6\\
188 &3651+1030  &0.410 &24  &12   &  6.1 &4.3  &$<$2.8 &$<$1.4 &--- & 83 &0.42  &0.6\\
246 &3708+1056  &0.423 &22  &7.5  &  4.1 &13   &9.8    &   5.3 &1.4 & 36 &0.20  &0.4\\
278 &3716+1512  &0.232 &22  &6.6  & 0.94 &8.7  &8.2    &   1.2 &1.4 &160 &0.23  &0.2\\
new &3644+1249  &0.557 &9.2 &3.4  &  3.5 &7.3  &3.8    &   3.9 &2.1 &31  &0.32  &0.7\\
new &3701+1146  &0.884 &6.7 &3.8  &  12  &6.9  &2.8    &   8.9 &3.4 &93  &3.0   &0.5\\
new &3638+1116  &1.018 &4.6 &2.8  &   13 &7.4  &2.8    &    13 &2.2 &92  &4.2   &1.1\\
new &3652+1354  &1.355 &3.8 &$<$2.4 &$<$23 &16 &6.2    &   59  &--- &20  &1.9   &---\\
%new &hd2-264_1 &0.475 &14  & 10  &  7.1 &7.2  & 3.9    &   2.8 &2.8 &44  &0.31  &---\\
\hline
\end{tabular}
%\object{NGC253} &1 &2 &3 &4 &5 &6 &7 &8 &9 &10\\
%\object{M82} &1 &2 &3 &4 &5 &6 &7 &8 &9 &10\\
%\multicolumn{3}{c}{\sc Local Sample}\\ 
%\hline\hline
%
%\object{M82}*     &\object{NGC\,2276} &\object{NGC\,4449}\\
%\object{M101}    &\object{NGC\,2403} &\object{NGC\,4631}\\
%\object{M108}    &\object{NGC\,2903} &\object{NGC\,4654}\\ 
%\object{NGG\,891}  &\object{NGC\,3310}&\object{NGC\,6946}\\
%\object{NGC\,1569} &\object{NGC\,3367} &\object{IC\,342}  \\
%\object{NGC\,2146} &\object{NGC\,3690} \\
%\\
%\multicolumn{3}{c}{\sc Supplementary Sample }\\
%\hline\hline
%\object{NGC\,55}       &\object{NGC\,1672} &\object{NGC\,3256}\\
%\object{NGC\,253}*     &\object{NGC\,1808} &\object{Antennae}\\
%\end{tabular}
%{\sc VLA98 Id.} &{\sc Chandra \#} &$z$ &{\sc Extr. Band} &{\sc Soft Counts} &{\sc Hard Counts} &L210 &F210 &L210 &S14 &L14 &{\sc Notes} \\
%4,6  22:30 -- rit: 8.6 la fra frp lor -- albergo: 50eur bb 85 mp 100 pc
%CXOHDFN J123634.4+621212
%CXOHDFN J123634.5+621241
%CXOHDFN J123637.0+621134
%CXOHDFN J123651.1+621030
%CXOHDFN J123653.4+621139
%CXOHDFN J123708.3+621055
%CXOHDFN J123716.3+621512
}
\caption{
 Data for deep field galaxies (counts, fluxes, luminosities). 
 $\Gamma$ is the
 best fit X-ray slope (photon index), $\alpha$ the radio slope (energy index).
 Sources are identified via their entry number
 in the \citet{bran01} catalogue (cfr.~Table \ref{deepnomi}) and their
 denomination in the VLA catalogue \citep{vla98}.
 X-ray fluxes in $10^{-17}$
  erg s$^{-1}$ cm$^{-2}$, radio fluxes in $\mu$Jy. X-ray luminosities in
 $10^{40}$ erg s$^{-1}$, radio luminosities in $10^{30}$ erg s$^{-1}$ Hz$^{-1}$. 
 X-ray counts are extracted in redshifted bands (soft band $:=0.5$--$2.0/(1+z)$ keV,
 hard band $:=2.0/(1+z)$--$8/(1+z)$ keV); the counts shown here are
 already background subtracted. X-ray fluxes and luminosities are in rest-frame
 0.5--2.0 and 2.0--10 keV bands, radio ones at rest-frame 1.4\,GHz.
 For source \#194, the absorbed flux and luminosity are quoted. The unabsorbed luminosity
 is $\lesssim 4\cdot 10^{42}$ erg s$^{-1}$. Upper limits on fluxes and
 luminosities were assigned whenever the estimated error on the flux
 was of the same magnitude of the flux.
\label{tab_deep}
}
\end{sidewaystable}

\chapter{The X-ray number counts and luminosity function of galaxies}
\label{fondox_chap}

An estimate of the contribution of star-forming galaxies to the cosmic
X-ray background (XRB) has been attempted several times (e.g.\ 
\citealt{bookbinder80}, \citealt{gp90}, \citealt{moran99}).  The main
purpose for the earlier studies was the possibility to explain the
flatness of the XRB spectrum via the X-ray binaries powering the X-ray
emission of these galaxies (High Mass X-ray Binaries have flat spectra
with slopes $\Gamma\sim 1.2$).  Although AGN have since a long time been
recognized to provide by far the most important contribution to the
XRB \citep{sw89,comastri95}, the ongoing deep \chandra\ and \XMM\ 
surveys offer unique opportunities to both test the AGN models and pin
down the contribution from other kind of sources.

A study of the X-ray number counts and of the luminosity function (LF)
of galaxies is also valuable in the perspective of studying the
evolution of the X-ray luminosity function (XLF) and the cosmic star
formation history.  In this chapter, the X-ray luminosity function and
number counts of star forming galaxies are determined making use of
different approaches, such as objects selection from X-ray surveys,
conversion of radio and FIR LFs and number counts. An estimate is also
performed of the contribution to the XRB. Finally,
the possibilities to use X-ray observations to determine the cosmic
star formation history are discussed.

\section{The X-ray luminosity function of galaxies}
\label{sect_norman}

The availability of multiwavelength photometry and spectral
information \citep{barger03,szokoly04} for the X-ray objects in the \chandra\ 
Deep Fields allows a direct identification of the star forming
galaxies in these surveys. This is however a difficult task, for two
reasons: i) at the limiting fluxes of current X-ray surveys, the
majority of detected sources are AGN; ii) it may prove to be
difficult, for objects at redshifts $z\sim 1$, to distinguish a bright
starburst galaxy from a low luminosity type-2 Seyfert galaxy (cfr.\ 
\S~\ref{sec:introX}).  Thus, in order to pick up normal galaxies in
the X-ray survey samples, extreme care has to be put in the choice of
the selection tools. We will not venture into a
comprehensive discussion of the selection tools (good starting points
for this might be \citealt{machalski99} and \citealt{colin}), but just
briefly illustrate a few of them that will be used throughout this chapter.

A direct determination of the X-ray luminosity function (XLF) of
galaxies has been recently attempted in \citet{colin}, where a sample
was defined containing 210 galaxies with known redshift from the
\chandra\ Deep Field catalogues \citep{alexander03,giacconi02}.  A
Bayesian approach was chosen to derive a selection probability from
the values of three different parameters:
\begin{itemize}
\item the 0.5--2.0 keV X-ray luminosity for star forming galaxies is usually
  less than $\sim 10^{42}$ erg/s;
\item star forming galaxies have a softer spectrum than AGN; this
  translates in selecting objects with an hardness ratio $HR\lesssim -0.8$
  (with $HR=(H-S)/(H+S)$, where $H$ and $S$ represent the 2.0--10 and
  0.5--2.0 keV fluxes respectively);
\item an X-ray/optical flux ratio $\Log (F_{{\rm X-ray}}/F_{{\rm
      opt}})<-1$ (see below).
\end{itemize}

The optical flux is defined as
\begin{equation}
  \Log\ F_{\rm opt} = -0.4 R -5.5
\end{equation}
were $R$ is the red magnitude in the Kron-Cousins system.  An
X-ray/optical flux ratio equal (in logarithm) to $-1$ is usually taken as an
approximate boundary between normal galaxies and Seyferts
(\citealt{maccacaro88}; see also \citealt{mchardy03}).

We will consider the local differential luminosity function
$\varphi(L) \de \Log L$, which is the comoving number density of
sources per logarithmic interval of luminosity.  A binned luminosity
function (LF) was derived with the method developed in
\citet{pageca00}, which is a variant on the classical $1/V_{\rm max}$
method by \citet{schmidt68}. In a given bin of redshift and
luminosity, the density of galaxies may be written as
\begin{equation}
  \varphi \sim \frac{N}{
\int_{L_{\rm min}}^{L_{\rm max}} \int_{z_{\rm min}}^{z_{{\rm max}(L)}}
\int_0^{4\pi}
\frac{\de V}{\de z\,\de \Omega}\, \de \Omega\, \de z\, \de L}
\end{equation}
where $\de V/\de z \de \Omega$ is the comoving volume per unit
steradian sampled by the surveys.  For a flat universe with non-zero
cosmological constant we may take
\begin{equation}
\frac{\de V}{\de z} = 4\pi \left(\frac{c}{H_0}\right)^3 
\frac{1}{E(z)} \left[ \int_0^z \frac{\de z^\prime}{E(z^\prime)} \right]^2
\end{equation}
with
\begin{equation}
E(z)=\sqrt{\Omega_{\rm M}(1+z)^3 + \Omega_\Lambda}.
\end{equation}

\begin{figure*}[t]    % NORMAN BAYESIAN - z DISTRIBUTION
  \begin{center}      % & XLF
  \includegraphics[width=0.49\textwidth,bb=26 34 545 540,clip]{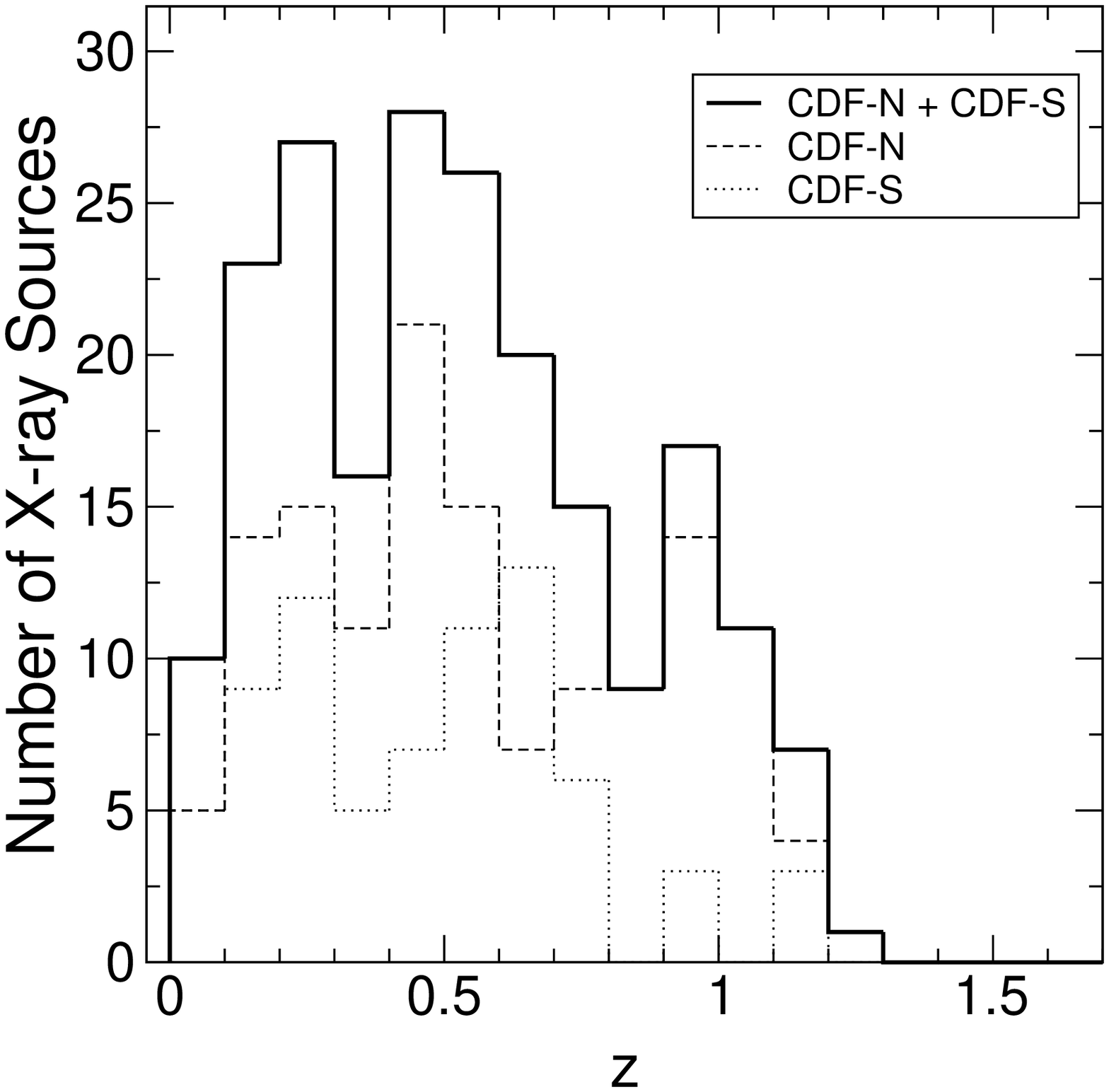}
  \includegraphics[width=0.49\textwidth]{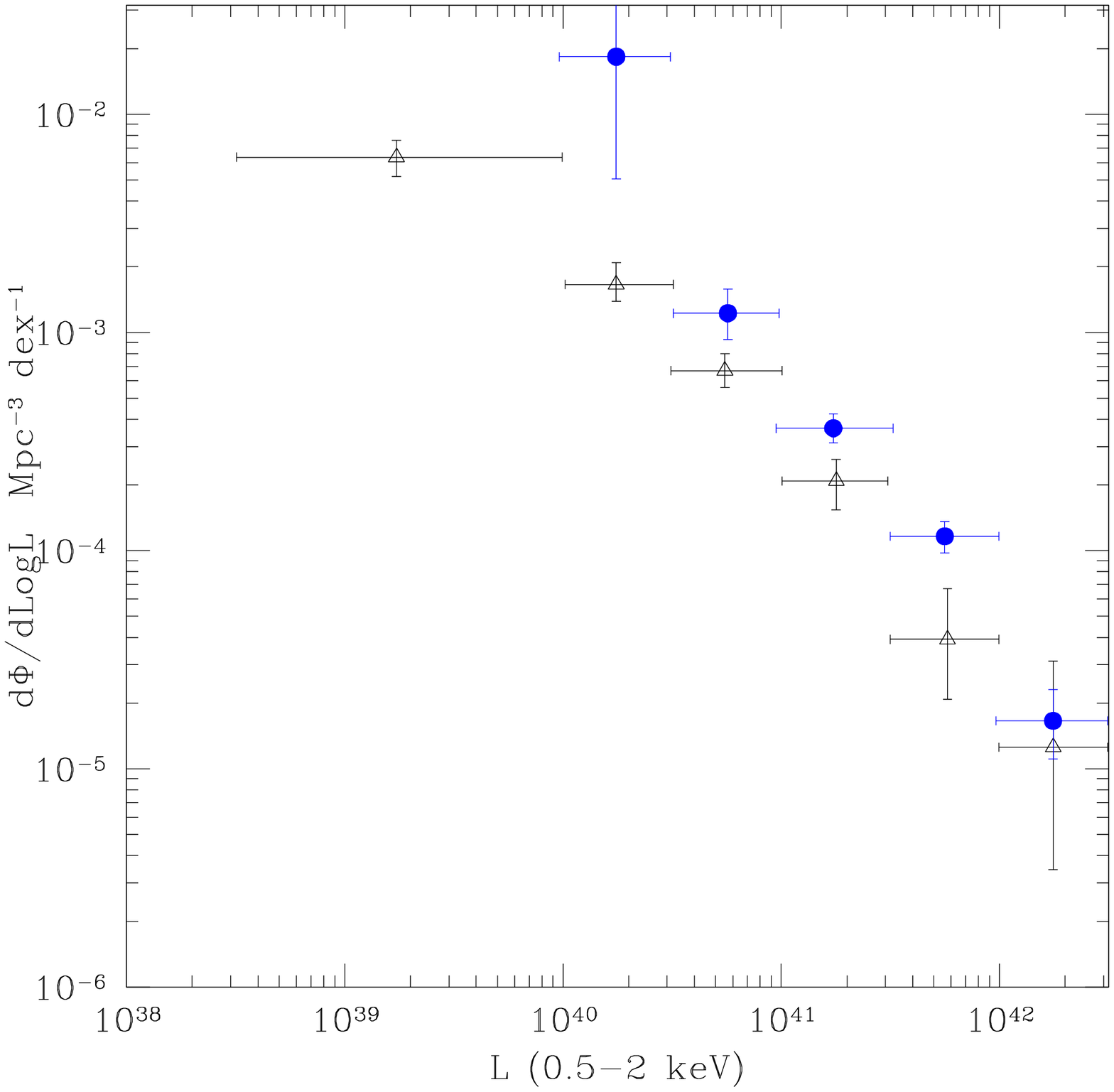}
  \end{center} 
  \begin{minipage}[t]{.49\textwidth}
   \caption { Redshift distribution for the Bayesian sample of X-ray
     selected normal galaxies defined in \citet{colin}. Dotted and
     dashed line: redshift distributions for galaxies in the \chandra\ 
     Deep Fields South and North, respectively. Continous line: sum of
     the North and South subsamples.
   \label{fi:nz-norman} }
  \end{minipage}
  \hfill
  \begin{minipage}[t]{.49\textwidth}
   \caption { X-ray luminosity function for the galaxies in the
     Bayesian sample by \citet{colin}. Triangles: LF of objects with
     $z\leq0.5$; circles: LF of objcts with $z>0.5$. The median
     redshift of the two bins are $z=0.27$ and 0.79 respectively, thus
     the two data sets may be regarded as the XLFs {\em at the median
       redshifts}; the evolution of the LF from $z=0.27$ to $z=0.79$
     is clearly visible.
   \label{fi:xlf-norman} }
  \end{minipage}
\end{figure*}

The redshift distribution of the \citet{colin} sample is shown in
Fig.~(\ref{fi:nz-norman}). Two redshift bins were considered, $z\leq
0.5$ and $0.5<z\leq 1.2$, with mean redshifts $z=0.27$ and 0.79
respectively, in order to have a comparable number of galaxies in both
bins.

The XLF is shown in Fig.~(\ref{fi:xlf-norman}). \citet{colin} also
determined the evolution of the LF, by fitting the XLFs in the two
redshift bins with two power laws. Considering a pure luminosity
evolution, they found $L(z)/L(0) = (1+z)^{2.7}$.
This result will be checked in the following sections against FIR-
and radio-based determinations of the XLF.

\section{Determining the XLF from the FIR and radio luminosity functions}
\label{sect_FIR2XLF}

Infrared surveys are a powerful method to select star forming, spiral galaxies,
since the bulk of the far and near infrared emission is due to
reprocessed light from star formation, with AGN representing only a minor
population \citep{dejong84,franceschini01,elbaz02}.
In the following, we summarize the current
determinations of the FIR luminosity functions, which are then
converted to the X-rays. 

We will only consider a pure density evolution of the form
$(1+z)^\eta$,  which is the simplest
and most used form by the authors of the papers from which we draw the
FIR LFs. This is equivalent, in a more rigorous formalism,
to a bivariate LF of the form $\varphi_{\rm b} (L,z)\, \de \Log
L\, \de z$ with the assumption that the LF is separable for $L$ and
$z$, so that we
may write $\varphi_{\rm b}(L,z)=\varphi(L) (1+z)^\eta$.

\begin{figure*}[t]    % REDSHIFT DISTRIBUTION
  \begin{center}
    \includegraphics[width=0.49\textwidth]{PSCz_nz1.ps}
    \includegraphics[width=0.49\textwidth,bb=93 389 482 678,clip]{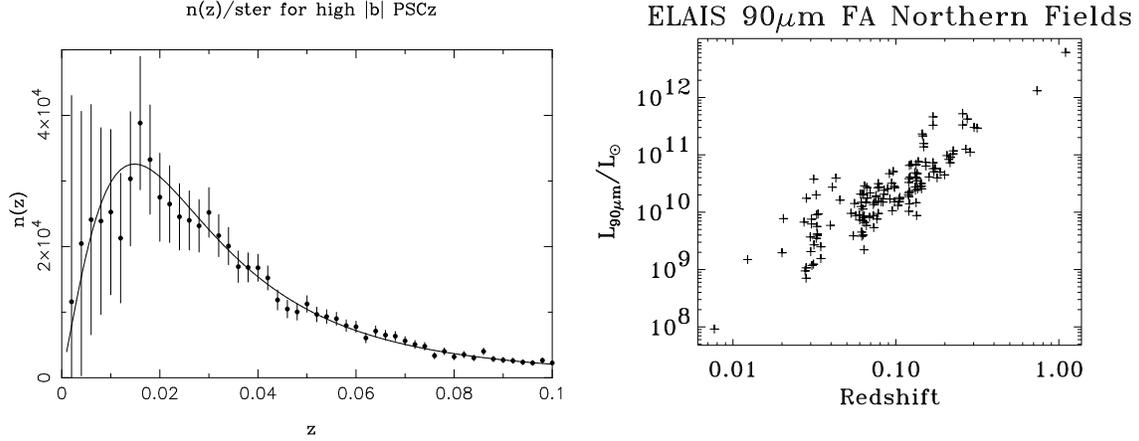}
  \end{center} 
  \caption { Left panel: Redshift distribution for galaxies in the IRAS PSC$z$
    survey, from \citet{saun00}. Right panel: Luminosity-redshift
    relation for galaxies in the ELAIS survey, from \citet{serje04}. 
  \label{fi:FIRsurveys_nz} }
\end{figure*}

\begin{figure}[t]    % % FUNZIONI DI LUMINOSITA' FIR & FIR -> X
  \centering
   \includegraphics[width=0.48\textwidth]{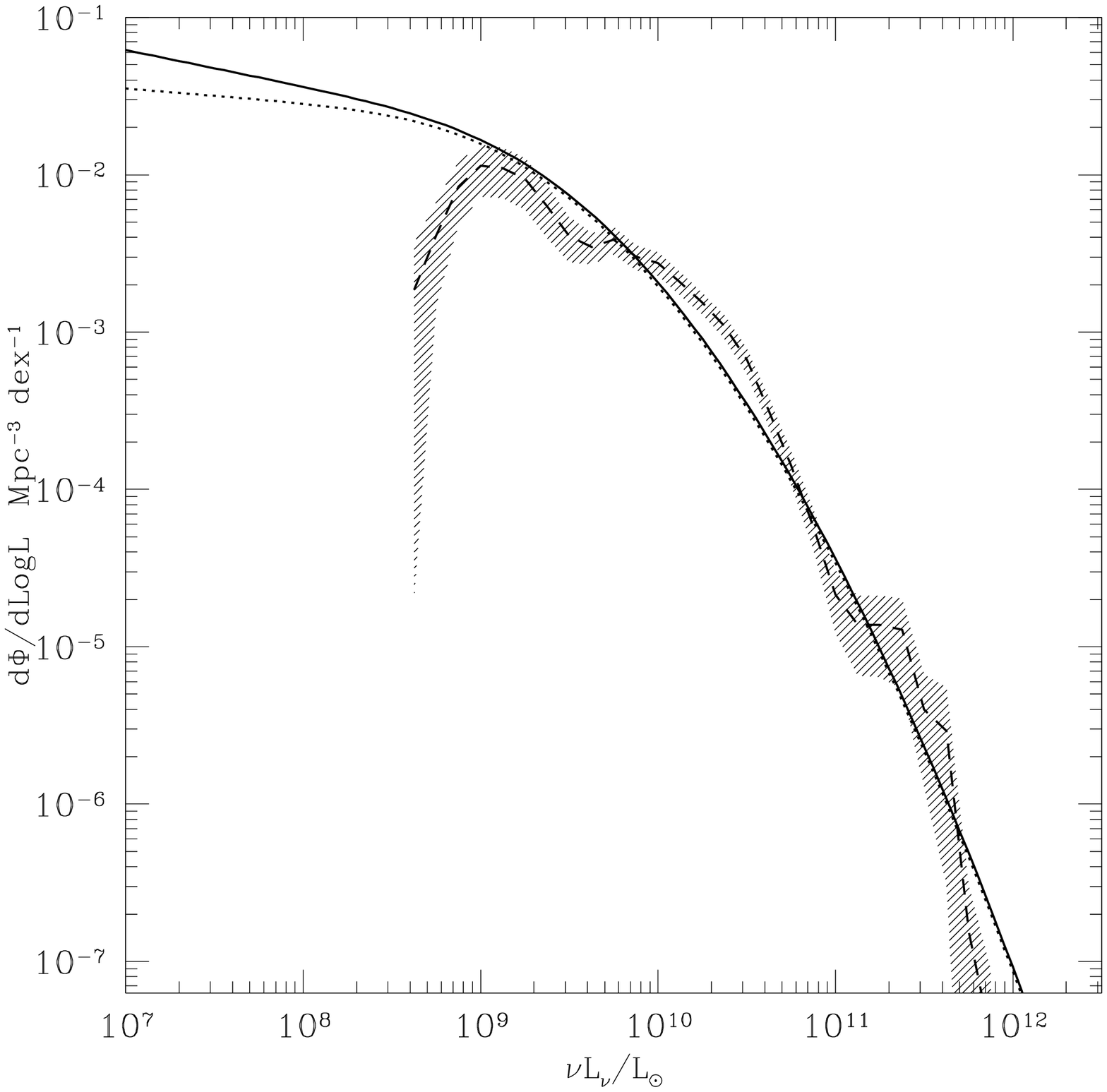}
   \hfil
   \includegraphics[width=0.48\textwidth]{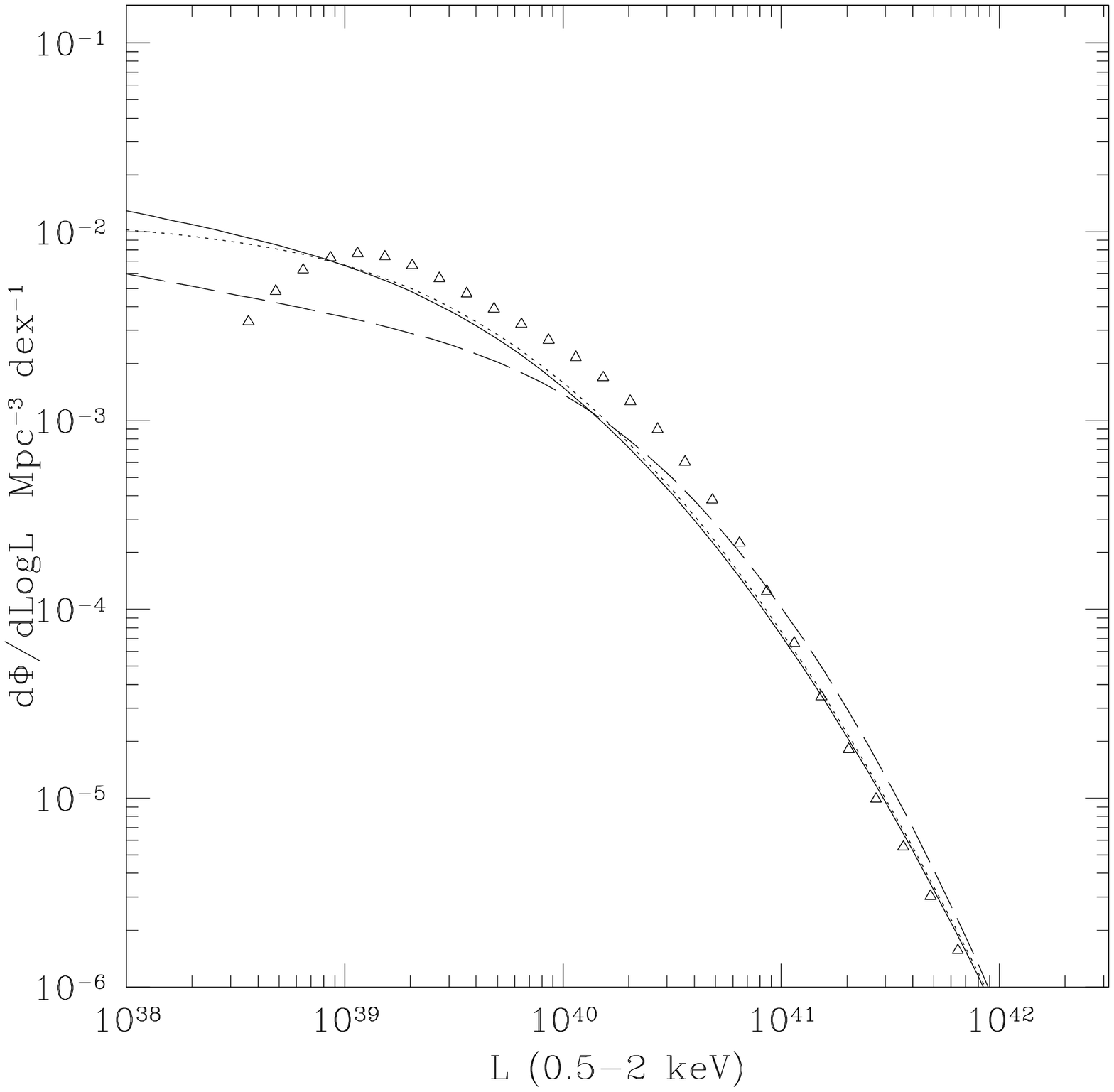}
  \begin{minipage}[t]{.49\textwidth}%
   \caption { IRAS and ISO local (i.e.\ $z=0$) luminosity functions. 
    Solid curve: IRAS 60$\mu$ LF from \citet{takeu03}. Dotted curve:
    IRAS 60$\mu$ LF from \citet{saun90}. The two IRAS LFs coincide
    at bright luminosities. Dashed curve: ISO 90$\mu$ LF
    from \citet{serje04}; the shaded area corresponds to the 1$\sigma$
    errorbar. While the IRAS LFs are well constrained in the whole
    range of luminosities shown here, the ISO LF suffers from
    incompleteness for $L\lesssim 10^9 L_\sun$.  The monocromatic
    luminosities have been converted to bolometric ones by assuming
    $\nu L_\nu = {\rm cost}$. $H_0=70$ is assumed.
    \label{fi:infratake} }
 \end{minipage}%
\hfill
\begin{minipage}[t]{.49\textwidth}
    \caption { IRAS, ISO, and radio luminosity functions converted to
      the X-rays. Solid curve: IRAS 60$\mu$ LF from \citet{takeu03}.
      Dotted curve: IRAS 60$\mu$ LF from \citet{saun90}. Points: ISO
      90$\mu$ LF from \citet{serje04}. Dashed curve: radio LF from
      \citet{machalski00}. \label{fi:xraytake} } \end{minipage}
\end{figure}

\citet{saun90} defined a sample of 2,818 galaxies matching common
selection criteria from different IRAS samples, with a flux limit
around 0.6 mJy at 60$\mu$, completeness at 98\% level, redshifts
$z\lesssim 0.05$, and derived a $60\mu$ luminosity function. The
best-fit model has the shape
\begin{equation}
  \varphi(L)=\varphi^* \left( \frac{L}{L^*}
  \right)^{1-\alpha}
  \exp \left[ -\frac{1}{2\sigma^2} \Log^2_{10} \left( 1+\frac{L}{L^*}
      \right) \right] \quad\rm{Mpc}^{-3}
\label{eq:saun-param}\end{equation}
with parameters at present epoch
\[
\varphi^*=(2.6\pm 0.8)\cdot 10^{-2}\, h^3\, {\rm Mpc}^{-3}, \alpha=1.09 \pm 0.12,
\]
\begin{equation}
\sigma=0.72\pm 0.03 \mbox{ and } L^*=(2.95_{-1.21}^{+3.06})\cdot
10^{8}\, h^{-2}\, L_\sun
\end{equation}
(shown as the dotted curve in Fig.~\ref{fi:infratake})
and the evolution (parameterized as pure density evolution)
\begin{equation} (1+z)^{6.7\pm 2.3}. \end{equation}

\begin{figure*}[p]    % NORMAN XLF COMPARED TO OTHERS
  \begin{center}
  \includegraphics[height=0.43\textheight]{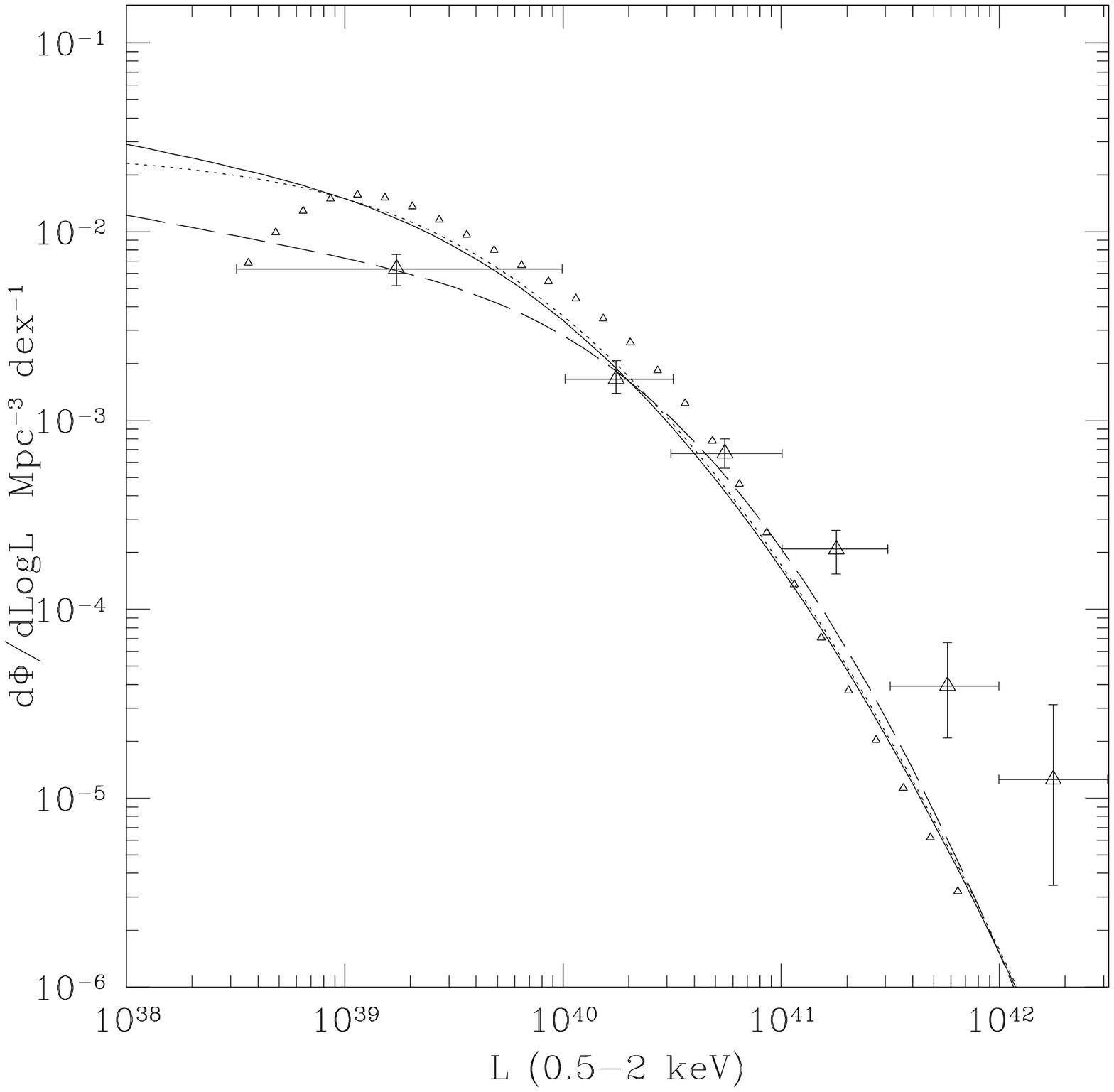}\\
  \includegraphics[height=0.43\textheight]{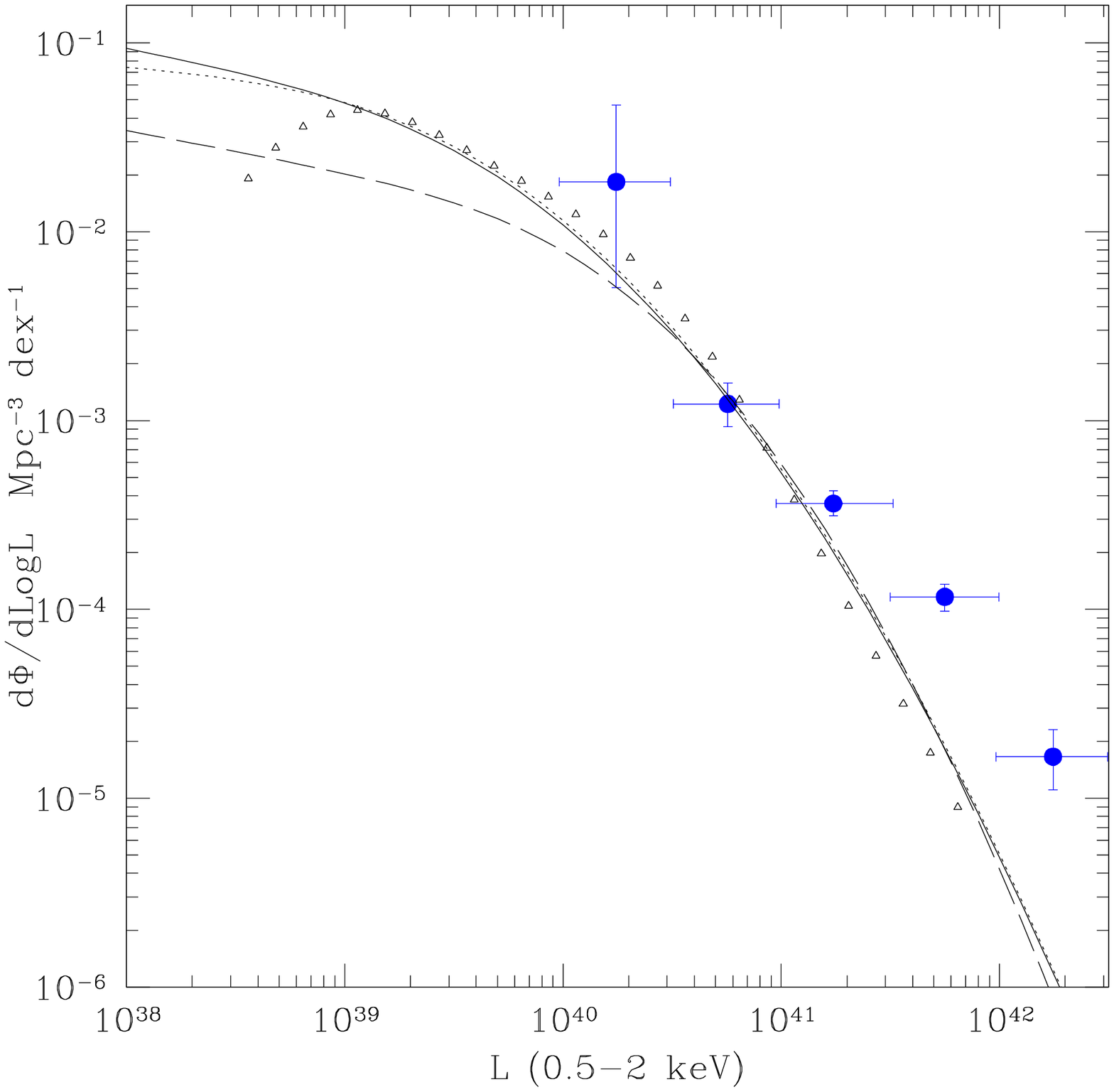}
  \end{center} 
  \caption { Comparison of the X-ray luminosity function with the FIR-
    and radio-derived ones. Symbols as in Figs.~(\ref{fi:xlf-norman},
    \ref{fi:xraytake}). The FIR- and radio-based LFs have been evolved
    (see text) to the median redshifts of the two bins in which the
    XLF was calculated (i.e.\ $z=0.27$ in the left panel, $z=0.79$ in
    the right panel). For the \citet{saun90} IRAS LF, the milder
    evolution determined by \citet{takeu03} was used.
  \label{fi:xlf-normanetal} }
\end{figure*}

The $60\mu$ luminosity function was revised by \citet{takeu03} by
enlarging the galaxy sample. 15,411 galaxies from the Point Source
Catalog Redshift (PSC$z$) were used, covering 84\% of the sky with
a flux limit of 0.6 mJy at $60\mu$. 
%\citet{takeu03} report, both in
%text and in their Fig.~1, that the
%main difference between their study and Saunders' one
%is a factor-of-$\sim 2$ overdensity of galaxies with
%$L_{60\mu}\lesssim 10^8 L\sun$; 
Using the same parameterization of Eq.~(\ref{eq:saun-param}),
\citet{takeu03} found\footnote{T. Takeuchi (priv.\ comm.) recently
  found an error in the value of $L^*$ reported in his paper. The one
  reported here is the correct one.}
$$
\varphi^*=(2.60\pm 0.30)\cdot 10^{-2}\, h^3\, {\rm Mpc}^{-3}, \alpha=1.23 \pm 0.04,
$$ 
\begin{equation}
\sigma=0.724\pm 0.01\mbox{ and }L^*=(4.4\pm 0.9)\cdot 10^8\, h^{-2}\,
L_\sun.   
\end{equation}

In Fig.~\ref{fi:infratake} it is shown as the dotted curve. 
%As can clearly be seen, there is also a difference between Takeuchi's
%and Saunders' LFs of almost one order of magnitude at the bright end. 
They also found that the $(1+z)^{6.7}$ density evolution proposed by
\citet{saun90} is not consistent with their sample. A milder
evolution $(1+z)^{3.4}$ is reported to be the best-fit description of
the data. Note however that the redshift distribution of the PSC$z$
galaxies only span a very limited range (Fig.~\ref{fi:FIRsurveys_nz}),
thus the evolution exponent is very poorly constrained; this may also
explain the factor-of-two difference in evolution between Takeuchi's
and Saunders' results.

The LF at $90\mu$ was recently determined by the ELAIS survey team
\citep{serje01,serje04} with the PHOT and CAM instruments onboard the
ISO satellite. It has not been fitted to a
parametric form. \citet{serje04} also found that the evolution, if
parameterized as a $(1+z)^\eta$ pure density form for $z\leq 1$ and no
evolution beyond $z\sim 1$, has an exponent $2.4<\eta<4.6$ at 68\%
confidence; in the following we will assume $\eta=3$.  The ELAIS LF is
shown in Fig.~\ref{fi:infratake} as the dashed curve, along with
$1\sigma$ errors.

\subsection*{Conversion of the LF to the X-rays}
The FIR luminosity functions may be converted to the X-rays using the
approach first developed in \citet{avnitan86} (see also \citealt{ioannis99}
and \citealt{colin}).
\begin{equation}
\varphi_{\rm X}(\Log\ \LX) = \int\limits_0^{+\infty}
\varphi_{60\mu}(\Log\ L_{60\mu}) \, P\left(\Log\ \LX|\Log\ L_{60\mu}\right)
\de\Log\ L_{60\mu}
\end{equation}
where $P\left(\Log\ \LX|\Log\ L_{60\mu}\right)$ is the probability
distribution for observing $\LX$ for a given $L_{60\mu}$. Considering
the radio/FIR relationship of \S\ref{correlazioni_section},
this may be taken as a gaussian, yielding (for the
0.5-2 keV band)
\begin{equation}
P\left(\Log\ \Lsx|\Log\ L_{60\mu}\right) =
\frac{1}{\sqrt{2\pi}\sigma} \,
e^{-\frac{{\rm Log}\ L_{60\mu} +9.05 - {\rm Log}\ \Lsx}{2\sigma^2}}
\end{equation}
where $\sigma\sim 0.30$ as estimated in
\S~\ref{correlazioni_section}. To convert the ELAIS LF, a
$S_{60\mu}/S_{90\mu}=0.66$ flux ratio may be derived by assuming $\nu
L_\nu={\rm cost}$. The same figure can be obtained by considering a
grey body law (Eq.~\ref{eq:greybody}) with $\beta=1$ and temperature $T=40$ K.
The converted luminosity functions are shown in Fig.~\ref{fi:xraytake}.

\subsection*{The XLF from the radio LF}
\label{sec:radiolf}

Information from the radio surveys may also be used to derive an XLF.
We consider the sample of radio detected galaxies in the Las Campanas
Redshift Survey \citep{machalski99}, which comprises 1157 galaxies
with magnitude $R\leq 18.0$ and detected in the NRAO VLA Sky Survey
(NVSS), i.e.\ brighter than 2.5 mJy beam$^{-1}$ at 1.4 GHz.
These galaxies were classified as ``starburst'' (502 galaxies) or
``AGN'' (655 galaxies) on the
basis of their FIR-radio flux ratios, 25$\mu$--60$\mu$ spectral
indices and radio-optical flux ratios. \citet{machalski00} derived the
LFs for both classes of objects; here we consider only the ``starbursts''.

For a LF with the functional form of Eq.~(\ref{eq:saun-param}),
\citet{machalski00} give the set of parameters
$$
\varphi^*=(7.9_{-2.6}^{+3.8})\e{-3}\, h^3\ {\rm Mpc}^{-3} {\rm dex}^{-1},
 \alpha=1.22 \pm 0.27,
$$ 
\begin{equation}
\sigma=0.61\pm 0.05\mbox{ and }\Log\ L^*=(28.13\pm 0.37)\, h^{-2}\,
{\rm \ erg\ s}^{-1}{\rm \ Hz}^{-1}.
\end{equation}

We converted the radio LF into an XLF following the same method
as in the previous paragraph, and considering the radio/X-ray relation
of \S~\ref{correlazioni_section}. The resulting XLF is shown in
Fig.~(\ref{fi:xraytake}) along with the ones derived from the infrared
surveys. The radio-based LF agrees well with the FIR-based ones for
$\LX\gtrsim 10^{40}$ erg/s, while it lies a factor of $\sim 2$ below
the FIR-based LFs at lower luminosities.

The redshifts of the Las Campanas ``starburst'' galaxies are in the
range $0.01\leq z\leq 0.075$. As in the case of the IRAS galaxies,
this redshift range is too small to allow a reliable estimation of the
evolution. The constraints on the evolution found by
\citet{machalski00} are too loose to either confirm of reject cosmic
evolution for the Las Campanas galaxies.

\section{Comparison with the XLF}

To compare the pure luminosity evolution determined by \citet{colin}
with the pure density forms used above, we assume that the XLF may be
well represented by a power law whose slope remains constant with
varying redshifts. We remind, in fact, that for a power
law shaped LF the two forms of evolution cannot be distinguished. In
fact, it may easily be shown that a LF with pure luminosity evolution
with the shape
\begin{equation}
\varphi_0 L(z)^{-\alpha}
=\varphi_0 \left[ L_0/ (1+z)^\eta \right]^{-\alpha} 
\end{equation}
transforms into a pure density form as
\begin{equation}
\varphi(z) L_0^{-\alpha}
=\varphi_0 (1+z)^{\eta\alpha} L_0^{-\alpha}.
\end{equation}
Since the XLF derived by \citet{colin} has $\alpha\sim 1.1$, their
$(1+z)^{2.7}$ luminosity evolution is equivalent to a $(1+z)^{2.9}$
density evolution --- which agrees well with the evolution exponents
found for the FIR LFs.

A comparison between the \citet{colin} LF and the FIR/radio derived
ones is made in Fig.~(\ref{fi:xlf-normanetal}). To make the comparison
the easiest possible, the FIR/radio derived XLFs were evolved with
their own evolution exponents, up to the mean redshifts of the two
bins in which the \citet{colin} was calculated.  We found that the
X-ray derived LF is in agreement with the FIR-radio derived LFs for
$L_{\rm X}\lesssim 10^{41.5}$, but it is significantly flatter at its
bright end ($L_{\rm X}\gtrsim 10^{41.5}$), where it has a density of
galaxies which is one order of magnitude larger than that of the
FIR-radio galaxies. This may be symptomatic of a residual fraction of
low luminosity AGN in the sample used by \citet{colin}.

\section{The observed X-ray \lognlogs of galaxies}
\label{sect_obslogn}

We turn now to an analysis of the X-ray number counts of galaxies,
since they represent the most direct testbed for both the selection
criteria for normal galaxies in X-ray surveys, and for the shape and
evolution of different LFs.

\begin{figure*}[tp]    % FUNZIONI DI LUMINOSITA' - ANCHE MACHALSKI
  \begin{center}       % & CONTEGGI
  \includegraphics[width=0.8\textwidth]{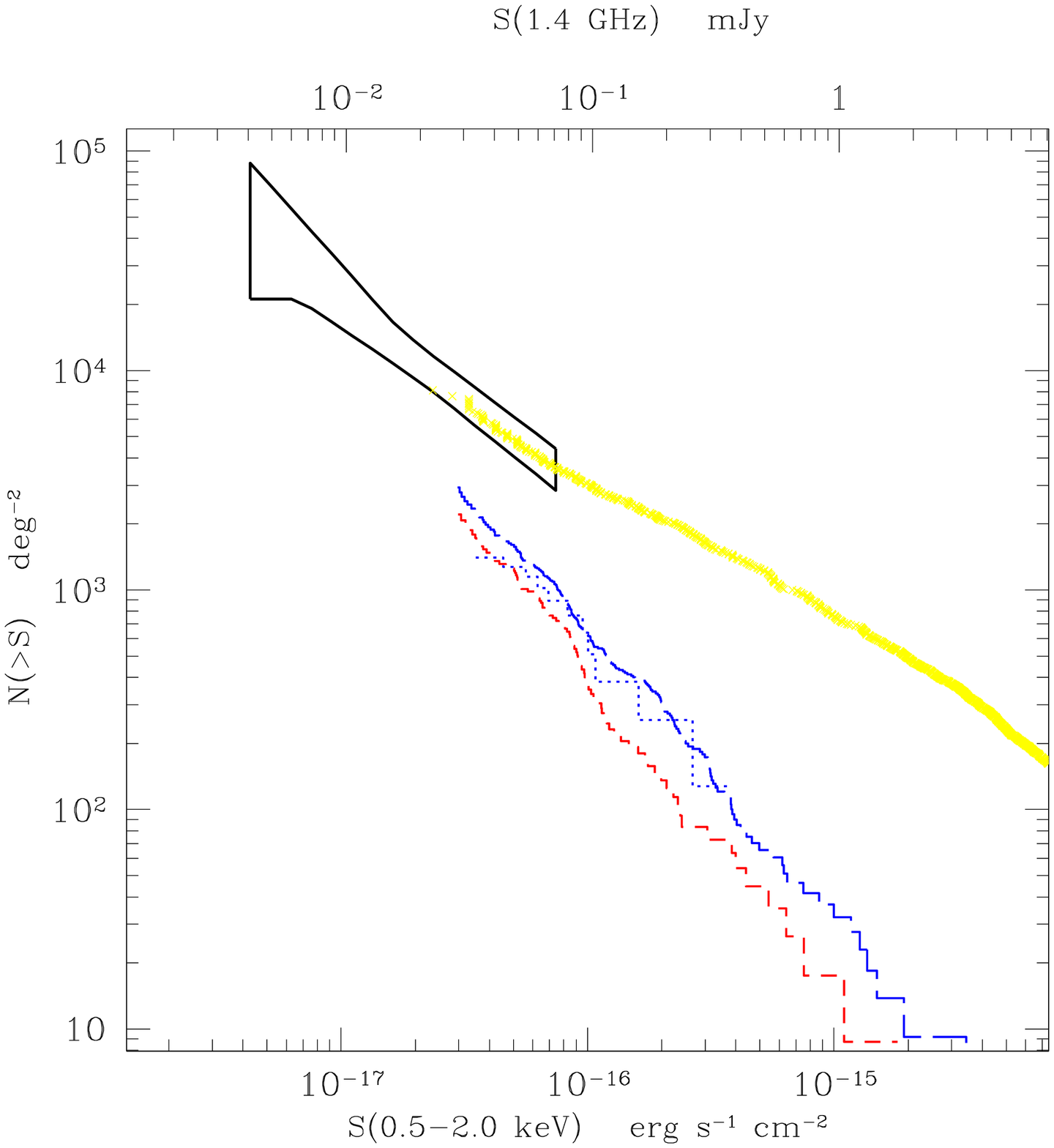}
  \end{center} 
   \caption { X-ray counts in the \chandra\ Deep Fields. The thick,
     upper line and the horn-shaped symbol show the observed \lognlogs
     for all X-ray sources in the Deep Fields \citep{moretti03} and
     the limits from the fluctuation analysis
     \citep{miyaji02a,miyaji02b}. The histograms show the \lognlogs
     for normal galaxies. Short-dashed (lower) histogram: sources with
     Log (X-ray/optical flux ratio) $<-1$. Dotted (middle) histogram: X-ray
     sources with a radio detection and an optical spectrum compatible
     with ion excition by \Hii regions \citep{bauer02}. Long-dashed
     (upper) histogram: sources from the Bayesian sample of \citet{colin}.
   \label{fi:xcts-baueretal} }
\end{figure*}

First we considered the \lognlogs of the sample defined in \citet{colin}
and discussed in \S~\ref{sect_norman}, which is shown in
Fig.~(\ref{fi:xcts-baueretal}) as the long-dashed histogram.

Then we considered a sample defined in
\citet{bauer02}, who investigated the link between the faint X-ray and
radio sources in the \chandra\ Deep Field North (CDFN); 11 sources
from their catalogue could be reliably selected as star formation
powered systems, on the basis of the reported optical spectrum (i.e.\ 
classified as ``emission line galaxies'').  By considering the sky
coverage of the CDFN survey, as reported in \citet{alexander03}, we
derived the \lognlogs for these sources, which is shown in
Fig.~(\ref{fi:xcts-baueretal}) as the dotted histogram.

% \begin{figure*}[tp]    % log N -- log S con BAUER & F_X/F_ott & NORMAN
%   \begin{center}
%   \includegraphics[width=0.49\textwidth]{xcts-bauer.ps}
%   \includegraphics[width=0.49\textwidth]{xcts-norman.ps}
%   \end{center} 
%   \begin{minipage}[t]{.49\textwidth}
%    \caption { BAUER X-ray counts.
%    \label{fi:xcts-bauer} }
%   \end{minipage}
%   \hfill
%   \begin{minipage}[t]{.49\textwidth}
%    \caption { NORMAN X-ray counts.
%    \label{fi:xcts-norman} }
%   \end{minipage}
% \end{figure*}

We also considered 97 CDFN sources from \citet{barger03}, which were
selected for having an X-ray/optical flux ratio $\lesssim -1$.
Although this selection criterium may be regarded as somewhat crude,
it is indeed useful since it allows an immediate object selection with
the least possible number of parameters.  The resulting \lognlogs is
plotted as the short-dashed histogram in Fig.~(\ref{fi:xcts-baueretal}).

We found that there is a quite good agreement (within a factor of 2--3)
between the three observed \lognlogs described above. In the following
sections, the X-ray \lognlogs will be checked against predictions made
on the basis of the radio counts for the sub-mJy population and from
the FIR and radio LFs.
%% . Here, we anticipate that
%% there is a very good agreement also between the observed X-ray
%% \lognlogs and the ones predicted from the radio counts
%% (\S~\ref{fondox_section}) and from the FIR and radio LF
%% (\S~\ref{sect_FIR2XLF} and \ref{sec:radiolf}).

\section{X-ray number counts from radio surveys}     \label{fondox_section}

We consider the radio sub-mJy population associated with faint blue
galaxies at high redshifts ($22\lesssim V\lesssim 27$, $0.5\lesssim
z\lesssim 1.5$; \citealt{windhorst90}) representing an early era of
star formation in the universe \citep{haarsma00}. This strongly
evolving population \citep{franceschini01} accounts for the
majority of the number counts below $\sim 0.5$ mJy \citep{windhorst85}
and contributes about half of the radio cosmic background at 1.4 GHz
\citep{haarsma00}.  The deepest radio surveys have been performed at
1.4 GHz \citep{vla00}, 5 GHz \citep{fomalont91} and 8.4 GHz
\citep{vla98}.
% <TDD>   via, che il fondo X non e' piu' l'ultima prospettiva
%Although a quick
%estimate of the contribution to the XRB may be worked out by simply
%applying Eqs.~(\ref{eq:radiox},\ref{eq:radiox2}) to the total radio
%fluxes obtained by integrating over the deepest radio \lognlogsa, it
%is instructive to first derive the X-ray source counts and compare
%them with the deepest counts obtained by {\em Chandra}.  
% </TDD>
In order to
derive the X-ray counts from the radio ones a full knowledge of the
redshift distribution and spectra of the sources would be
required. Under the simplifying assumption that the sub-mJy population
lies at a redshift $\bar{z}$, so that the K-correction term is the
same for all sources, the differential counts are obtained as
\begin{equation}
n(S_{\rm X})=n(S_{1.4}\cdot \kappa^{\rm X}_{1.4})
\end{equation}
where $n(S)$ are the differential number counts, and $\kappa^{\rm
X}_{1.4}=\kappa^{\rm X}_{1.4}(\bar{z},\alpha,\Gamma)$ is the X-ray
band/1.4 GHz luminosity ratio (eqs.\ref{eq:radiox},\ref{eq:radiox2})
which depends on the redshift and on the radio and X-ray spectral indices
via the K-correction:
\begin{equation}
\kappa^{\rm X}_{1.4}(\bar{z},\alpha,\Gamma) =
\left.\kappa^{\rm X}_{1.4}\right|_{z=0} \cdot (1+\bar{z})^{-(\Gamma-1)+\alpha}
\end{equation}
where, as customary, $\alpha$ denotes the radio spectral index
(`energy index', defined by $S(\nu)\propto \nu^{-\alpha}$) and
$\Gamma$ is the X-ray spectral index (`photon index', $S(\nu)\propto
\nu^{1-\Gamma}$).

Concerning the sub-mJy \lognlogsa, \citet{vla00} gives $n(S)=(2.51\pm
0.13)\cdot 10^{-3} S^{-2.38\pm 0.13}$ deg$^{-2}$ Jy$^{-1}$ as a
best-fit to the differential number counts at 1.4 GHz in the range
45--1000$\mu$Jy, while \citet{fomalont91} find $n(S)=1.2\cdot 10^{-3}
S^{-2.18\pm 0.19}$ at 5 GHZ in the range 16--1500 $\mu$Jy. The number
density at 4 $\mu$Jy (as estimated from fluctuation analysis) is
consistent with extrapolation of the 16--1500 $\mu$Jy slope.
%
% <TDD>
%{\todo fare magari una censored data analysis sul richards???
% e un confronto con le galassie locali? quel paper sull'indice
% spettrale che ancora non ho letto? }
%

The mean radio spectral index is in the range $\alpha=0.3$--0.7.
\citet{fomalont91} report their distribution as having a moda of
$\alpha=0.5$, a median of $\alpha=0.38$, and an average of
$\alpha=0.28$ (indices measured between 1.5 and 5 GHz). From
\citet{vla00} data we find an average $\alpha=0.47$ (between 1.4 and
8.4 GHz), when we consider detections at both frequencies, and
$\alpha=0.67$ when we consider only detections at 1.4 GHz and treat
upper-limits at 8.4 GHz as detections. The latter, steeper slope is
also more consistent with the average index of our HDFN sample (Table
\ref{tab_deep}).  Here we assume $\alpha=0.5$, and estimate the
uncertainty in the K-correction due to the radio spectral index to be
around a $15-20\%$.  

     According to the results of \S\ref{hdf_chap}, we further assume an
average X-ray spectral slope $\Gamma = 2.1$. For the sake of simplicity
we also assume that the objects are placed at $\bar{z}\sim 1$, the mean redshift
for the sub-mJy galaxies which, according to \citet{windhorst90},
are distributed in the redshift interval 0.5--1.5 with a peak at
$z\sim 1$. To estimate the effect on the counts due to the actual distribution
of the sources in this redshift interval we consider a simplified case
in which they are equally distributed at redshifts $z=0.5$, 1.0 and 1.5. Since
the effect enters in the computations via the K-correction term, we find
that the predicted number counts would increase by only $\sim 10\%$. As a
last remark we notice that a fraction, as yet undefined, of the X-ray
spectra might steepen at energies $\gtrsim 10$ keV, thus entailing a decrease in the
predicted source counts; however, given the redshift range under consideration,
the predicted soft X-ray source counts should not be affected.

%%%%%%%%%%%%%%%%%%%%%%%%%%%%%%%%%%%%%%%%%%%%%%%%%%%
% FIGURA: log N log S
%%%%%%%%%%%%%%%%%%%%%%%%%%%%%%%%%%%%%%%%%%%%%%%%%%%
\begin{figure*}[tp]    % log N -- log S
  \begin{center}
  \includegraphics[height=0.4\textheight]{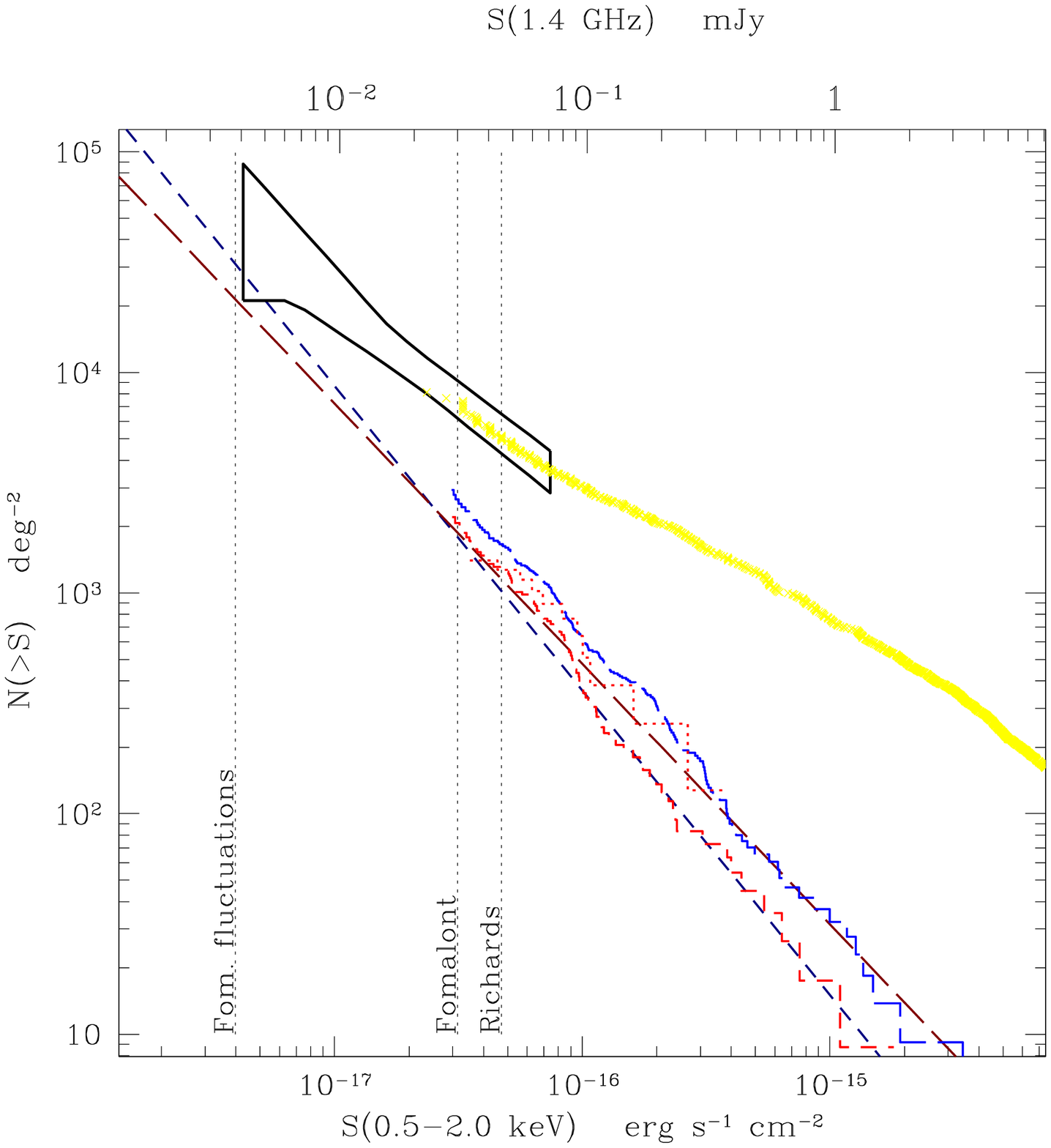}
  \bigskip
  \includegraphics[height=0.4\textheight]{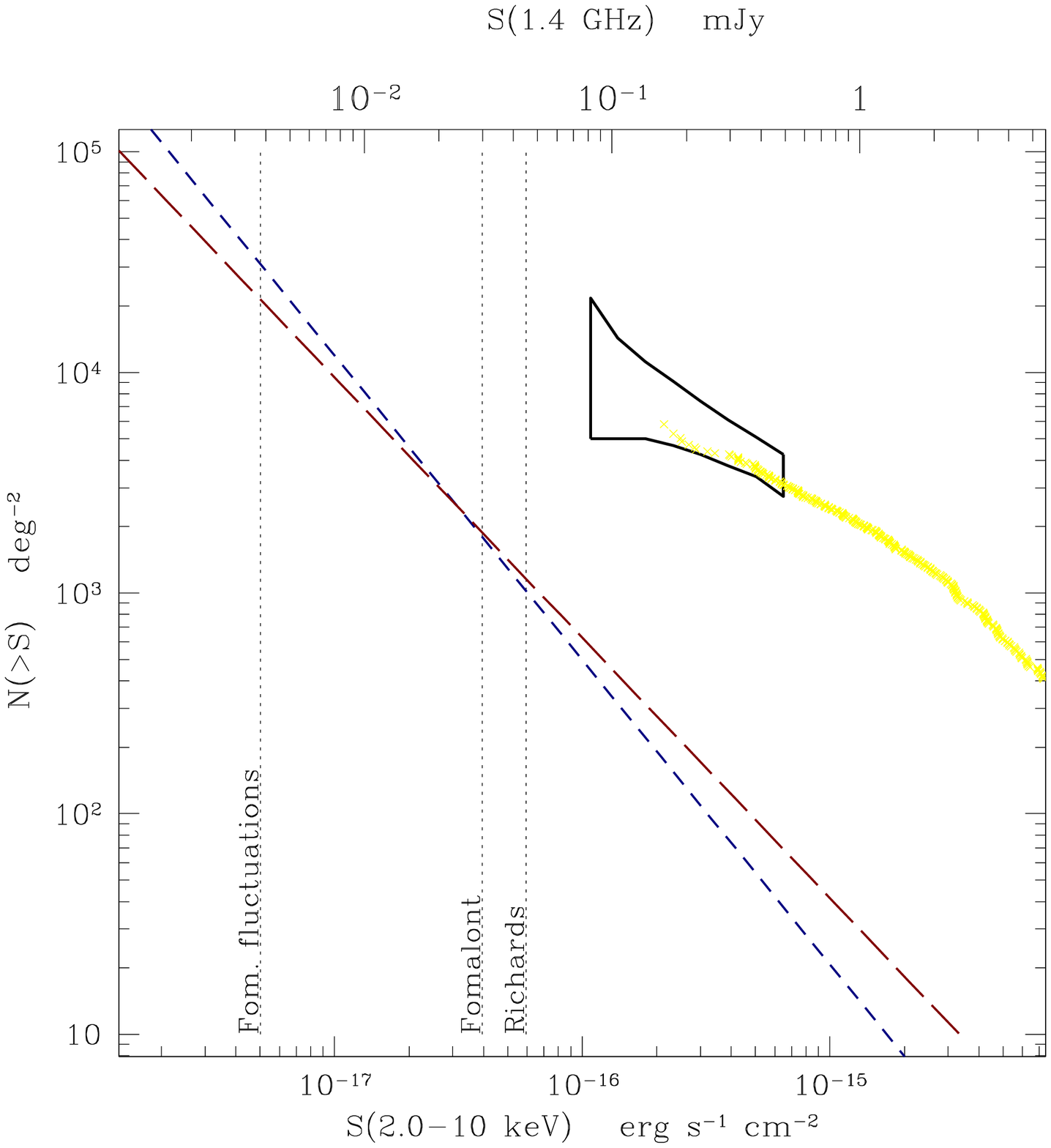}
  \end{center} 
  \caption {\small X-ray counts derived from deep radio \lognlogs in the
  0.5--2.0 (upper panel) and 2.0--10 keV (lower panel)
    bands. Histograms as in previous figures.
    The short-dashed lines
  represent the 1.4 GHz \lognlogs \citep{vla00} with 
  K-corrections assuming $\bar{z}=1$, $\Gamma=2.1$ and
  $\alpha=0.5$.  The long-dashed lines represent the
  \lognlogs by \citet{fomalont91} reported at 1.4 GHz with $\alpha=0.5$
  (K-corrections like for the 1.4 GHz
  one). Thick line and horn-shaped symbols: total X-ray number counts
  and resuls from fluctuation
  analysis, respectively  (see Fig.~\ref{fi:xcts-baueretal}). 
  The vertical dotted lines show the limiting
  sensitivities for the radio surveys. For easier reading, the
  horizontal scale is shown both in radio and X-ray fluxes, assuming
  the conversion factors of \S~\ref{correlazioni_section}.
  \label{fig_lognlogs} }
\end{figure*}

The X-ray number counts predicted with the above assumptions
(i.e.~$\alpha=0.5$, $\Gamma=2.1$) from \citet{vla00} and
\citet{fomalont91} \lognlogs are shown in Fig.~(\ref{fig_lognlogs}),
along with observed counts from \citet{moretti03} and limits from
fluctuation analysis \citep{miyaji02a,miyaji02b}.  The extrapolations
of the radio \lognlogs below $\sim 50 \mu$Jy and down to a few $\mu$Jy
at 1.4 GHz ($\sim 6\cdot 10^{-18}$ erg s$^{-1}$ cm$^{-2}$ in the
0.5--2.0 keV band) are consistent with the limits from fluctuation
analysis \citep{miyaji02b} and do not exceed the X-ray expected number
counts.   It should be noted, however, that
eqs.(\ref{eq:radiox},\ref{eq:radiox2}) may not apply to the entire
sub-mJy population.
%
%{\todo questo andrebbe discusso un po' di piu'. magari quel
% paper sul rapporto sb/agn al variare del flusso che diceva la gregorini? }
For instance, a
fraction (up to a $15-20\%$, \citealt{haarsma00}) of the sources may
still present a sizable contribution from an AGN. It follows that the
derived \lognlogs should be regarded as an upper limit to the X-ray
counts from star-forming galaxies.

As a final remark it should be pointed out that our results depend on
the basic assumption of a strict linearity between the radio and X-ray
luminosities. Had we assumed a non-linear relationship, such as
Eq.~(\ref{eq:radiox1}) for the soft X-rays or
Eq.~(\ref{bf210b}) for the 2-10 keV band, we would have
found an increase or decrease, respectively, of about $50\%$ in the
predicted counts at a flux level around $10^{-17}$ erg s$^{-1}$
cm$^{-2}$. 
% A larger, well defined sample of star-forming galaxies down
% to the faintest radio and X-ray fluxes would obviously be of great
% importance to better constrain the X-ray vs.~radio luminosities
% relationship; this will be attempted in \S~\ref{sect_obslogn}.

%% It might be argued that a more direct derivation of the
%% source counts could be made by adopting the observed fluxes, rather
%% than the luminosities of the objects. However, this procedure would
%% entail an arbitrary extrapolation of the X-ray/radio flux ratio for
%% almost two orders of magnitude at fluxes fainter than the \chandra\ 
%% deep field. On the other hand, the radio luminosity interval of
%% Fig.~(\ref{fig_deeplumin}) essentially encompasses the radio power of
%% the sub-mJy population if placed in the redshift interval mentioned
%% above.

\section{X-ray counts from integration of the FIR and radio LFs}

\begin{figure}[p]    % log N -- log S con SAUNDERS
  \begin{center}
  \includegraphics[width=0.49\textwidth]{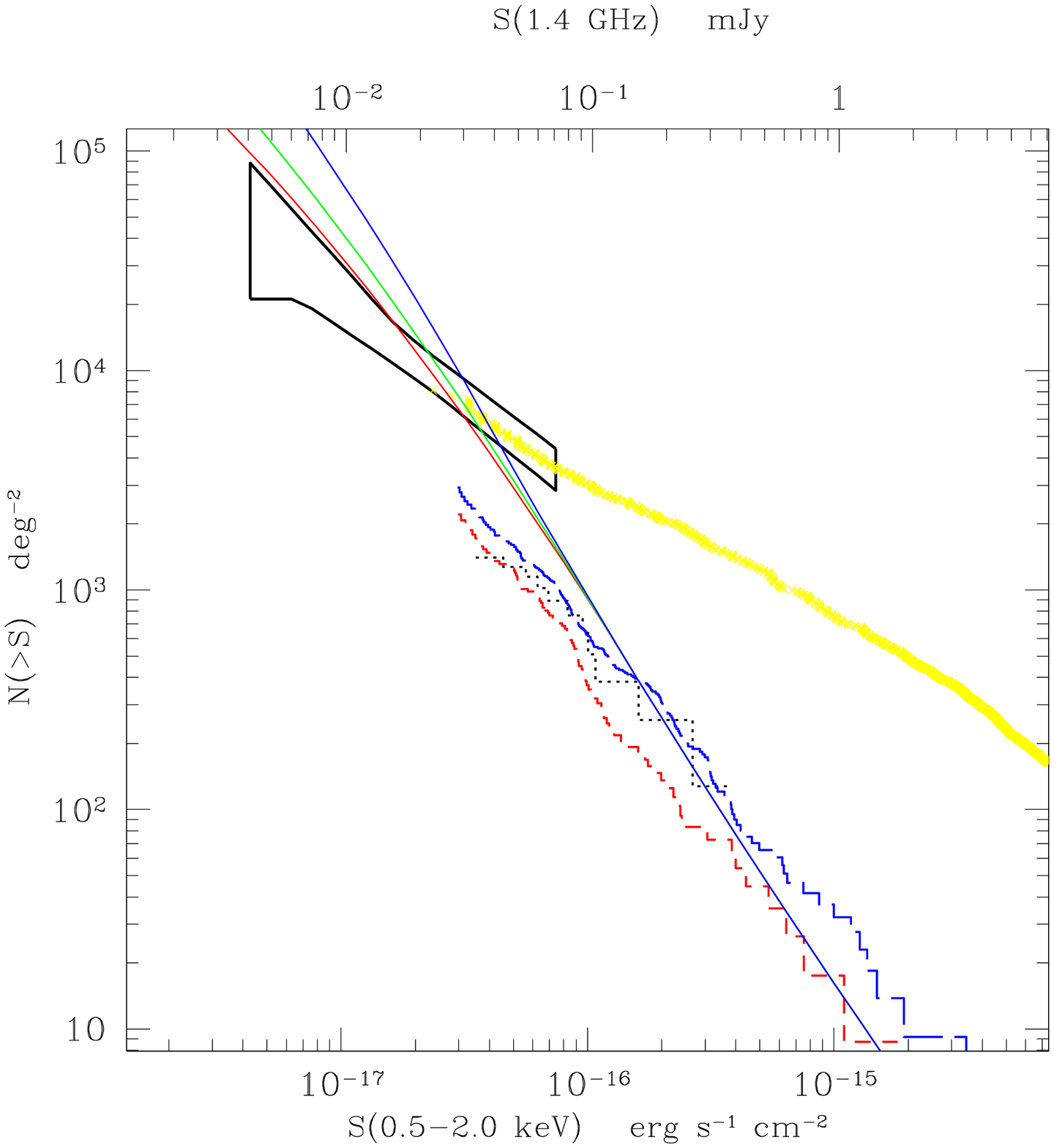}
  \includegraphics[width=0.49\textwidth]{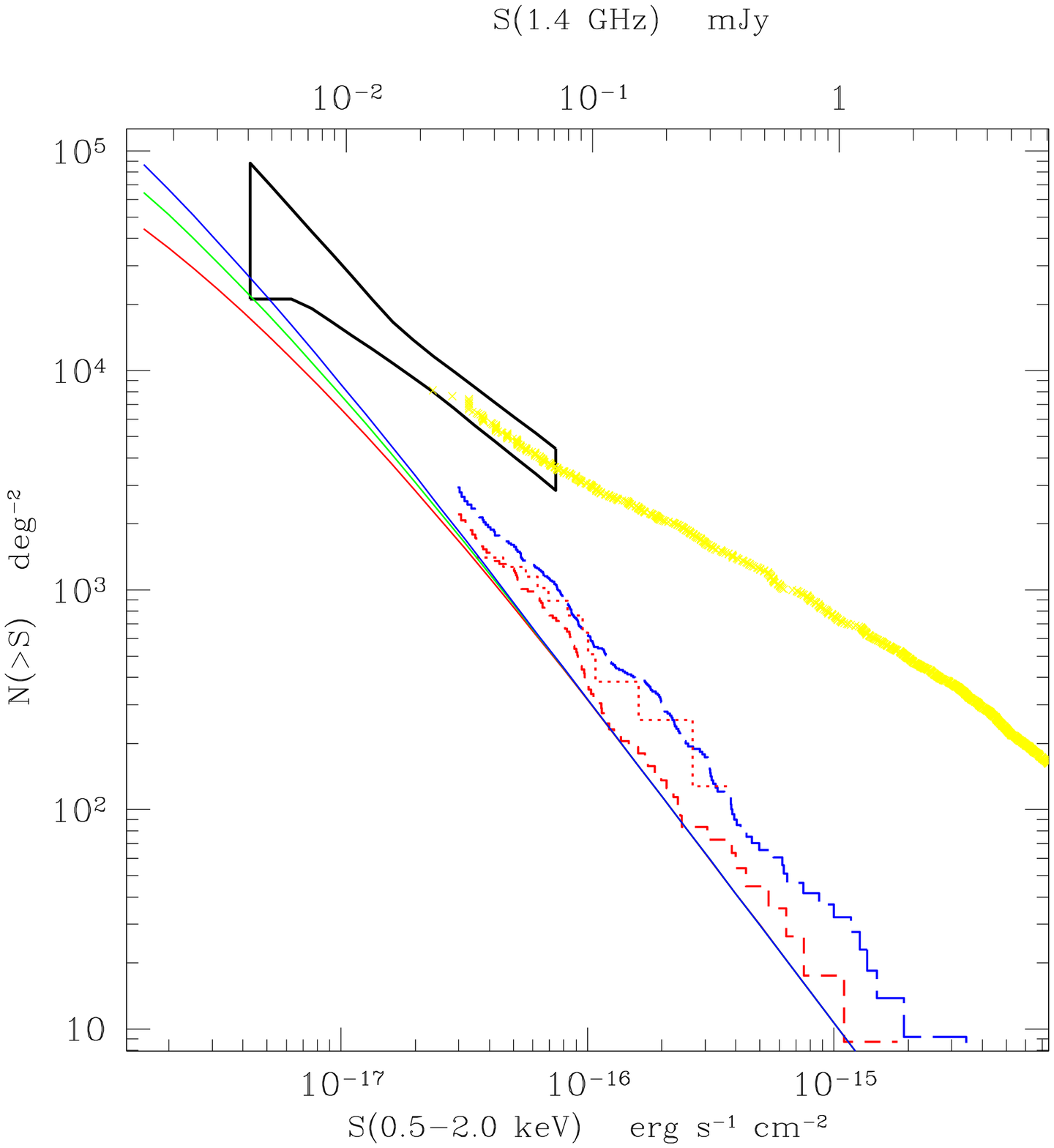}
  \end{center} 
  \caption { X-ray counts derived from the IRAS luminosity function by
    \citet{saun90}. Left panel: $(1+z)^{6.7}$ density evolution; right
    panel: $(1+z)^{3.4}$ density evolution (see text).  Thick line,
    horn shaped symbol and histograms as in the previous figures. The
    three thin curves, which converge at fluxes $\gtrsim 10^{-16}$
    \ergscmq, show the number counts from the integral of the LF;
    lower curve: the LF was integrated with $z_{\rm max}=1.1$ and
    evolution as described in the text; upper curve: $z_{\rm max}=2$;
    middle curve: $z_{\rm max}=2$ but evolution stopped at $z=1$.
  \label{fi:xcts-saund} }
  % TAKEUCHI
  \begin{center}
  \includegraphics[width=0.49\textwidth]{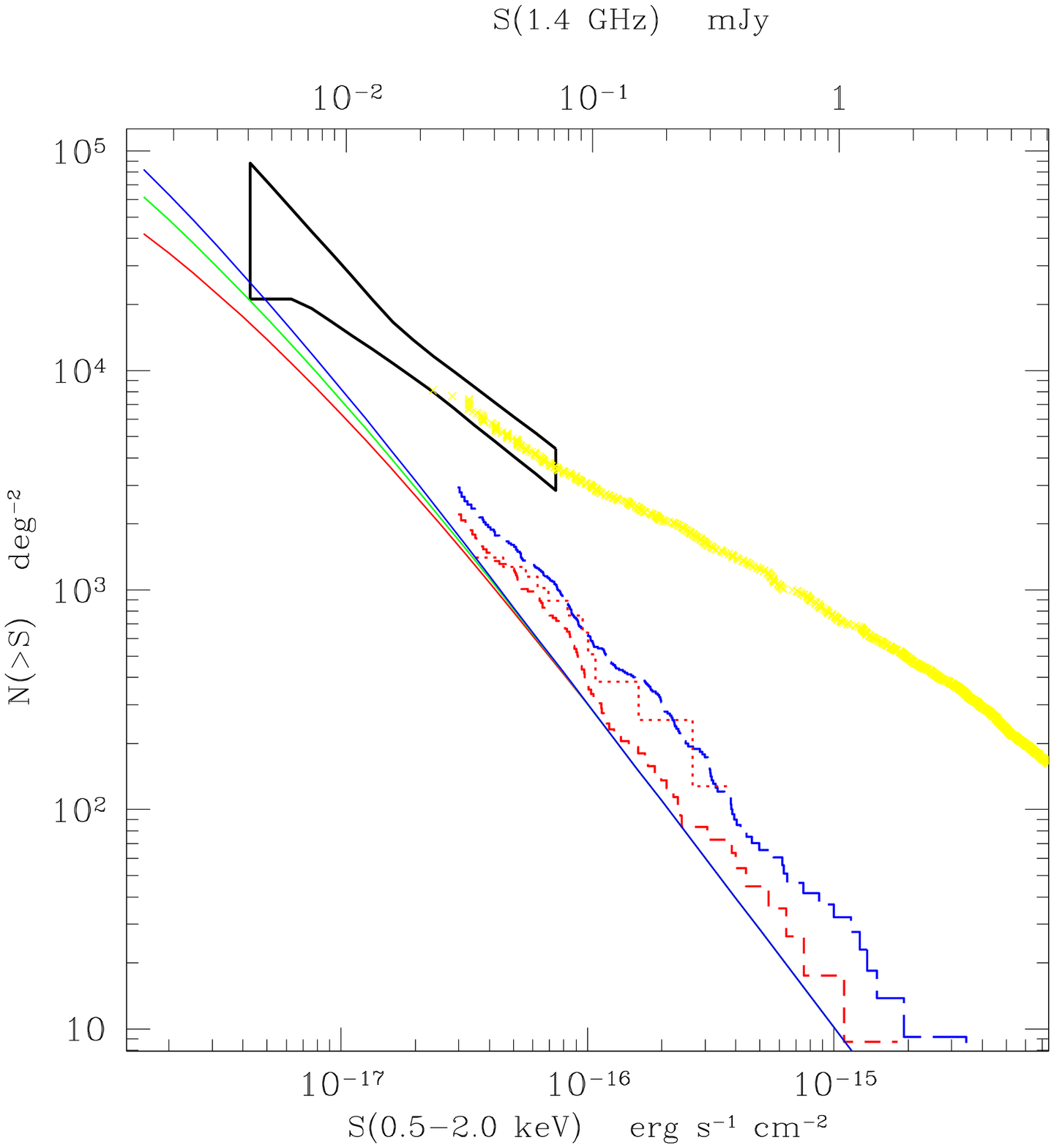}
  \includegraphics[width=0.49\textwidth]{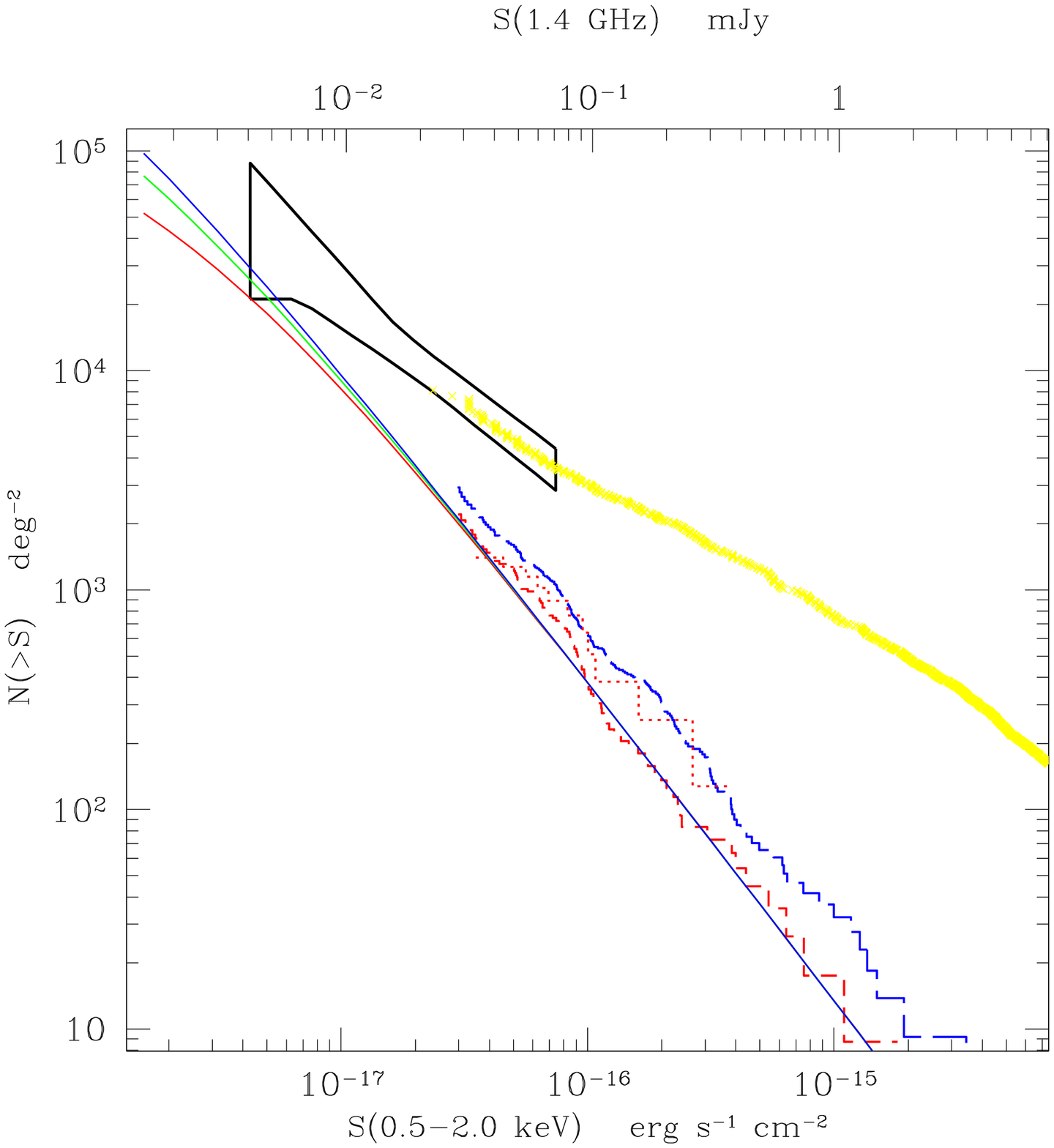}
  \end{center} 
 \begin{minipage}[p]{.49\textwidth}
   \caption { X-ray counts derived from the IRAS luminosity function
  \citet{takeu03}. Symbols as in Fig.~(\ref{fi:xcts-saund}).
   \label{fi:xcts-takeu} }
 \end{minipage}\hfill
 \begin{minipage}[p]{.49\textwidth} % ELAIS
  \caption { X-ray counts derived from the ELAIS luminosity function
    by \citet{serje04}. Symbols as in Fig.~(\ref{fi:xcts-saund}).
  \label{fi:xcts-elais} }
 \end{minipage}
\end{figure}

The observed X-ray counts may also be checked against the number
counts obtained by integrating the FIR- and radio based luminosity
functions.  %This is the simplest test that may be performed on the
%FIR- and radio-derived XLFs, since it does not require any assumption on the
%nature of the X-ray sources.
However, it should be noted that the galaxies considered in the IRAS
and ELAIS surveys lie at redshifts much lower than the \chandra\ Deep
Fields ones. The redshift distributions for the PSC$z$ and ELAIS
surveys (Fig.~\ref{fi:FIRsurveys_nz}) show that the bulk of the
galaxies considered in these surveys is placed at $z\lesssim 0.05$
(PSC$z$) or $z\lesssim 0.3$ (ELAIS), while the X-ray faint galaxies
are usually placed at $z\sim 1$ (cfr.\ Chap.~\ref{hdf_chap} and
\S~\ref{fondox_section}).  Thus a comparison of the X-ray counts with
the FIR-derived ones involves an extrapolation of the luminosity
function at higher redshifts. Although this introduces a further source of
uncertainty, it may be regarded as a possibility to test for the
maximum sustainable evolution.

The number counts are given by:
\begin{equation} 
N(>S) = \int\limits_{z_{\rm min}}^{z_{\rm max}}
\int\limits_{L_{\rm min}(z)}^{L_{\rm max}} 
\varphi(L,z) \, \frac{{\rm d}V}{{\rm d}z} \,\, {\rm d}L \, {\rm d}z
\end{equation}
where $\de V/\de z$ is the comoving volume between $z$ and $z+\de z$.
In the following the integration will performed in the luminosity interval
$10^{39}\leq \LX\leq10^{42}$ erg/s and for two different intervals in $z$:
$0\leq z\leq 1.1$ (cfr.\ \S~\ref{sect_norman}) and over the enlarged
redshift range $0\leq z\leq 2$. We also considered as an intermediate
possibility to stop the evolution at $z=1$.

The \lognlogs resulting from the integration of the infrared luminosity
functions are shown in Fig.s~(\ref{fi:xcts-takeu}),
(\ref{fi:xcts-saund}), and (\ref{fi:xcts-elais}). 

In the left panel of Fig~(\ref{fi:xcts-saund}) we plot the counts
obtained from the \citet{saun90} LF with the evolution $(1+z)^{6.7}$. It
is immediately clear that this evolution is too strong, since the
predicted counts lie much above
the observed \lognlogs at fluxes $\lesssim 4\e{-17}$
\ergscmq. If we assume a milder evolution (e.g.\ the one found by
\citealt{takeu03}, $(1+z)^{3.4}$), then the predicted counts are
consistent with the observed ones. In this case, a good agreement is
also found with the predictions from the radio \lognlogs of
\S~\ref{fondox_section} (cfr.\ Fig.~\ref{fig_lognlogs}, upper panel). 

Integrating the ELAIS LF gives number counts which are very similar to
the \citet{takeu03} case (Fig.~\ref{fi:xcts-elais}) The X-ray number
counts obtained by integrating the Las Campanas LF (not shown) are
also very similar to the IRAS \citep{takeu03} and ELAIS LFs (for the
radio LF we considered a pure density evolution with $(1+z)^3$, i.e.\ 
the same evolution of the ELAIS galaxies).

\section{Final remarks on the X-ray counts}

We have derived estimates for the X-ray number counts at faint fluxes
($5\e{-18}\lesssim S_{\rm X}\lesssim 10^{-15}$ \ergscmq) with various
methods (selection of galaxies in X-ray deep surveys, conversion of
FIR and radio LFs, conversion of radio counts). We note that all of
them agree in two points:
\begin{itemize}
\item {\em the fraction of normal galaxies} at the current flux limit
  ($\sim 5\e{-17}$ \ergscmq in the 0.5--2.0 keV band) of the \chandra\ 
  Deep Field surveys {\em is about the $20\%$ of the total number of
    objects};
\item  {\em the X-ray number counts of normal
  galaxies should overcome the counts from AGN at fluxes below
  $1$--$2\e{-17}$ \ergscmq}. Since \citet{miyaji02b} suggested the
emergence of a new population (beyond that of AGN which dominates at
brighter fluxes) in the 0.5--2.0 keV band at fluxes around $10^{-17}$
erg s$^{-1}$ cm$^{-2}$, it is tempting to identify this new population
with the sub-mJy galaxies. 
\end{itemize}

We also note that it is unlikely that the X-ray \lognlogs of normal galaxies
could sustain its slope much below 2--3$\e{-18}$ \ergscmq, for
two reasons: i) because of radio/X-ray correlation, the integrated
radio emission from weak sources would diverge \citep[cfr.][]{fomalont91}; ii)
a much stronger evolution of the luminosity functions would be
required for the counts not to flatten below $\sim 10^{-18}$ \ergscmq;
but this would be at odds with the limits from the fluctuation
analysis \citep{miyaji02b}.

\section{The contribution to the X-ray background}

The derived X-ray number counts can be integrated to estimate
the contribution to the XRB. For the observed 2--10 keV background we
take the \XMM\ value of $2.15\cdot 10^{-11}$ erg s$^{-1}$ cm$^{-2}$
\citep{lumb02}, which is comprised between the ASCA \citep{gendreau95}
and BeppoSAX \citep{vecchi99} figures.  The integration of counts
derived from the 1.4 GHz \lognlogs \citep{vla00}, performed in its
validity range ($5.9\cdot 10^{-17}-1.3\cdot 10^{-15}$ erg s$^{-1}$
cm$^{-2}$, corresponding to 45-1000 $\mu$Jy at 1.4 GHz), yields a
contribution to the XRB of $4.0\cdot 10^{-13}$ erg cm$^{-2}$ s$^{-1}$
($\lesssim 2\%$ of the observed background). By extrapolating to
$10^{-18}$ erg s$^{-1}$ cm$^{-2}$ ($\sim 1 \mu$Jy) the
contribution would increase to $2.3\cdot 10^{-12}$ erg cm$^{-2}$ s$^{-1}$
($11\%$).  Integration of counts from the flatter
\citet{fomalont91} \lognlogs in the 1-1000 $\mu$Jy range yields a
contribution of $1.4\cdot 10^{-12}$ erg cm$^{-2}$ s$^{-1}$
($6.4\%$). We also note that 1 $\mu$Jy at 1.4 GHz is a limit $\sim
3-5$ times fainter than the constraint from radio fluctuation
analysis, and it is unlikely that the radio \lognlogs could sustain its
slope below this limit (cfr.\ previous section).

\section{An X-ray estimate of the cosmic star formation history}
\label{sec:SFH}

\begin{figure*}[t]    % NORMAN XLF
  \begin{center}
  \includegraphics[width=0.49\textwidth]{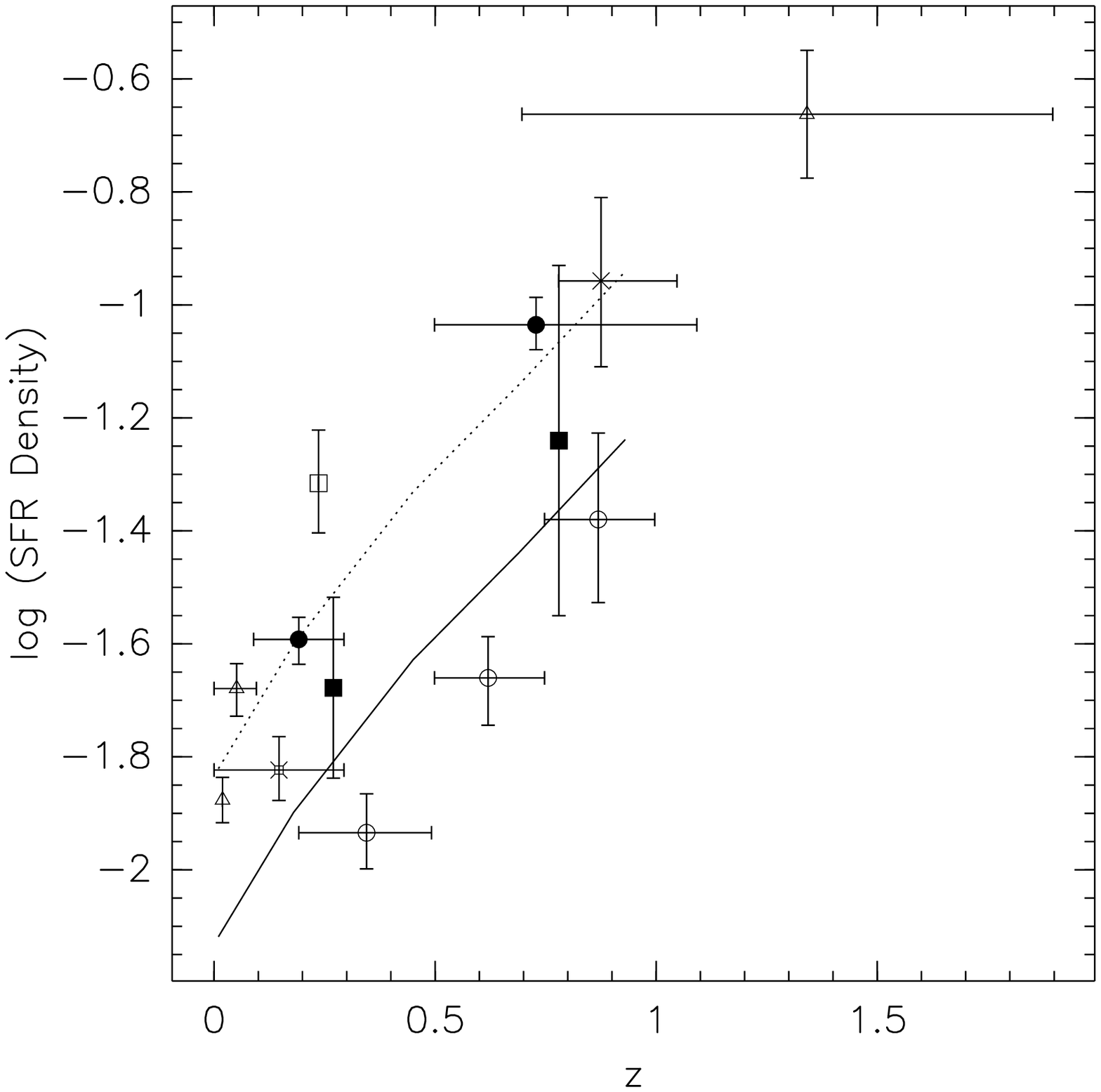}
  \end{center} 
  \caption { Estimates of the cosmic star formation history based on
different SFR idnicators, after \citet{colin}. The black squares
represent the X-ray estimates derived in \citet{colin} at the median
redshifts $z=0.27$ and $z=0.79$. The triangles represent
  the H$\alpha$ SFR values from \citet{gallego1995} at z $\sim 0$,
  \citet{gronwall1999} at z$\sim 0.05$, and \citet{hopkins2000} at z$\sim
  1.3$.  The empty square represents the UV-selected H$\alpha$
  \citet{pascual2001} value at z$\sim 
  0.24$.  The filled circles show the SFR densities from
  \citet{tresse02}. The star gives the UV-selected z=0.15 SFR density from
  \citet{sullivan2000}.  The open circles give 2800\AA~ CFRS points
  from \citet{lilly96}, without dust correction. We also plot the
  SFR history based on the 60 $\mu$m LF luminosity density at z $\sim
  0$, including (1+z)$^3$ evolution.  The 60 $\mu$m luminosity density
  was computed from \citet[solid line]{saun90} and \citet[dotted
    line]{takeu03}.
  \label{fi:sfh-norman} }
\end{figure*}

The current determinations of the cosmic star formation rate density
(CSFRD) show a rapid increase of the size of one order of magnitude
from $z=0$ to $z\sim 1$ \citep{lilly96,madau96}. At larger redshifts, either a peak
at $z\sim 1$--2 and a slow decrease, or a {\em plateau} are found, if UV
and H$\alpha$ data are taken at their face value, or if they are
corrected for extinction, respectively.

The main advantage in estimating the CSFRD from X-ray surveys would be
that the X-ray emission from star forming galaxies does not suffer
from absorption, so that one of the main sources of uncertainty in the
CSFRD determinations is removed. However, the current X-ray surveys do
not detect galaxies beyond redshifts $\sim 1.2$, nor any deeper survey
is currently planned. Thus the most interesting part of the cosmic
time ($z\sim 3$--5) is precluded to these studies. Nonetheless, it is
still a good exercise to check if an X-ray estimate of the CSFRD is
consistent with the determinations from data taken at other
wavelengths.

A determination of the cosmic star formation rate in a redshift
interval can be obtained by (cfr.\ eq.~\ref{eqsfrsoftX2})
\begin{equation}
{\rm SFR}(z) = 2.2\cdot 10^{-40} \int_{z_{\rm min}}^{z_{\rm max}}
\varphi(L)L\, \de L.
\end{equation}

In Fig.~(\ref{fi:sfh-norman}) we show the results for the cosmic SFR
at the mean redshift considered for the \citet{colin} sample, along
with compilation of SFR estimates from \citet{tresse02}. The overall
agreement of the X-ray derived points with the ones estimated from
infrared and H$\alpha$ observations is very good. This may not be
surprising, because the X-ray and FIR LFs agree quite well and the
CSFRD conveys the same information contained in a luminosity function,
but this result indeed confirms that X-ray observations may
successfully be used to derive an estimate of the CSFRD and of its
evolution.

% METALLICITY ENHANCEMENT IN STARBURSTS: THE CASE OF M82
\chapter{Metallicity enhancement in starbursts: the case of M82}
\label{sec:m82intro}

The signature of the star formation (SF) history of a galaxy is
imprinted in the abundance patterns of its stars and gas.  Determining
the abundance of key elements released in the interstellar medium
(ISM) by stars with different mass progenitors and hence on different
time scales, will thus have a strong astrophysical impact in drawing
the global picture of galaxy formation and evolution \citep{mwi97}.

Determining the abundances of SB galaxies is not only relevant to
constrain their SF history, but also offers the unique chance of
directly witnessing the enrichment of the ISM \citep{mc94}.  Metals
locked into stars give a picture of the enrichment just prior to the
last burst of SF, while the hot gas heated by SNe~II explosions and
emitting in the X-rays should trace the enrichment by the new
generation of stars.  Nebular metallicities from the cold gas should
potentially trace the stellar ones if the cooling and mixing
timescales are slow (up to 1 Gyr), as suggested by some models
\citep{tt96} since the gas enriched by the ongoing burst is still too
hot to be detected in the optical lines.  On the contrary, if rapid (a
few Myr) cooling and mixing occur as suggested by other models
\citep{rmde01}, the nebular abundances could trace the gas enriched by
the new population of stars (similarly to X-rays).  Hence, important
constraints on the cooling and mixing timescales of the gas can be
provided by comparing the metallicity inferred from stars and hot
plasma with those of the cold gas.

So far, the metallicities of distant starburst (SB) galaxies have been
mainly derived by measuring the nebular lines associated to their
giant HII regions \citep{sck94,coz99,hek00}.  Direct determination of
O and N abundances can be obtained once electron temperature ($T_e$)
is well established \citep{mca84}.  At high metallicities $T_e$
decreases due to strong lines cooling and its major diagnostic (the
weak [OIII] $\lambda$4636) is no longer observable.  Several
alternative methods based on other strong lines were proposed in the
past years \citep{pag79,de86,mcg91} to derive $T_e$, but all of them
rely on a large number of physical parameters and model assumptions
for the geometry and the nature of the ionizing populations
\citep{sta01}.  Depending on the adopted calibrations, quite different
abundance sets can reproduce the observed line ratios.
Moreover, stellar abundances can be poorly constrained from
optical spectra, since the nebular emission strongly dilutes the
absorption lines and dust can heavily obscure the central regions
where most of the burst activity is concentrated.

However, abundances in SB galaxies can be also measured from
absorption stellar features in the near IR and/or emission lines in
the hot X-ray gas. In the near IR the stellar continuum due to red
supergiants (RSGs) usually largely dominates over the possible gas and
dust emission \citep{oo98,oo00}, while the X-ray emission from SB
galaxies at energies lower than $\sim 2$ keV is mainly due to hot
plasma heated by SN explosions \citep{dahlem98} (\S~\ref{sec:introX}).

In order to test the above discussed enrichment scenario we have
started a pilot project to measure the metallicity enhancement in a
sample of starburst galaxies, for which we obtained high resolution
infrared (J and H band) spectra with the 3.6 m Italian Telescopio
Nazionale Galileo (TNG) and archival data from the \xmm\ and \chandra\ 
missions.  Our sample comprises M82, NGC253, NGC4449 and the {\em
  Antennae}, sampling a difference of two orders of magnitude in star
formation, as it ranges from the 0.3 $M_\sun$/yr of NGC4449 to the 30
$M_\sun$/yr of the Antennae.

The preliminary results which have been achieved for M82 with the
available \xmm\ archival data (a 20 ks exposure which provides high
resolution data, although with a low signal to noise ratio) will be
presented throughout this chapter.  \S~\ref{sec:m82lit} briefly
reviews the nebular abundance estimates of M82, as published so far in
the literature. \S~\ref{sec:m82ir} shows our near IR spectra of the
nuclear region of M82 and the derived stellar abundances.
\S~\ref{sec:m82xray} presents a re-analysis of the \xmm\ and \chandra\ 
nuclear X-ray spectra and the hot gas-phase abundances.
\S~\ref{sec:m82disc} summarized the overall chemical enrichment
scenario in M82 as traced by the stellar, cold and hot
gas-phase abundances. %, while in \S~\ref{sec:m82conc} we draw our
%conclusions.

\section{Nebular abundances of M82 from the literature}
\label{sec:m82lit}

M82 is a prototype of SB galaxies \citep{rie80}, experiencing a major
SF formation episode in its nuclear regions, with strong super-wind
and SN activity \citep{wil99}.  Despite the many multi-wavelength
studies \citep[][and references therein]{gak96}, only few
measurements of nebular abundances, mainly derived from mid/far IR
forbidden lines, have been published so far.  From the analysis of the
[SiII] 35$\mu$ line in the central 300~pc in radius at the distance
of M82 ($\sim 3.6$~Mpc), \citet{l96} suggest an indicative
$\rm [Si/H]\sim -0.4$~dex relative to the Solar value of
\citet{grev98}.

\citet{fs01}, by analyzing the nuclear spectra of M82
taken with the Short Wavelength Spectrometer (SWS) onboard the
Infrared Space Observatory (ISO) within an aperture corresponding to
$100\times200$~pc on the sky, give cold gas-phase abundances of
[Ne/H]$=+0.1$, [Ar/H]$=+0.3$ and [S/H]$=-0.7$ dex relative to the Solar
values of \citet{grev98}, with an overall uncertainty between 0.2 and
0.3 dex.

By taking the line ratios listed by \citet{acsj79} which refer to the
central $\sim 100$ pc of M82 \citep{ps70}, and the
empirical calibrations of \citet{mcg91} which give N and Fe abundances
as a function of [O/H] (see e.g.\ their Figure~11, equations~9 and
10a), one can also obtain rough estimates of [O/H]$=+0.1$, [N/H]$=0.0$ and
[Fe/H]$=-0.3$ dex relative to the Solar values of \citet{grev98}, with
an overall uncertainty of $\pm 0.3$~dex.

The first attempt to measure hot gas-phase abundances in M82 has been
made quite recently by \citet{pta97} and slightly revised by
\citet{ume02}, by analyzing ASCA X-ray (0.5--10.0~keV) spectra.  Fe and
O abundances as low as 1/20 Solar and significantly higher (by a
factor between 3 and 10) Si, S, Mg, Ca ones have been obtained, also
consistent with BeppoSAX results \citep{cappi99}.

More recently, \citet{rs02} by analyzing the spectra obtained with the
Reflection Grating Spectrometer (RGS) onboard \xmm\ found near-Solar
Fe and O and super-Solar (between a factor of 2 and 5) Mg, Si, Ne, N
abundances.  While the abundance ratios are in reasonable agreement
within $\pm 0.3$~dex with \citet{pta97}, there is one order magnitude
difference between the zero-point of the two calibrations.

\section{IR spectra and stellar abundances}
\label{sec:m82ir}

Near-IR spectroscopy is a fundamental tool to obtain accurate
abundances of key elements like Fe, C, O and other metals (e.g.\ Si,
Mg, Ca, Na, Al, Ti) in cool stars ($\rm T_{eff}\lesssim 5000$~K).
Several atomic and molecular lines are strong, not affected by severe
blending, hence also measurable at low-medium resolution
\citep{kh86,o93,wh97,joy98,fro01}, making them powerful abundance
tracers not only in stars but also in more distant stellar clusters
and galaxies and in a wide range of metallicities and ages
\citep{ori97,oo98}.  Abundance analysis from integrated spectra of
galaxies requires full spectral synthesis techniques to properly
account for line blending and population synthesis to define the
dominant contribution to the stellar luminosity.

In SB galaxies the stellar IR continuum usually dominates over the
possible gas and dust emission.  This represents a major, conceptual
simplification in population and spectral synthesis techniques, making
possible and easier the interpretation of integrated spectra from
distant stellar clusters and galaxies.

\begin{figure}[t]\centering 
\includegraphics[width=.85\textwidth]{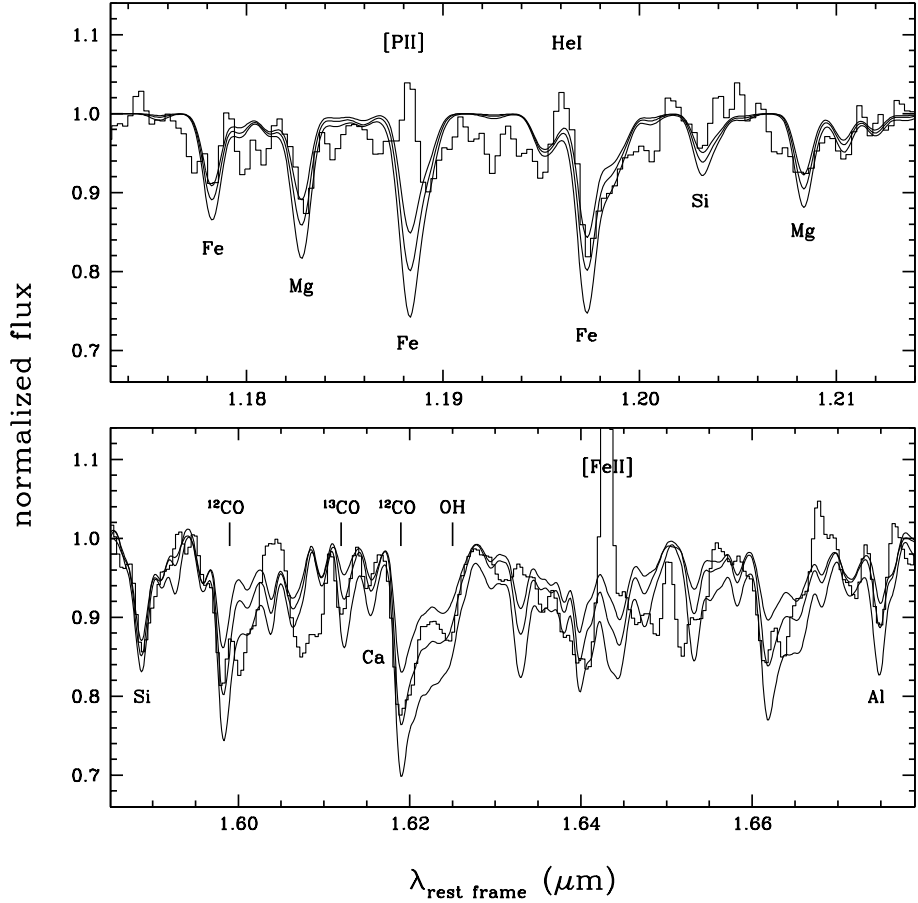} 
\caption {Near-IR spectra of the nuclear region of M82.
  Histograms: observed spectra; solid lines: synthetic stellar
  best-fit solution and two other models with $\pm$0.3 dex abundance
  variations.  A few major stellar and nebular lines are also marked.
\label{IR}}
\end{figure}

Since the near IR continuum of SB galaxies is dominated by luminous
RSGs \citep{oo00}, their integrated spectrum can be modeled with an
equivalent, average star, whose stellar parameters (temperature
$T_{\rm eff}$, gravity $\Log\ g$ and microturbulence velocity $\xi$)
mainly depend on the stellar age and metallicity.  Both observations
and evolutionary models \citep[][and references
therein]{k99,o99,ori03} suggest that RSGs of ages between $\sim $6
and 100 Myr and metallicities between 1/10 and Solar are characterized
by low gravities ($\Log\ g<1.0$), low temperatures ($\le 4000$~K) and
relatively high microturbulence velocity ($\xi\ge 3$ km/s).  A
variation in the adopted stellar parameters for the average RSG
population by $\Delta $T$_{\rm eff}\pm200$~K, $\Delta \Log\ g=\pm 0.5$
and $\Delta \xi\mp 0.5$~km~s$^{-1}$ implies a $\le \pm 0.1$~dex change
in the abundances estimated from atomic lines, and $\le \pm 0.2$~dex
in those estimated from the molecular lines, which are more sensitive
to stellar parameters.

The 1.0--1.8$\mu$ longslit spectra (see Fig.~\ref{IR}) at R=2,500 of
the nuclear region of M82 were obtained with the Near IR Camera
Spectrometer (NICS) mounted at the Italian Telescopio Nazionale
Galileo (TNG) on December 2002.  The spectra were sky-subtracted,
flat-fielded and corrected for atmospheric absorption using an O-star
spectrum as reference.  They were wavelength calibrated by using a
Ne-Ar lamp and the monodimensional spectra were extracted by summing
over the central $0.5\arcsec\times 3.0\arcsec$, corresponding to an
aperture projected on the source of $9\times 52$~pc at the distance of
M82.  The spectra have been normalized to the continuum, which was
determined applying a low-pass smoothing filter to each spectrum.
Total integration times of 72 and 32 min in the J and H bands,
respectively, were used, providing a final signal to noise ratio
$\ge$40.

By measuring the absorption line broadening a stellar velocity
dispersion $\sigma \sim 105 \pm 20$ km/s has been derived, in
perfect agreement with the estimate by \citet{glt93} and in the
typical range of values measured in other massive SB galaxies
\citep{oli99}.

\begin{figure}[t]\centering
\includegraphics[width=.85\textwidth,height=.33\textheight]{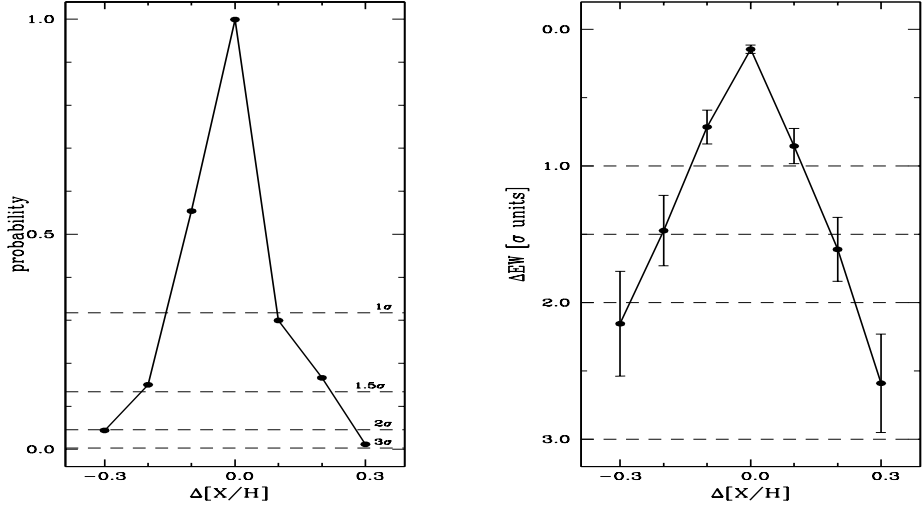} 
\caption{{\it Left panel}:  probability of a random realization
of our best-fit full spectral synthesis solution with varying the
elemental  abundances $\Delta[X/H]$ of $\pm$0.2 and 0.3 dex with
respect to the best-fit (see Sect.~\ref{sec:m82ir}). 
{\it Right panel}: average difference between the model and the
observed equivalent widths of a few selected lines (see Sect.~\ref{sec:m82ir}
and Fig.~\ref{IR}).
\label{IRtest}}
\end{figure}

A grid of synthetic spectra of red supergiant stars for different
input atmospheric parameters and abundances have been computed, using
an updated \citep{o02,o03} version of the code described in
\citet{o93}.  Briefly, the code uses the LTE approximation and is
based on the molecular blanketed model atmospheres of \citet{jbk80}.
It includes several thousands of near IR atomic lines and molecular
roto-vibrational transitions due to CO, OH and CN.  Three main
compilations of atomic oscillator strengths are used, namely the
Kurucz's database\footnote{cfr.\ {\tt
  http://cfa-www.harward.edu/amdata/ampdata/kurucz23/sekur.html}}, and
those published by \citet{bg73} and \citet{mb99}.
 
The code provides full spectral synthesis over the 1-2.5 $\mu$m range.
Given the high degree of line blending, the abundance estimates
are mainly obtained by best-fitting the full
observed spectrum and by measuring the equivalent widths
of a few selected features (cf. Fig.~\ref{IR}), dominated by 
a specific chemical element, as a further cross-check. 
The equivalent widths have been measured by performing a Gaussian fit 
with $\sigma$ equal to the measured stellar velocity dispersion, 
typical values ranging between 0.5 and 3 \AA, 
with a conservative error of $\pm$200~m\AA\ to also account for a $\pm$2\% 
uncertainty in the continuum positioning.

\begin{table}[t]
  \centering
  \begin{tabular}{lcccc}
     &{\sc Stellar abundances} & & &{\sc hot gas-phase}\\
\hline
$\rm [Fe/H]$ &  $-0.34\pm0.20$ &&& $-0.37\pm0.11$\\
~Fe/Fe$_{\odot}$ &$0.46^{+0.26}_{-0.17}$  &&& $0.43^{+0.12}_{-0.08}$\smallskip \\
$\rm [O/H]$  &  $+0.00\pm0.16$ &&& $-0.58\pm0.19$\\
~O/O$_{\odot}$   &$1.00^{+0.46}_{-0.32}$  &&& $0.26^{+0.15}_{-0.09}$\smallskip\\
$\rm [Ca/H]$ &  $+0.05\pm0.28$ &&& ---              \\
~Ca/Ca$_{\odot}$ &$1.11^{+1.01}_{-0.52}$  &&& ---\smallskip\\
$\rm [Mg/H]$ &  $+0.02\pm0.15$ &&& $+0.13\pm0.09$  \\
~Mg/Mg$_{\odot}$~~~~~ &$1.06^{+0.44}_{-0.31}$  &&& $1.36^{+0.32}_{-0.26}$\smallskip\\
$\rm [Si/H]$ &  $+0.04\pm0.28$ &&& $+0.17\pm0.08$  \\
~Si/Si$_{\odot}$ &$1.09^{+0.99}_{-0.52}$  &&& $1.49^{+0.32}_{-0.26}$\smallskip\\
$\rm [Al/H]$ &  $+0.23\pm0.20$ &&& ---\\
~Al/Al$_{\odot}$ &$1.69^{+0.97}_{-0.62}$  &&& ---\smallskip\\
$\rm [C/H]$  &  $-0.60\pm0.10$ &&& ---\\
~C/C$_{\odot}$   &$0.25^{+0.06}_{-0.05}$  &&& ---\smallskip\\
$\rm [Ne/H]$ &  ---             &&& $-0.35\pm0.14$\\
~Ne/Ne$_{\odot}$ &---                      &&& $0.45^{+0.17}_{-0.12}$\smallskip\\
$\rm [S/H]$  &  ---             &&& $+0.15\pm0.13$\\
~S/S$_{\odot}$   &---                      &&& $1.42^{+0.48}_{-0.40}$\\
\hline
  \end{tabular}

  \caption{%
Element abundances in Solar units as derived from our analysis of the
IR and X-ray spectra. The solar reference from \citet{grev98} was
used. Stellar abundances from TNG/NICS near IR absorption spectra of
red supergiants (see Fig.~\ref{IR} and Sect.~\ref{sec:m82ir}).  Hot
gas-phase abundances from XMM/RGS and {\it pn} spectra (see
Figs.~\ref{XR}, \ref{fig-abun} and Sect.~\ref{sec:m82xray}). }

  \label{ab}
\end{table}

%\sloppypar 
By fitting the full observed IR spectrum and by measuring
the equivalent widths of selected lines, we obtained the following
best-fit stellar parameters and abundance patterns for M82:
$T_{\rm eff}=4000$, $\Log\ g=0.5$, $\xi=3$, 
[Fe/H]$=-0.34$; [O/Fe]$=+0.34$, [$<$Si, Mg, Ca$>$/Fe]$=+0.38$; 
[Al/Fe]$=+0.5$; [C/Fe]$=-0.26$;
$^{12}$C/$^{13}$C$<$10.
Table~\ref{ab} lists the derived abundances and their associated
random errors at 90\% confidence. Reference Solar abundances are from
\citet{grev98}.

Synthetic spectra with lower element abundances are {\em
  systematically} shallower than the best-fit solution, while the
opposite occurs when higher abundances are adopted.  In order to check
the statistical significance of our best-fit solution, as a function
of merit we adopt the difference between the model and the observed
spectrum (hereafter $\delta$).  In order to quantify systematic
discrepancies, this parameter is more powerful than the classical
$\chi ^2$ test, which is instead equally sensitive to {\em random} and
{\em systematic} scatters \citep{o03}.

Since $\delta$ is expected to follow a Gaussian distribution, we
compute $\overline{\delta}$ and the corresponding standard deviation
($\sigma$) for the best-fit solution and 6 {\it test models} with
abundance variations $\Delta[X/H]=\pm$0.1, 0.2 and 0.3 dex with
respect to the best-fit.  We then extract 10000 random subsamples from
each {\it test model} (assuming a Gaussian distribution) and we
compute the probability $P$ that a random realization of the
data-points around a {\it test model} display a $\overline{\delta}$
that is compatible with the {\em best-fit} model.  $P\sim 1$
indicates that the model is a good representation of the observed
spectrum.
 
The left panel of Fig.~\ref{IRtest} shows the average results for the
observed J and H band spectra of M82.  It can be easily appreciated
that the best-fit solution provides in all cases a clear maximum in
$P$ ($>$99\%) with respect to the {\it test models}.  More relevant,
{\it test models} with an abundance variation $\Delta[X/H]\ge\pm
0.2$~dex lie at $\sim 1.5\sigma$ from the best-fit solution, while
{\it test models} with $\Delta[X/H]\ge\pm 0.3$~dex lie at $\sigma \ge
2$ from the best-fit solution.

The analysis of the line equivalent widths provide fully consistent
results.  The right panel of Fig.~\ref{IRtest} shows the average
difference between the model and the observed equivalent width
measurements.  Models with $\pm 0.2$~dex abundance variations from the
best-fit solution are still acceptable at a $\sim 1.5\sigma$
significance level, while those with $\pm 0.3$~dex variations are only
marginally acceptable at a $2-3 \sigma$ level.

Models with stellar parameters varying by $\Delta T_{\rm
  eff}\pm200$~K, $\Delta \Log~g=\pm0.5$ and $\Delta
\xi\mp 0.5$~km~s$^{-1}$ and abundances varying accordingly by
0.1-0.2 dex, in order to still reproduce the deepness of the observed
features, are also less statistical significant (on average only at
$\ge2 \sigma$ level) with respect to the best-fit solution.  Hence, as
a conservative estimate of the systematic error in the derived
best-fit abundances, due to the residual uncertainty in the adopted
stellar parameters, one can assume a value of $\le \pm 0.1$~dex.

By taking into account the overall uncertainty in the definition of
the average population and the statistical significance of our
spectral synthesis procedure, we can safely conclude that the stellar
abundances can be constrained well within $\pm0.3$~dex and their
abundance ratios down to $\sim 0.2$~dex, since some (if not all) of
the stellar parameter degeneracy is removed.

\section{X-ray spectra and hot gas-phase abundances}
\label{sec:m82xray} 

The determination of the abundances of the hot X-ray emitting gas in
SB galaxies has traditionally suffered from large uncertainties.
Indeed, the low angular and spectral resolution of the various X-ray
telescopes in the pre-\xmm/{\it Chandra} era did not allow to
disentangle between point sources and hot gas emission, making
abundance determinations severely model-dependent \citep{dah00}. The
problem lies in the fact that the X-ray spectrum of SB galaxies
contains two major components: the emission from hot diffuse gas and
the integrated contribution of point sources. The first is described
by an optically thin thermal spectrum {\it plus} emission lines, the
latter by a power-law.  If the angular resolution is not good enough,
it is not possible to reliably subtract the point sources from the
total spectrum, so that the equivalent widths of the emission lines
(and thus the element abundances) are not unambiguously defined.

On the other hand, high angular resolution alone is not enough: recent
{\it Chandra } studies (\citealt{stric02}; \citealt{mart02})
demonstrated that the relatively low spectral resolution of the
  ACIS detector makes individual element abundance analysis still
problematic. The high spectral resolution of the {\it Chandra}
gratings rapidly degrades for sources which are much more extended
than the instrumental Point Spread Function (PSF) and thus these
gratings are almost useless for the study of nearby SB galaxies.
However, by coupling the high angular resolution ($0.5\arcsec$ PSF)
of {\it Chandra} with the high spectral resolution
($\lambda/\Delta\lambda\sim 400$) of the XMM/RGS, which is not
sensitive to the source extension, it is possible to overcome the
above described difficulties.

M82 was observed several times by \chandra; we consider the two
longest exposures, of 33 ks and 15 ks respectively, both obtained on
September 1999. A cumulative spectrum of the brightest point sources
was extracted from the two \chandra\ ACIS-I observations; the
best-fit model was an absorbed power-law with $\rm
N_{{H}}=7.9\pm0.7\times 10^{21}$ cm$^{-2}$, $\Gamma=0.84\pm0.07$,
$\chi_{{\rm r}}^{2}=0.4$, which is typical of High Mass X-ray
Binaries \citep[see][and reference therein]{pr02}.  The 0.5--10 keV
flux was $4.2\times 10^{-12}$ erg s$^{-1}$ cm$^{-2}$, which
represented $\sim25\%$ of the total (point source {\it plus} diffuse)
flux (but notice that M82 is a variable source, \citealt{ptak99},
\citealt{rephaeli02}).

\begin{figure}[tp] \centering
\includegraphics[height=.8\textwidth,angle=-90]{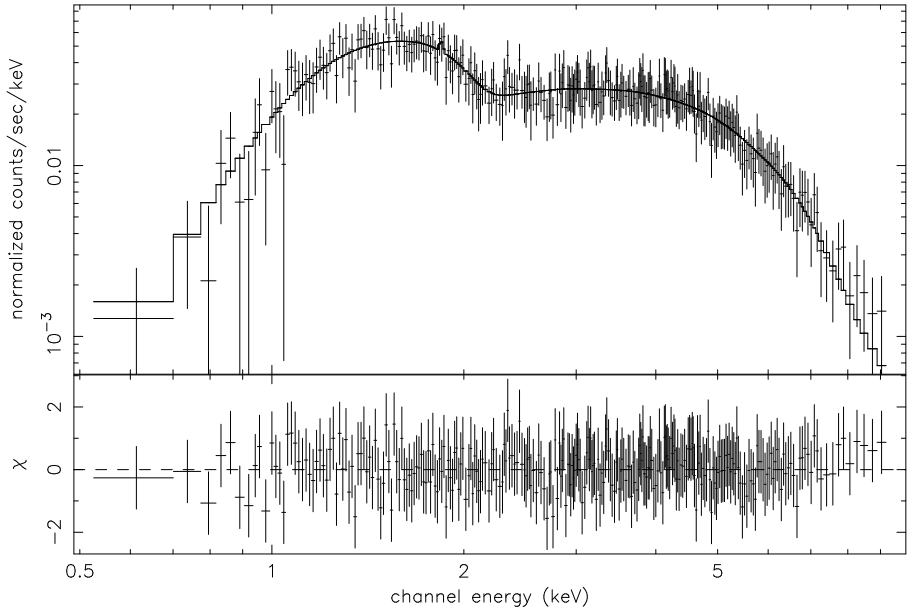} 
\caption{ Total spectrum of the most luminous point sources in M82,
  obtained with \chandra.
\label{fig:m82pointsrc}}
\end{figure}

\begin{figure}[tp] 
\includegraphics[width=\textwidth]{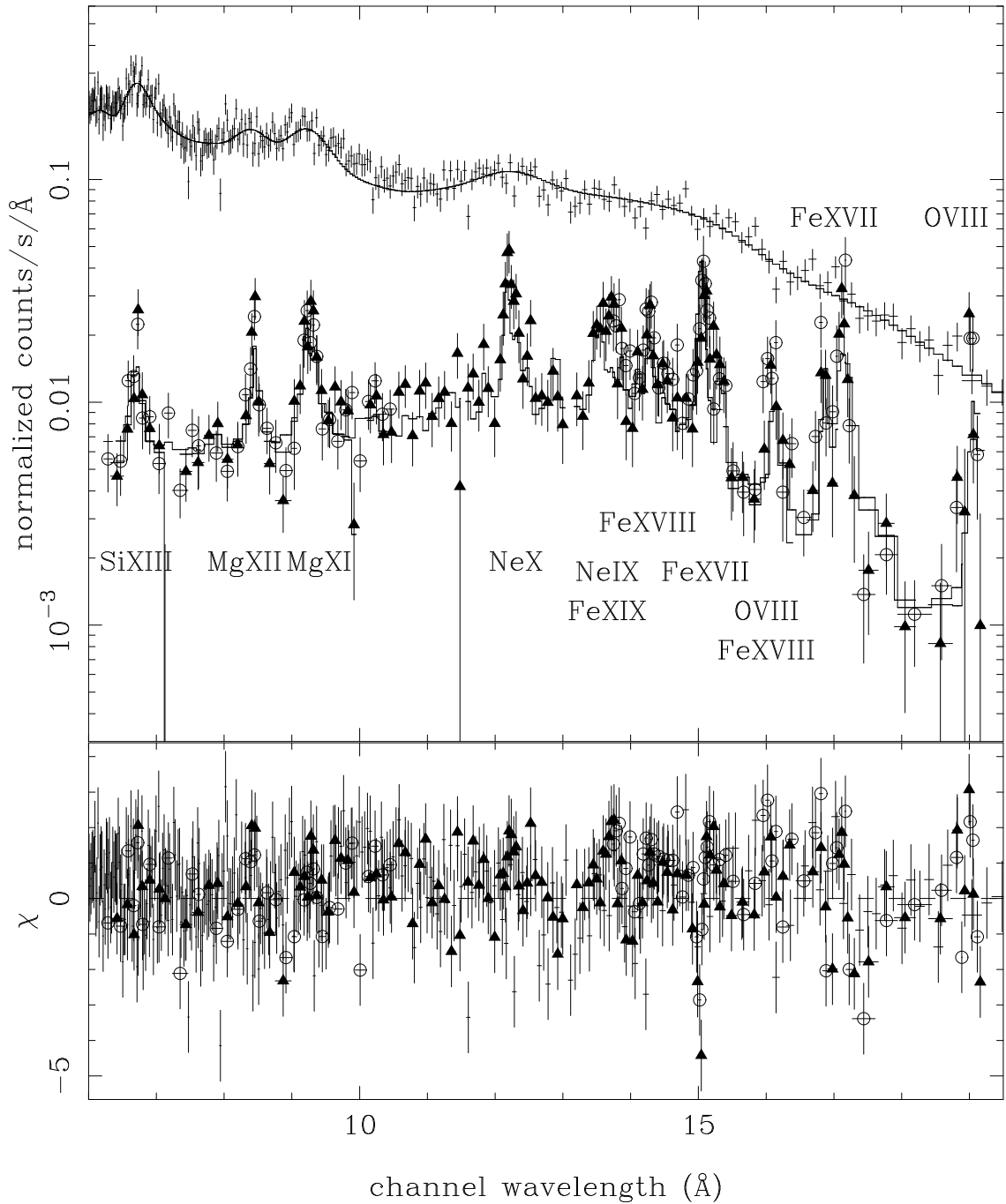} 
\caption{
  XMM-RGS spectrum of the nuclear region of M82.  Upper panel: data
  points (lower spectrum - open circles: RGS1; filled triangles: RGS2;
  upper spectrum: {\it pn}) along with the best-fit model. Lower
  panel: residuals in units of $\sigma$.
\label{XR}}
\end{figure}

M82 was observed by \xmm\ for 30 ks in May 2001; in the \xmm\ 
observation, after light-curve cleaning, $\sim 20$~ks of data were
available for scientific analysis. We extracted the EPIC {\it
  pn} spectrum from a circular region centered on the starbursting
core of M82 and with a $15\arcsec$ radius, which approximately matches
the RGS PSF; the standard recipe for EPIC spectra extraction
suggested in the XMM web pages was followed.  The XMMSAS
software version 5.3.0 was used.  A background spectrum was taken from
standard background files provided by the \xmm\ SOC. RGS spectra were
extracted with standard settings (i.e.\ including events within 90\%
confidence and rejecting as background events those outside 95\% of
the PSF figure); this procedure is similar to the one used by
\citet{rs02}\ but with more recent calibrations. After visual
inspection, we discarded the 2nd-order spectra because of low
signal/noise ratio. We included in the fit only the 1st-order channels
comprised in the 6--20 \AA\ interval, since outside the S/N ratio
rapidly drops. A systematic error up to 10\% in the 14--18 \AA\ 
channels was included (XMM calibration document XMM-SOC-CAL-TN-0030
Issue 2).

The hard-band (2--7 keV) \chandra\ images show that the point sources
emission largely exceeds that of the hot gas, while the hard part
(3--7 keV) of the {\em pn} spectrum can be well described by a
power-law whose best-fit slope resulted in the same value previously
determined for the \chandra\ point sources.  Thus, making use of
information from both \chandra\ and \xmm\ observations, we have been
able to constrain the point source slope, absorption and normalization
at the moment of the \xmm\ observation.

We jointly fitted the {\it pn} and RGS spectra with XSPEC
(Fig.~\ref{XR}) with a two component model which
is the sum of: i) an absorbed power-law (accounting for point
sources) with the slope and absorption parameters fixed at the best
fit values obtained with \chandra\ while normalization was left free
to account for variability; ii) an absorbed, optically-thin
thermal plasma emission, with variable line intensities and
differential emission measure (DEM) distribution.  The DEM
distribution was modeled with a 6$^{\textrm{th}}$ order Chebyshev
polynomial (\textsc{c6pvmkl} model in XSPEC); with the best-fit
parameters it resembles a bell-shaped curve with a peak at $kT\sim0.7$
keV and FWHM $\sim0.65$ keV (i.e.\ the plasma temperatures range from
$\sim$0.35 keV to $\sim$1 keV), in agreement with \citet{rs02}. The
thermal component is absorbed by cold gas with best-fit column density
$N_H \sim 3.8 \times 10^{21}$ cm$^{-2}$.

\begin{figure}[tp] \centering
\includegraphics[width=.75\textwidth]{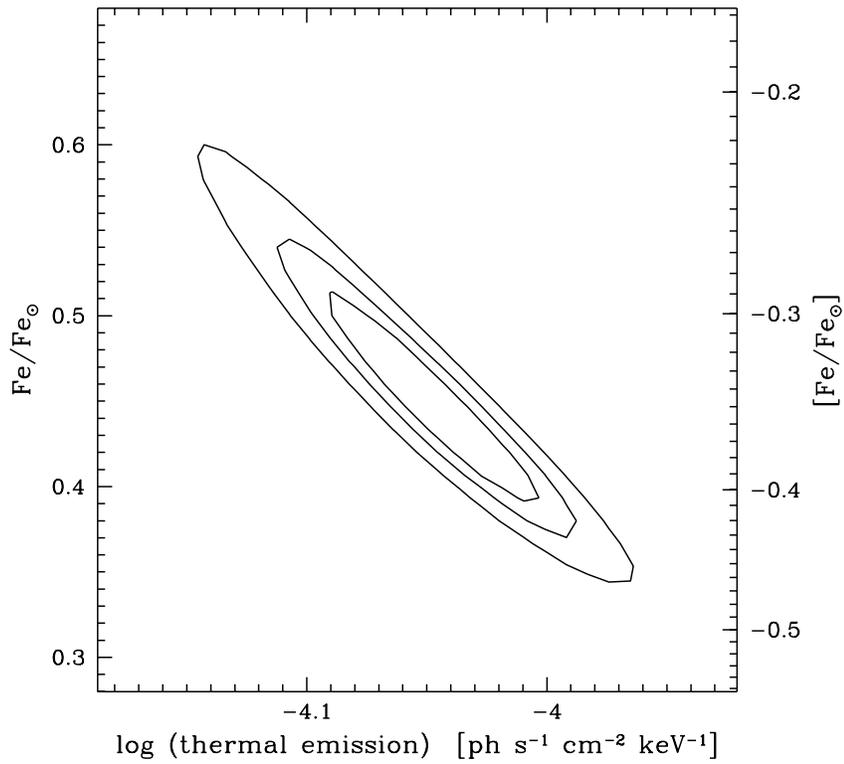}
\caption {68, 90 and 99\% confidence contours of the 
  Iron abundance {\it vs} intensity of the thermal component.  For
  ease of reading the y-axis is shown both in linear and logarithmic
  scale.
\label{fig-fevsnorm}}
\end{figure}

The confidence contours of the Iron abundance versus normalization of
the thermal component are reported in Fig.~\ref{fig-fevsnorm}. As
expected the two parameters are well correlated as the higher is the
normalization of the continuum spectrum the lower is the intensity of
the emission lines and in turn the element abundances.  The derived
abundances and their associated 90\% confidence error are shown in
Table~\ref{ab}. We notice that these abundances are about a factor 2
below those reported in \citet{rs02}; we attribute the discrepancy to
the different modeling of the continuum emission (\citealt{rs02}\ did
not include {\it pn} data, neither considered the contribution from
point sources).  However, it is worth of mentioning that both sets of
abundances indicate a low O abundance compared to the other
$\alpha$-elements.

\begin{figure}[tp] \centering
\includegraphics[width=.75\textwidth]{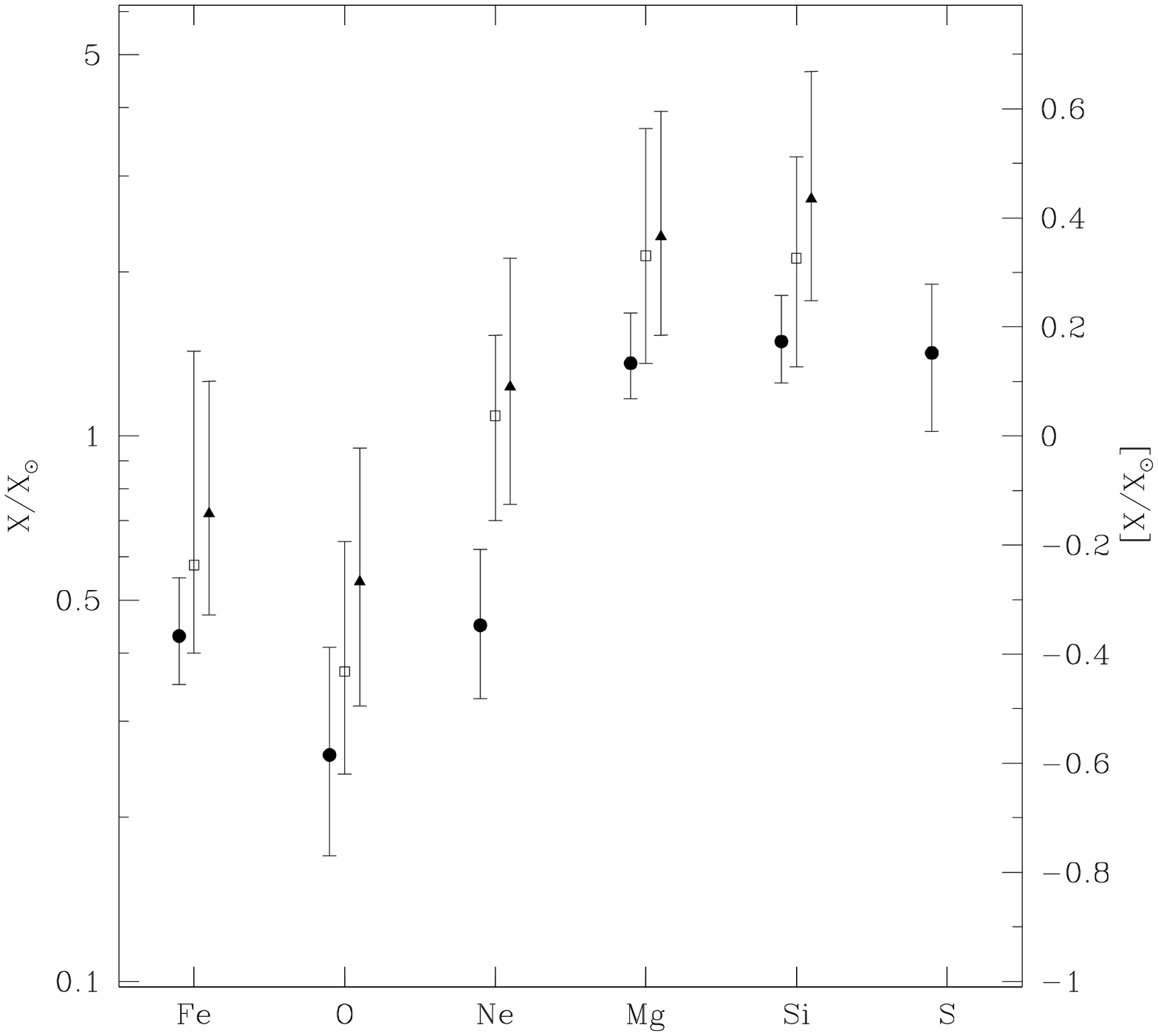}
\caption {Best-fit element abundances of the hot gas for different models. 
  Filled circles: fit of both RGS and {\it pn} data, thermal {\it
    plus} power-law model.  Open squares: RGS data alone, thermal {\it
    plus} power-law model.  Filled triangles: RGS data alone, thermal
  model alone.  For ease of reading the y-axis is shown both in linear
  and logarithmic scale.
\label{fig-abun}}
\end{figure}

To further check the consistency and robustness of our estimates, we
repeated the fit two times considering only the RGS data and using
i) the above described thermal {\it plus} power-law model and
ii) a thermal model alone.  This last check, in particular,
most closely matches the analysis previously made by \citet{rs02} to
the point that the discrepancies are within the errors and can be
attributed to the differences in the used instrument calibration.  The
abundances derived from the three different models are plotted in
Fig.~\ref{fig-abun}. Those obtained combining {\it pn} and RGS have
significantly smaller uncertainties.  One can see that different
models have mainly the effect to almost rigidly scale all the
abundances, with minor impact on the abundance ratios.

\begin{figure}[tp] \centering
\includegraphics[width=.75\textwidth]{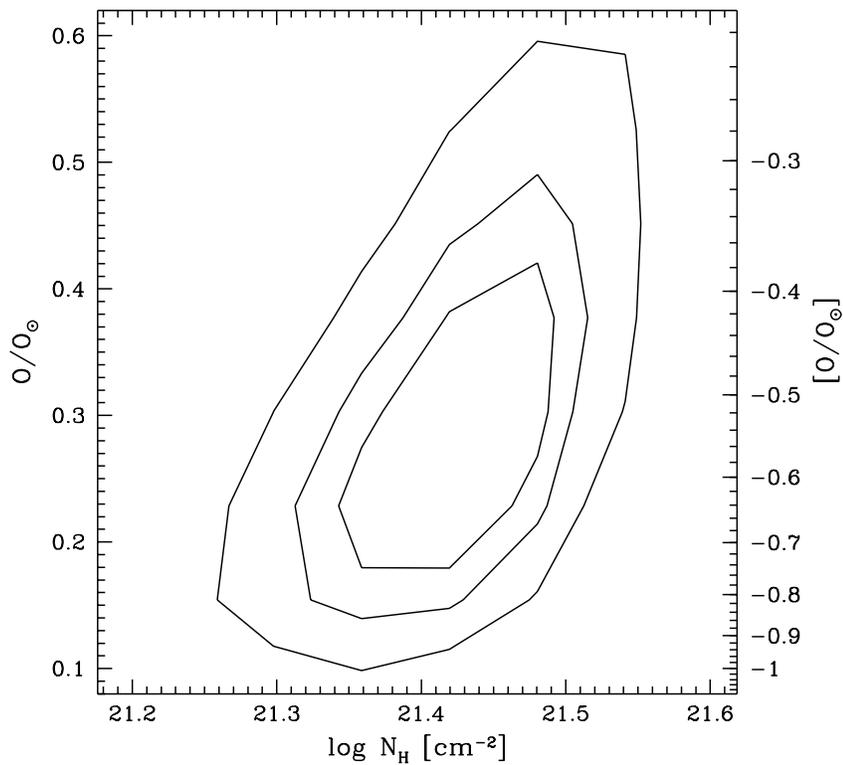}
\caption {Hot gas-phase Oxygen abundance {\it vs} absorption. 
  For ease of reading the y-axis is shown both in linear and
  logarithmic scale.  The correlation, if any, is rather weak, and the
  Oxygen abundance can be well constrained.
\label{fig-ovsnh}}
\end{figure}

The low Oxygen abundance is somehow difficult to explain in the
framework of $\alpha$-element enhancement by type II SN explosions.
Thus, to check its reliability we included in the fit also the
channels corresponding to the 20--23 \AA\ interval, where the
\ion{O}{vii} triplet is observed (although with very few counts). We
find no significant difference neither in Oxygen abundance nor in
DEM distribution (the \ion{O}{viii} $\lambda$19/\ion{O}{vii}
$\lambda$22 ratio is extremely sensitive to the plasma temperature).
Ne is also somewhat under-abundant. We checked for possible
instrumental effects, such as variations in the RGS effective area in
proximity of the Ne and O lines; however the effective area in this
region is a rather smooth function of the wavelength, so that there
should be no instrumental issues and these low abundances are likely
to be real.

Since M82 is observed edge-on, the X-ray emission of its nuclear
region is subject to quite heavy absorption which could affect the O
lines detected at low energies.  However, as shown in
Fig.~\ref{fig-ovsnh}, the contour plot of the Oxygen abundance towards
the foreground absorption, suggest a rather weak correlation.  In the
next section we discuss other physical possibilities to explain the O
under-abundance issue.

\begin{figure}[tp] \centering
\includegraphics[width=.75\textwidth]{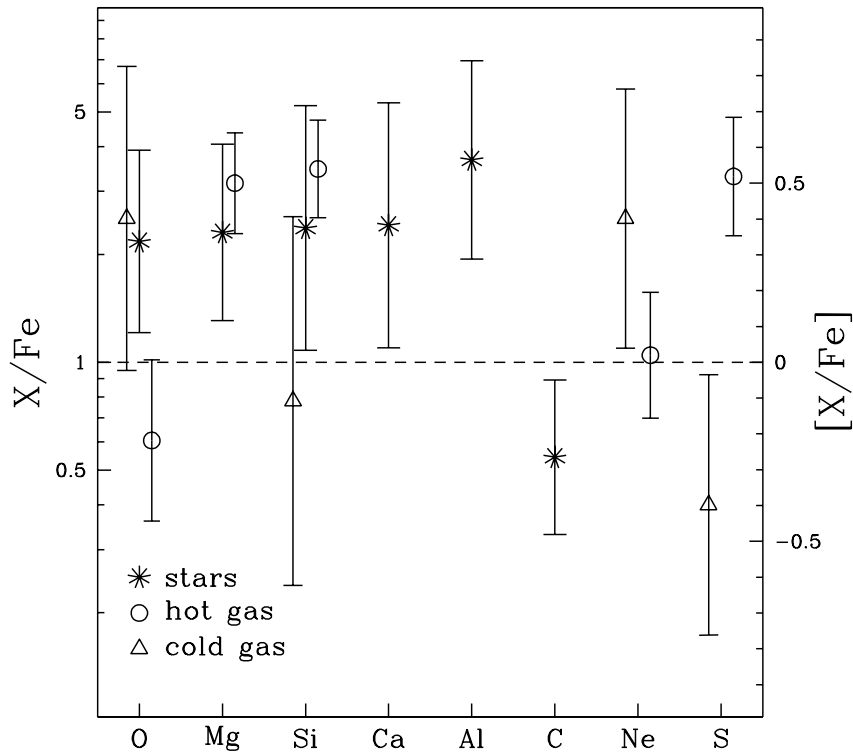}
\caption {Stellar, hot and cold gas abundance ratios relative to Iron.  
  Cold gas-phase abundances are from different data sets in the
  literature, hence they are not homogeneously determined.  The dashed
  line indicates the solar values.  For ease of reading the y-axis is
  shown both in linear and logarithmic scale.
\label{abtot}}
\end{figure}

\section{Discussion}
\label{sec:m82disc}

In M82 all the three components, namely hot and cold gas-phases and
stars trace a very similar Iron abundance, the average value being
[Fe/H]$\sim -0.35$~dex.  Indeed, since Iron is mainly produced by
SNe~Ia, it is expected to be released in the ISM only after $\sim
1$~Gyr from the local onset of SF.

At variance, $\alpha -$elements (O, Ne, Mg, Si, S, Ca, Ti) are
predominantly released by SNe~II with massive progenitors on much
shorter timescales.  Stars trace an average [$<$Si,Mg,Ca$>$/Fe]$\sim
0.4$~dex, while the hot gas suggests an average
[$<$Si,Mg,S$>$/Fe]$\sim 0.5$~dex.  Such an overall [$\alpha$/Fe]
enhancement in M82 is fully consistent with a standard chemical
evolution scenario, where the ISM in the nuclear regions of massive SB
galaxies is mainly enriched by the products of SNe~II explosions (see
e.g.\ \citealt{arn95} for a review) occurring in recursive bursts of
SF of relatively short duration.

Fig.~\ref{abtot} shows the various abundance ratios relative to Iron,
as measured in stars, hot and cold gas-phases.  Since the cold
gas-phase abundances are from different data sets in the literature,
hence not homogeneously determined, the inferred abundance ratios
should be regarded with caution.

When comparing the hot and cold gas-phase abundances, one finds Ne
over-abundant, Si and especially S significantly under-abundant in the
cold gas.  \citet{fs01} suggests S and to a lower level Si depletion
onto interstellar dust grains.  However, other metals like Fe in
particular, with even higher degree of incorporation into dust
\citep{ss96}, should be severely depleted in the cold gas, which does
not seem the case.  A possible explanation is that the optical lines
used to infer Fe and O abundances \citep{acsj79} and the the mid/far
IR lines used to infer Ne and S abundances \citep{l96,fs01} trace
different nebular sub-structures within the central few hundreds pc,
somewhat chemically dishomogeneous or with different dust content.

The [O/Fe] abundance ratio in M82 is even more puzzling to interpret.
Before the onset of the current burst of SF, the ISM was enriched in O
as well as in the other $\alpha$-elements, as suggested by our stellar
and cold gas-phase abundances.  The O under-abundance measured in the
hot gas cannot be modeled with a standard nucleosynthesis from the
present generation of SNe~II.  \citet{ume02} suggested explosive
nucleosynthesis in core-collapse hypernovae but they predicted a large
under-abundance of Ne and Mg as well, which is not observed.  [O/Fe]
under-abundance can be also explained with a major Fe enrichment by
current SN-Ia explosions from previous generations of stars, but in
this case all the [$\alpha$/Fe] ratios should be low, again contrary
to what observed.  O might be also locked into dust grains.  Indeed,
SNe~II are known to produce significant amounts of dust \citep{tf01}
and O, being a basic constituent of dust grains, is expected to suffer
of severe dust depletion.  However, other metals like Fe in
particular, but also Si and Mg should be depleted into dust
\citep{ss96}, leaving almost unchanged or eventually further enhancing
the [$\alpha$/Fe] ratio.

\chapter{Summary}

%The basic ideas in the interpretation of the radio/FIR relation lie
%in the mechanisms of star formation and evolution, leading to cosmic
%ray acceleration and dust heating.  Our results suggest that the 2--10
%keV X-ray emission of galaxies is closely correlated with the radio
%and far infrared ones. The origin of the hard X-ray emission must thus
%be closely linked to star formation too. There are a few
%possibilities, i.e.  X-ray binaries, stellar mass black holes, hot
%($\sim 6$ keV, see \cite{cappi}) plasma, diffuse
%emission; however a basic picture could be sketched. Massive stars
%which heat the dust by means of their UV radiation, also end up with a
%supernova explosion in which a remnant and a compact object are
%formed. The remnant could be the site of cosmic ray acceleration and
%plasma heating. Compact objects which were in a binary system before
%the explosion can become X-ray binaries if the bynary system has
%survived. A population of X-ray binaries can thus be
%formed, which number should be proportional to the SFR.

We have analyzed a small, but well defined sample of 17 star-forming
galaxies, extracted from the HFS97 catalogue, for which there is a
homogeneous information on optical, FIR, radio and X-ray bands (local
sample). In agreement with previous work \citep{djf92} we find that
the logarithms of the soft (0.5--2 keV) X-ray luminosities (corrected
for Galactic absorption only) are linearly correlated with the
logarithms of both radio (1.4 GHz) and FIR luminosities.  We have
extended our analysis to the harder X-ray band, essentially free from
internal absorption which may affect the soft X-ray fluxes, and found
that there is a tight linear correlation between the X-ray
luminosities in the 2--10 keV interval with both the radio and the FIR
luminosities, normally assumed as the indicators of the star formation
rate. The addition of 6 galaxies (supplementary sample) homogeneous
with, but not included in HFS97, does not modify these results.
We conclude that the origin of the hard X-ray emission must be
closely related to star formation and calibrate an X-ray SFR
indicator.

Candidate starburst galaxies have been selected in the HDF North, with
redshifts up to $z\,\sim\,1.3$, and their rest-frame X-ray
luminosities are computed by extracting counts in redshifted bands
from the \chandra\ observation of the HDFN. With this approach we have
shown that the 2--10 keV/radio linear correlation holds up to $z\sim
1.3$, encompassing five orders of magnitude in luminosity, up to
$L_{2-10} \sim$ a few $10^{42}$ erg~s$^{-1}$ and a corresponding
star formation rate $\sim 1000~ {\rm M}\sun$ yr$^{-1}$ --- but at
such large luminosities some contribution from nuclear activity
might be present.
The fit to the 0.5--2.0 keV/radio data is also linear up to $z\sim 1.3$. 
% However,
% the count ratio for the highest redshift galaxy at $z=1.275$ requires
% significant absorption if a spectral slope of $\Gamma\sim 2.1$ is
% assumed; thus this galaxy shows, on a larger scale, the same behaviour
% as that of the southern core in NGC\,3256 (see the Appendix).
% Therefore, while the linearity of the relations involving soft X-ray
% luminosities remains statistically significant, at high redshift
% (where galaxies are supposed to have more gas at their disposal to
% form stars, and so their X-ray emission is more likely to be absorbed)
% the 2--10 keV luminosity is a more secure indicator of the SFR.

As an additional investigation we have also analyzed a sample of LLAGN
(LINERs and Seyfert's) included in HFS97 \citep{terashima}: while, as
expected, the X-ray luminosities are generally in excess with respect
to star-forming galaxies for the same FIR luminosity, the distribution
of the objects in the X-ray vs.~FIR luminosity diagram is bounded from
below from the region occupied by the star-forming galaxies,
indicating that the X-ray emission of LLAGN falling in this
border-line region could be mainly due to star formation processes,
rather than being of nuclear origin.

We used the radio/FIR/X-ray correlations to convert the local ($z=0$)
FIR and radio luminosity functions (LFs) into X-ray LFs, which
were compared with the one recently derived in \citet{colin} for
objects at intermediate redshifts ($0.1\lesssim z\lesssim 1.2$). A
quite good agreement was found for $\LX\lesssim 10^{41}$ erg/s, while
at larger luminosities there is almost an order of magnitude of
difference.  Tho possible explanations may be proposed: i) a fraction of
AGN is contaminating the high-luminosity tail of the sample of
galaxies used in deriving the X-ray LF; ii) the infrared and radio
surveys do not sample a sufficiently large volume to find rare, high
luminosity objects.

X-ray number counts for the normal galaxies were predicted both by
integrating the radio- and FIR-based luminosity functions, and by converting
the radio \lognlogs for the sub-mJy population.
The expected counts extend much below the limiting fluxes
of the deepest X-ray surveys (about one order of magnitude in the soft
band, one and a half for the hard band), and are within the limits set
by the fluctuations analysis in the \chandra\ deep fields
\citep{miyaji02a,miyaji02b}. They are also consistent with the
predictions based on the evolution of the cosmic SFR density by
\cite{ptak01}. Since the results from fluctuations analysis in the
soft X-rays suggest an excess of sources with respect to AGN synthesis
models at fluxes below $\sim 10^{-17}$ erg s$^{-1}$ cm$^{-2}$, it may
be possible that the sub-mJy galaxies represent the dominant
population in the X-rays at very faint fluxes.
Our analysis of the X-ray number counts also allows a determination of
the fraction of normal galaxies among the X-ray sources detected in
the deepest X-ray surveys. Namely, we estimate that $20\%$ of the
X-ray sources with fluxes larger than $5\e{-17}$ \ergscmq\ in the
0.5--2.0 keV band are star forming galaxies.

The contribution to the cosmic X-ray background in the 2--10 keV band
is estimated by integration of the derived X-ray number counts.  The
contribution from galaxies detected in the deepest radio surveys is
$\lesssim 2\%$. This estimate may rise up to 11$\%$ by extrapolating
the radio counts down to 1~$\mu$Jy, or $\sim10^{-18}$ erg s$^{-1}$
cm$^{-2}$ in the X-ray band. However, since a fraction of the sub-mJy
objects may not be star-forming galaxies, these figures for the time
being should be regarded as upper limits.

Making use of our calibration of the X-ray luminosity as a SFR
indicator, it was also possible to rescale the X-ray luminosity
function into a cosmic star formation rate density. This definitely
assesses the possibility of using X-ray observations to obtain an
absorption-free estimate of the cosmic star formation history.
However, the deepest surveys available do not detect normal galaxies
beyond $z\sim 1.3$; the most interesting period of the cosmic history
(from $z\sim 2$ to $z\sim 4$, where it is still debated if the cosmic
star formation rate density declined or stayed constant) will remain
precluded to study until future X-ray telescopes, such as XEUS, will
be launched.

% M82

\label{sec:m82conc}

\smallskip
Our abundance analysis of the stellar and hot gas-phase components in
the nuclear region of M82 indicate an Iron abundance about half Solar
and an overall $\alpha-$enhancement by a factor between 2 and 3 with
respect to Iron.
These abundance patterns can be easily explained within a standard
nucleosynthesis scenario where the ISM is mainly enriched by SNe~II on
relative short timescales and with a star formation process occurring
in recursive bursts.

Oxygen behaves in a strange fashion. It is over-abundant in stars and
cold gas, similarly to the other $\alpha$-elements, while is
significantly under-abundant in the hot gas. Major calibrations and/or
modeling problems seem unlikely.  Hypernovae nucleosynthesis, dust
depletion, SN-Ia enrichment can somehow explain an Oxygen
under-abundance but other metals should follow the same pattern, while
they do not.  Somewhat exotic threshold effects and/or depletion
mechanisms preferentially affecting Oxygen could be at work in the
nuclear region of M82, but presently this issue remains controversial.
The situation will be probably clarified in the next future, since we
were recently granted a five-times-deeper reobservation of M82 with
\xmm.

%\addchap{Bibliography}

\addchap{List of publications}
\noindent{\it Refereed papers}
\begin{itemize}\item
{\bf P. Ranalli}, A. Comastri, G. Setti ``The 2-10 keV luminosity as a
Star Formation Rate indicator'', 2003, A\&A 399, 39

\item
L. Origlia, {\bf P. Ranalli}, A. Comastri, R. Maiolino ``Stellar and gaseous
abundances in M82'', 2004, ApJ 606, 862

\item
C. Norman, \dots, {\bf P. Ranalli}, et al.\ ``The X-ray derived cosmological
star formation history and the galaxy X-ray luminosity functions in
the Chandra Deep Fields North and South'', 2004, ApJ 607, 721

\end{itemize}

\noindent{\it Conference proceedings published on refereed journals}
\begin{itemize}\item
{\bf P. Ranalli} ``The faintest star forming galaxies'', 2003, Proc.\
Workshop ``X-ray surveys in the light of the new observatories'',
Santander. Astron.~Nachr. 324, 143
\end{itemize}

\noindent{\it Other conference proceedings}
\begin{itemize}
\item {\bf P. Ranalli}, A. Comastri, G. Setti ``The 2-10 keV
luminosity as a Star Formation Rate indicator'', Proc.\ Symp.\ ``New
visions of the X-ray universe in the XMM-Newton and Chandra era'', 2001,
ESTEC, Noordwijck (NL), {\em astro-ph}/0202241

\item {\bf P. Ranalli}, A. Comastri, G. Setti ``X-ray number counts of
  star forming galaxies'', Proc. Symp. ``Multiwavelength AGN
  Surveys'', 2003, Cozumel, Mexico, {\em astro-ph}/0404087

\item A. Comastri, {\bf P. Ranalli}, M. Brusa ``Beyond the X-ray
background with XEUS'', Proc. Workshop "XEUS - studying the evolution
of the hot universe", {\em astro-ph}/0211309

\item N. Cardiel, \dots, {\bf P. Ranalli} ``Comparing SFR
indicators from multiwavelength data in galaxies at intermediate
redshifts'' V Scientific Meeting of the ``Sociedad Espa\~nola de
Astronomia'', held in Toledo, September 9-13, 2002

\end{itemize}

\addchap{Colour plates}

\cleardoublepage

\begin{plate}[p]
%  \ifthenelse{\value{altaris}=1}{%
%      \includegraphics[width=\textwidth]{spitzer-m81-compo.ps}  
%    }{
\includegraphics[width=\textwidth]{spitzer-m81-compo-lowres.eps}
%  }
\caption{\spitzer\ observation of M81. The $3.6\mu$ emission is mainly
    due to red supergiant stars, while the emission at longer
    wavelenghts is progressively dominated by dust in \Hii regions.}
\label{plate:m81multiwave}
\end{plate}

%\MyProcessPlates

\begin{plate}[p]
%  \ifthenelse{\value{altaris}=1}{%
%      \includegraphics[width=\textwidth]{spitzer-m81-longw.ps}
%    }{
\includegraphics[width=\textwidth]{spitzer-m81-longw-lowres.eps}
%  }
\caption{\spitzer\ observation of M81. The $24\mu$ and $70\mu$ emission
    is mainly due to warm ($\sim 40$K) dust in \Hii regions, while the
    $160\mu$ emission traces a colder ($\sim 20$K) dust component.}
\label{plate:m81longw}
\end{plate}

%\MyProcessPlates

\begin{plate}[p]
%% %  {\centering
%% %  \ifthenelse{\value{altaris} = 1}{%
%% %      \includegraphics[width=0.98\textwidth]{antennaeROSAT-ASCA.eps}\\
%% %  }{
   \includegraphics[width=0.98\textwidth]{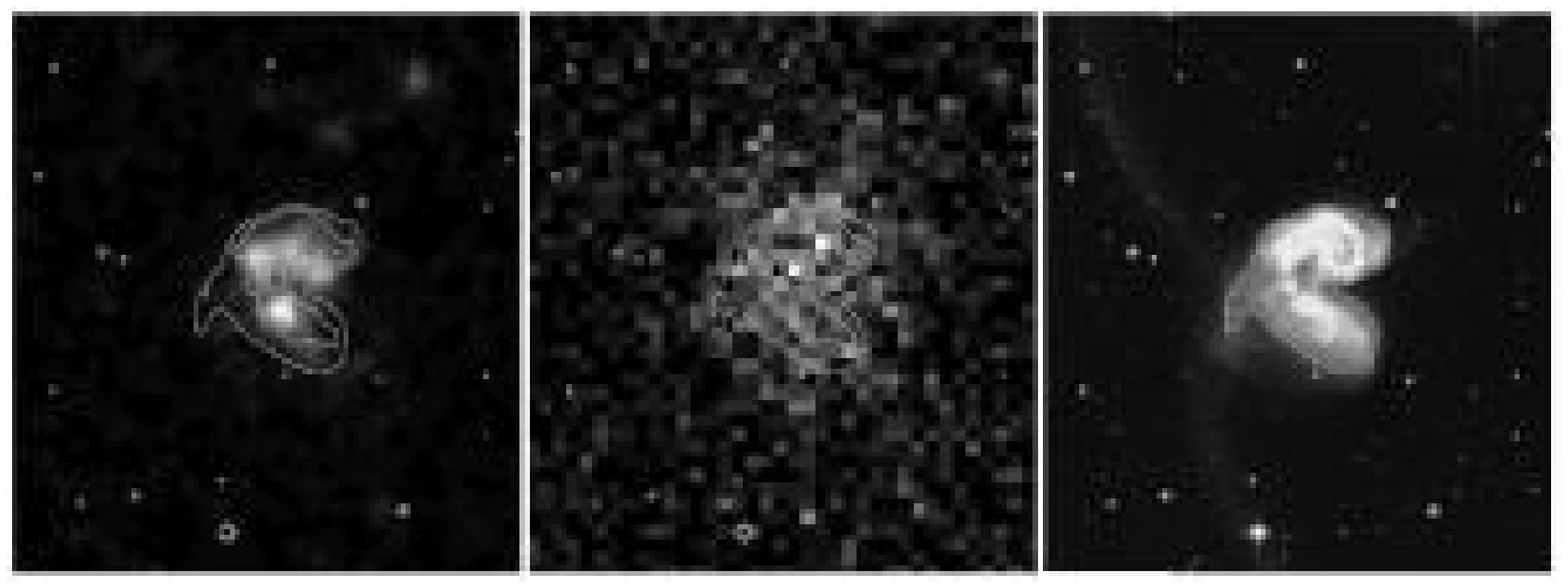}\\
%% %  }}
   %\caption{} 'sto comando non e' veramente necessario, serve
   % principalmente ad aumentare il contatore delle figure
   (a) X-ray emission from the Antennae, showing the low angular
   resolution of past X-ray telescopes. Right panel: optical image
   from the {\em Digitized Sky Survey}, with countours
   superimposed. Left panel: ROSAT PSPC image with optical contours
   superimposed. Middle panel: ASCA GIS image with optical contours
   superimposed. \\
   \vbox{\ }\\
   \centering
%% %  \ifthenelse{\value{altaris} = 1}{%
%% %      \includegraphics[height=0.53\textheight]{antennae_panel.ps}
%% %  }{
   \includegraphics[height=0.53\textheight]{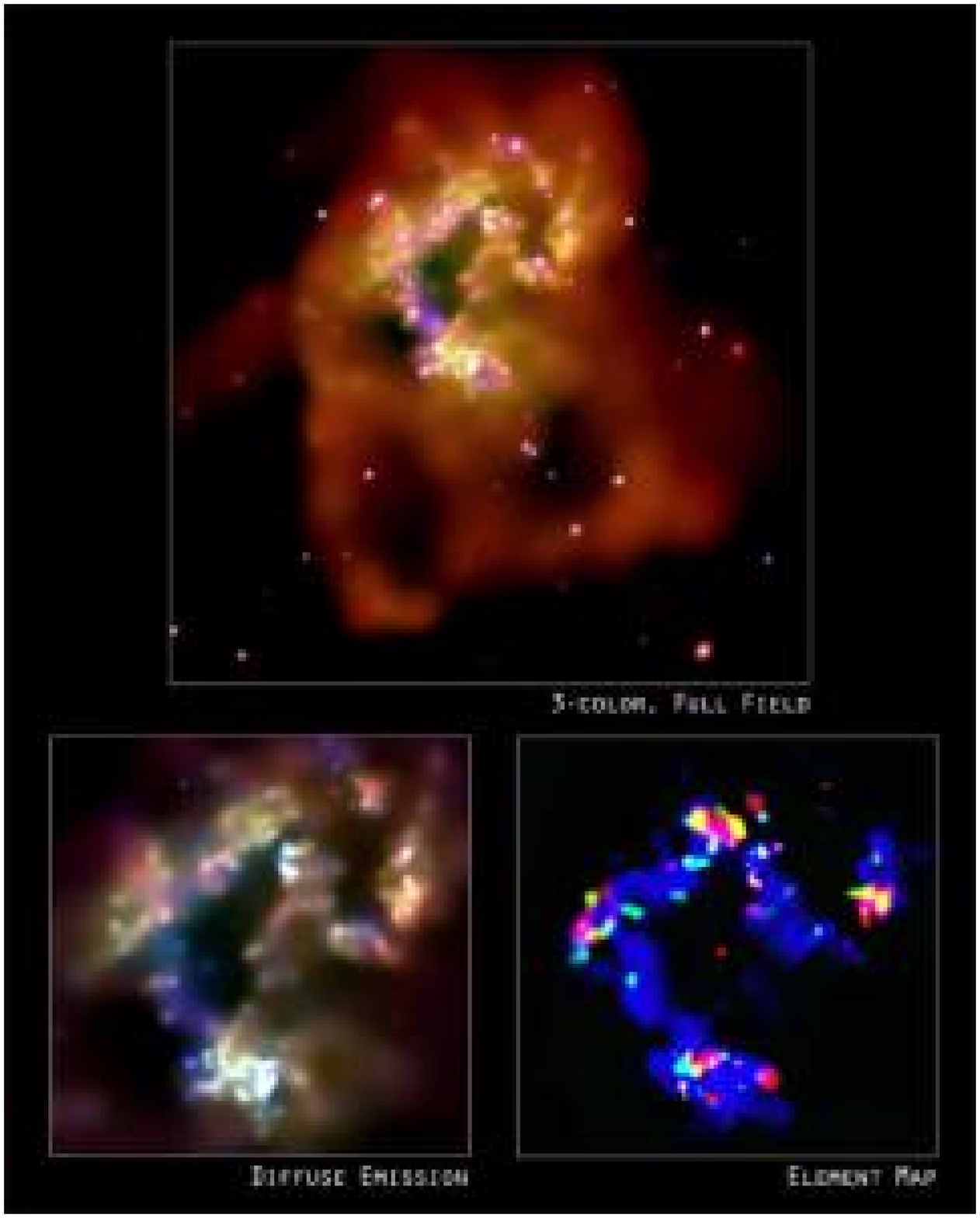}
%% %  }
   \caption{(b) The Antennae galaxy observed with \chandra, from
     \citet{fabbiano04}.  In the upper and lower left panels the colour
     coding represents the energy of the incoming photons (red: 0.5--1
     keV; green: 1--2 keV; blue: 2--7 keV). In the lower right panel,
     the color coding is referred to the different chemical elements
     present in the spatially resolved spectrum (red: Fe; green: Mg;
     blue: Si).  }
   \label{plate:antennaeX-ray}
\end{plate}

%\MyProcessPlates

\begin{plate}[p]
   \centering
  \ifthenelse{\value{altaris} = 1}{%
       \includegraphics[width=0.98\textwidth]{m82enhanced.ps}\\
  }{        \includegraphics[width=0.98\textwidth]{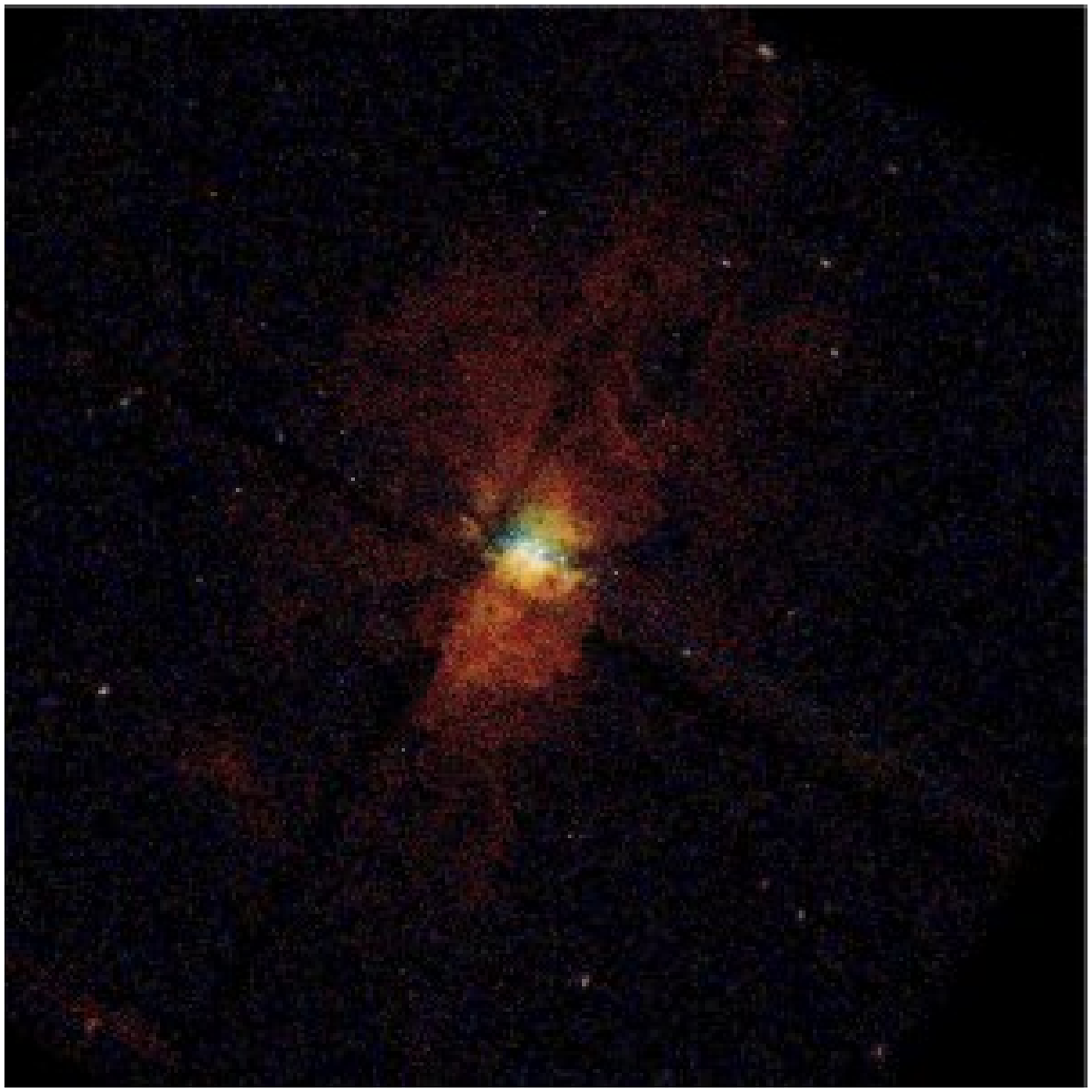}\\
  }
  \caption{\chandra\ image of M82.  The colour
    coding represents the energy of the incoming photons (red: 0.5--1
    keV; green: 1--2 keV; blue: 2--7 keV). The field size of the upper
    panel is $\sim 15\arcmin\times 15\arcmin$.  } 
  \label{plate:M82chandra}
\end{plate}

\begin{plate}[p]
   \centering
  \ifthenelse{\value{altaris} = 1}{%
       \includegraphics[height=0.45\textheight]{m82true_smoothed.ps}
       \includegraphics[height=0.45\textheight]{m82colimg-mos12.ps}\\
  }{   \includegraphics[height=0.45\textheight]{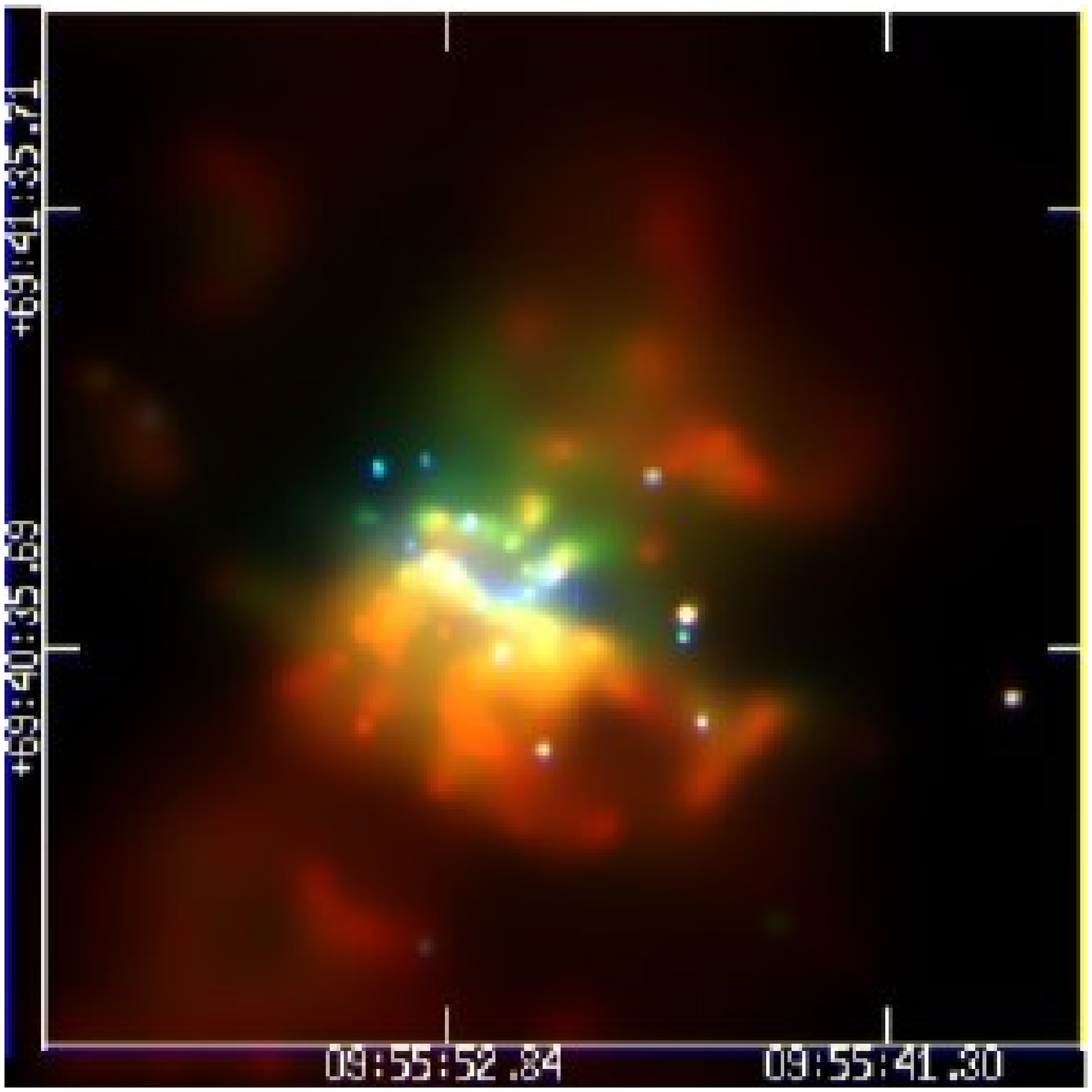}
       \includegraphics[height=0.45\textheight]{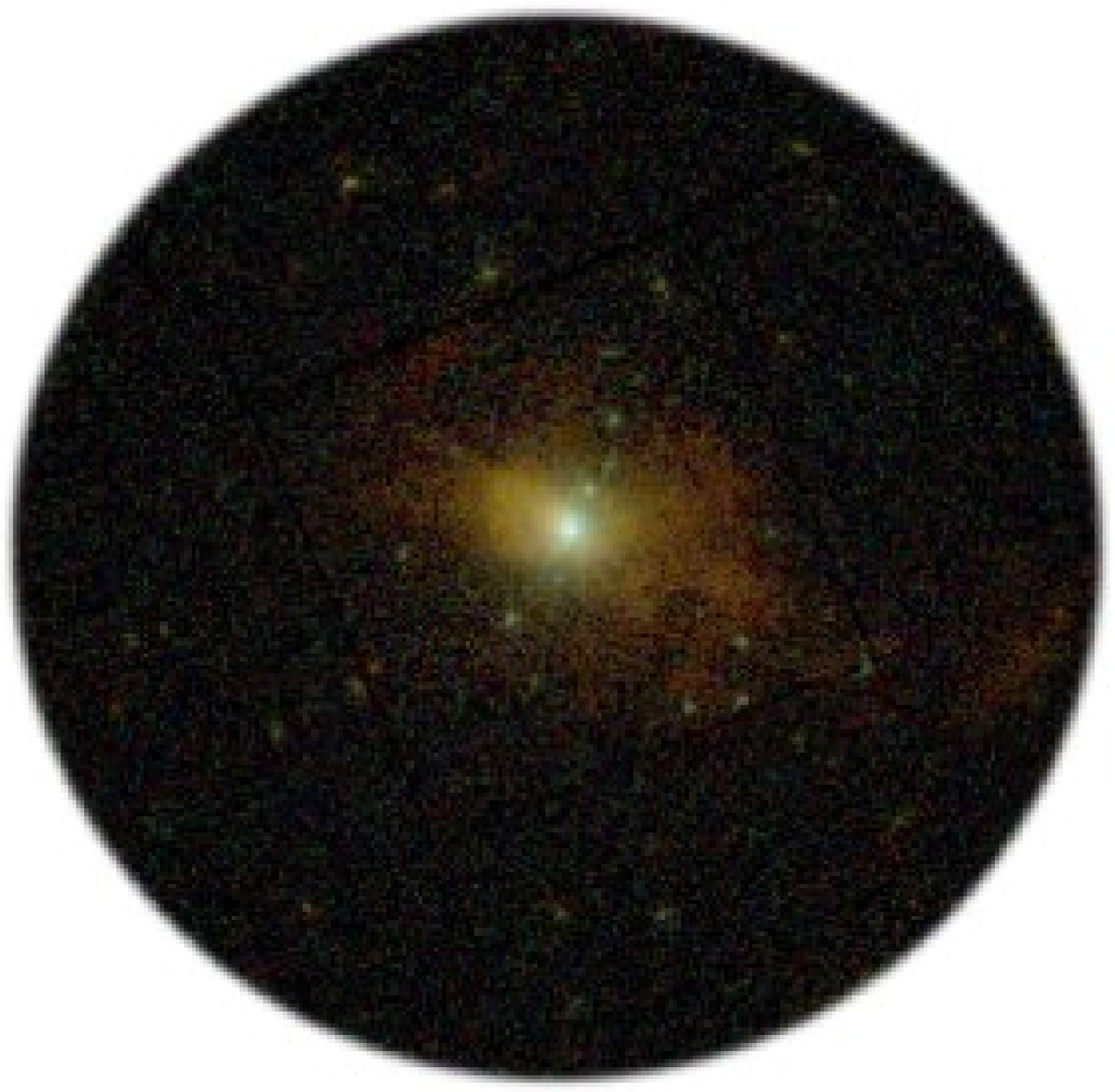}\\
  }
  \caption{\chandra\ (upper panel) and \xmm\ (lower panes) images of
    M82. The (smoothed) image in the upper panel shown the inner
    region of M82. The lower panel has a field diameter of $\sim
    20\arcmin$ and shows the different angular resolution of \xmm\ 
    (note that the image is shown with a different position angle).
    The colour coding represents the energy of the incoming photons
    (red: 0.5--1 keV; green: 1--2 keV; blue: 2--7 keV).}
  \label{plate:M82chandra}
\end{plate}

\begin{plate}[p]
  \centering
  \ifthenelse{\value{altaris} = 1}{%
     \includegraphics[height=0.55\textheight]{ngc253_ca.ps}\\
 }{  \includegraphics[height=0.55\textheight]{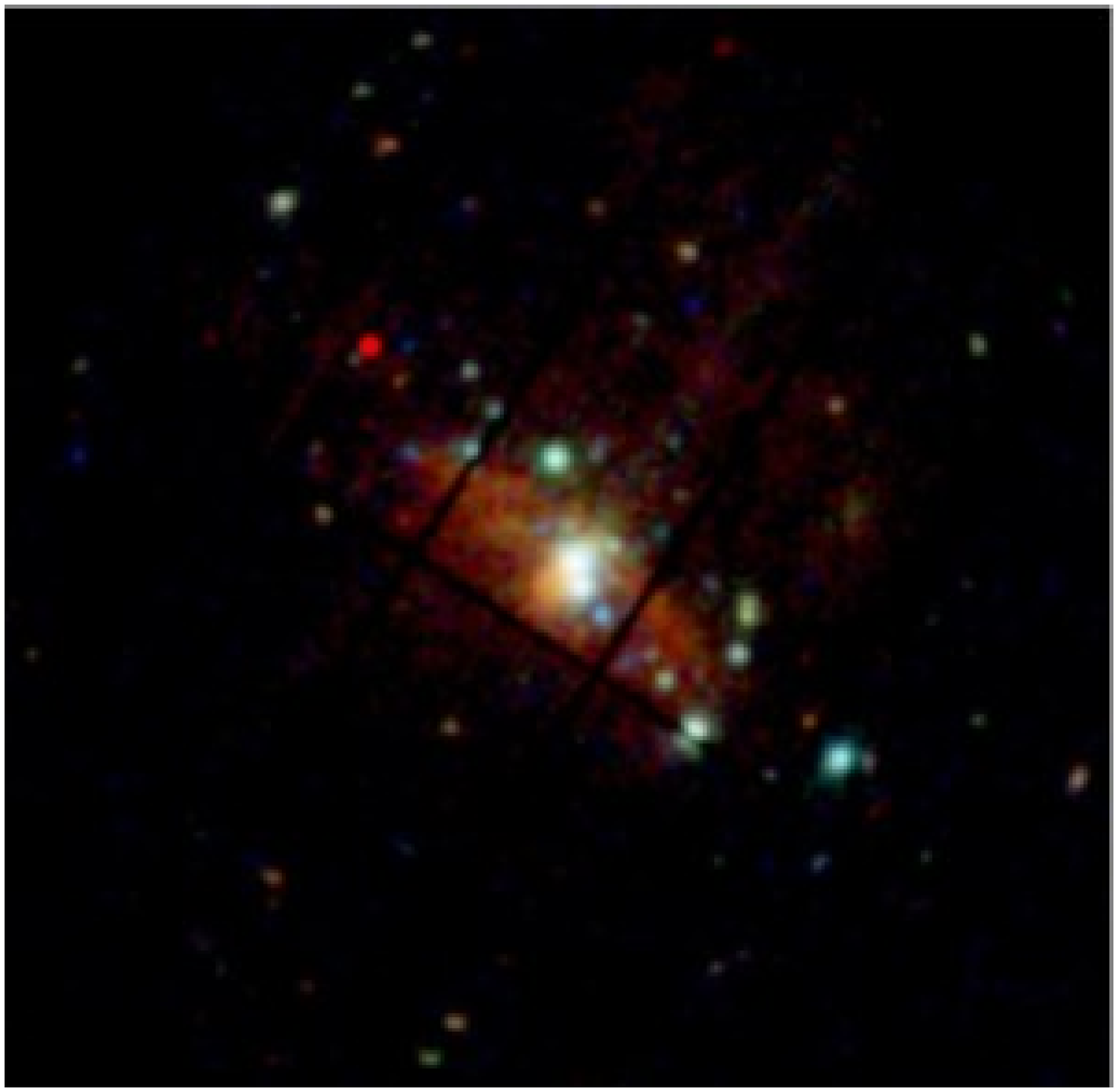}\\
 }
   (a) \xmm\ image of NGC253. The central starburst is clearly visible
     (the large white spot at the centre of the image). Also the
     galactic disc is visible (red diffuse emission). Colour coding as
     in the previous plates.\bigskip\\
%   \vbox{\ }\\
%   \centering
%   \ifthenelse{\value{altaris} = 1}{%
%       \includegraphics[height=0.6\textheight]{antennae_panel.eps}
%   }{  \includegraphics[height=0.6\textheight]{antennae_3panel_m_lowres.eps}
%   }
  \ifthenelse{\value{altaris} = 1}{%
     \includegraphics[width=0.8\textwidth,bb=51 289 492 512,clip]{strickland.fig1.eps}\\
 }{  \includegraphics[width=0.8\textwidth,bb=31 121 197 206,clip]{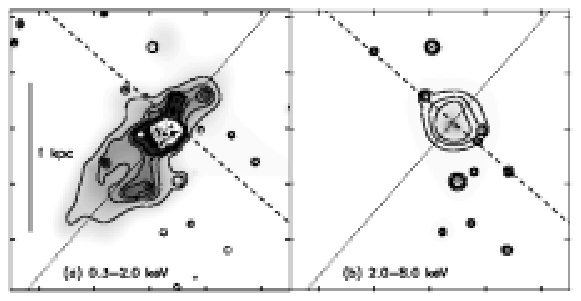}\\
  }
  \caption{(b) \chandra\ images of NGC253, from \citet{strickland00},
    showing the different morphology of the soft and hard X-ray
    emission. The galaxy outflow is clearly visible in the left panel,
    while the hard X-ray emission is dominated by sources concentrated
    in the galaxy centre.}
  \label{plate:NGC253xmmchandra}
\end{plate}

%\MyProcessPlates

\begin{plate}[p]    % 3256
  \begin{center}
  \ifthenelse{\value{altaris} = 1}{%
       \includegraphics[width=0.8\textwidth]{h3889f7.ps}
  }{   \includegraphics[width=0.8\textwidth]{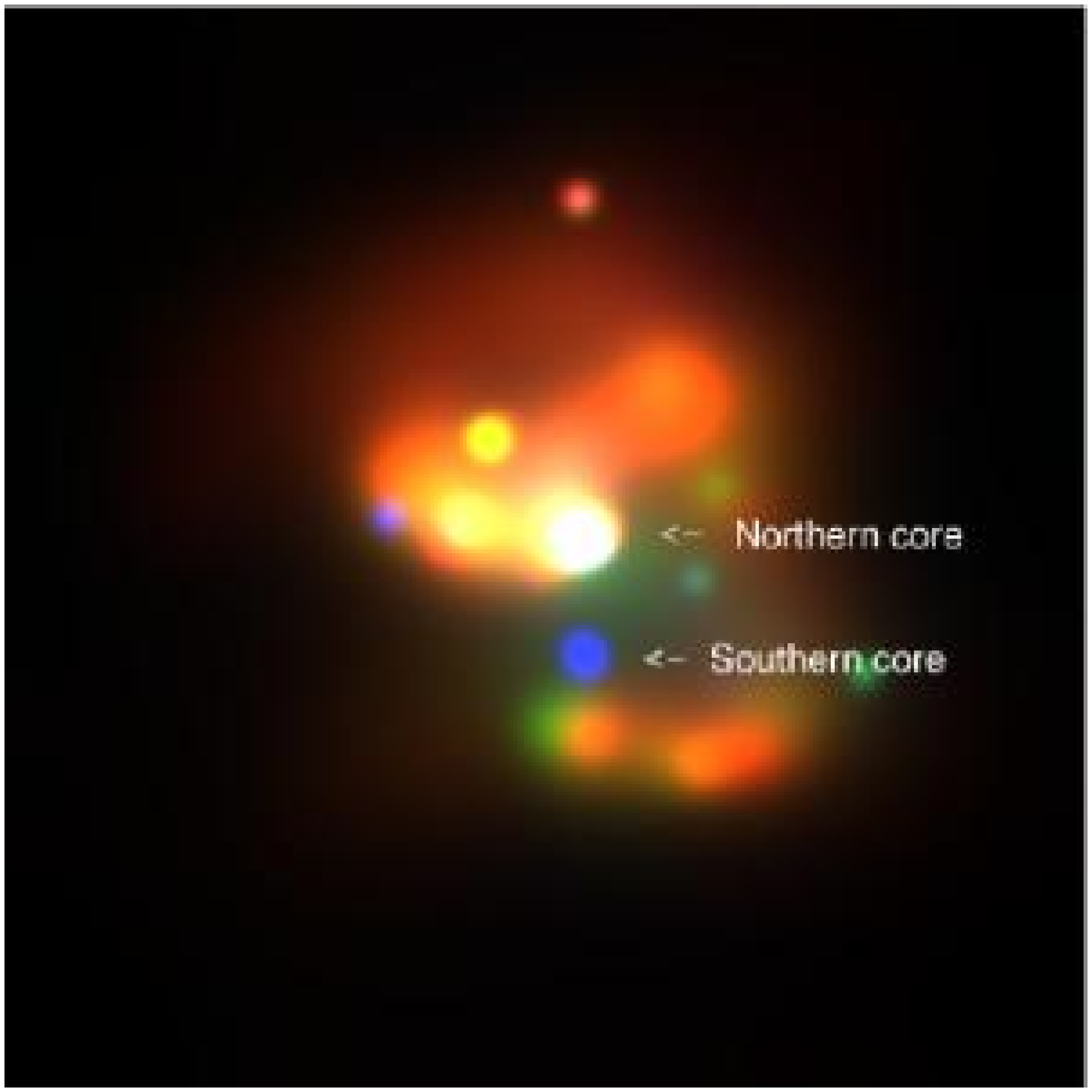}
  }
%     \framebox[.5\textwidth]{ figura }
  \end{center}
  \caption { True color, smoothed \chandra\ image of the centre of NGC 3256
  (the actual colour is the sum of red, green and blue, the intensity
  of each one representing the flux in the 0.3--1.0 keV, 1.0--2.0 keV
  and 2.0--8 keV respectively); north is up, east
  is left; the distance between the two cores is $\sim 5\arcsec$.
   The two cores are the brightest sources both at 1.4 GHz and in the
  2--10 keV band, but since the southern one lies behind a dust lane,
  its soft X-ray emission is completely absorbed. While the northern
  core follows both the radio/soft X-ray and radio/hard X-ray relations
  (Eq.~\ref{eq:radiox},\ref{eq:radiox2}), the southern one only follows
  the radio/hard X-ray relation.
  \label{fig_3256} }
\end{plate}
%\MyProcessPlates

\end{document}